\documentclass[review]{elsarticle}

\usepackage{amsmath,amssymb,amsfonts}
\usepackage{graphicx}
\usepackage{booktabs}
\usepackage{siunitx}
\usepackage{subcaption}
\usepackage{tikz}
\usetikzlibrary{arrows.meta,calc,positioning,decorations.pathreplacing}
\usepackage{hyperref}

\usepackage{placeins}
\graphicspath{{Figs/}}

\journal{Computers \& Fluids}

\begin{document}

\begin{frontmatter}

\title{A Riemann-invariant contact-capturing scheme with WENO
reconstruction and RH-motivated dual-speed dissipation for
compressible two-phase mixture flows}

\author[aff1]{Ankit Ruhi}

\author[aff2]{Ramesh Kolluru\corref{cor1}}
\ead{rameshkolluru43@gmail.com}
\ead{kollurur@alum.iisc.ac.in}
\cortext[cor1]{Corresponding author}

\author[aff3]{Shrinath KS}

\affiliation[aff1]{organization={Department of Mathematics and Statistics,
Dr.\ Harisingh Gour Vishwavidyalaya},
city={Sagar},
country={India}}

\affiliation[aff2]{organization={Independent researcher; Senior CFD consultant to BQP},
country={India}}

\affiliation[aff3]{organization={Chief Manager -- Aerodynamics, RWRDC, HAL},
country={India}}

\begin{abstract}
This work develops and benchmarks centred (Riemann-solver-free) finite-volume fluxes for a hyperbolic two-phase mixture model with velocity non-equilibrium. The baseline method combines a Riemann-invariant contact-capturing algorithm (RICCA) with dual-speed dissipation and primitive-variable WENO3/WENO5 reconstruction, avoiding full eigen-decomposition while retaining a conservative update.

An RH-motivated companion flux (RICCA--RH) is introduced, in which the acoustic weight in $\alpha = |V_n| + Ba$ is controlled by a binary pressure/Rankine--Hugoniot switch. On one-dimensional gas--magma Riemann suite, RICCA--RH tracks legacy RICCA and is sharper than first-order FORCE/Lax--Friedrichs at material contacts; a linearised spectral study clarifies how its dual-speed dissipation differs from FORCE. With WENO5, RICCA, RICCA--RH and FORCE produce essentially identical smooth profiles, indicating that most differences are confined to first-order contact metrics.

In two dimensions (gas--liquid benchmarks), the production flux implemented on CPU and Apple Metal reproduces expansion-tube, collision and forward-facing-step patterns and yields intermediate cut-based roughness between RICCA and FORCE. High order is obtained using an approximate face-normal acoustic characteristic reconstruction, a pragmatic multi-variable hybrid sensor, and a Zhang--Shu-style invariant-domain face limiter on $(\rho,\alpha,c)$. An under-dissipative soft-switch variant that rings severely is also documented, and a resident 2D Metal kernel on Apple Silicon is reported to deliver up to $3.9\times$ wall-clock speedup on completed CPU/Metal comparisons (peak on the $101\times101$ expansion tube).
\end{abstract}


\end{frontmatter}

%==============================================================================
\section{Introduction}
\label{sec:intro}

Compressible multiphase flows with velocity non-equilibrium remain demanding for both modelling and numerics \cite{Zeidan2019, Zeidan2014,Romenski2007}. Mixture formulations that retain a relative-velocity equation while preserving hyperbolicity and a fully conservative flux are attractive because they admit standard finite-volume treatments without non-conservative products. Centred schemes that avoid a complete eigenstructure decomposition are particularly suitable for such models and are practical for CFD because they remain conservative while mapping cleanly to GPU-friendly, stencil-based kernels. The Riemann-invariant based contact-capturing algorithm (RICCA \cite{KolluruRICCA,KolluruZeidan1D} belongs to this class: artificial dissipation is built from generalised Riemann invariants with separate contact and acoustic wave speeds, and a pressure sensor that increases dissipation only where needed. FORCE \cite{ToroFORCE,Zeidan2019} provides a well-established centred baseline.

The present paper places \textbf{RICCA with WENO reconstruction} at the centre of the methodology and develops an RH-motivated dual-speed companion (\textbf{RICCA--RH}) that retains the same contact/acoustic equation split while basing the acoustic weight on a pressure/Rankine--Hugoniot switch.

\paragraph{Contributions and scope.}
The main contributions  are:
\begin{enumerate}
\item A centred, Riemann-solver-free mixture flux (RICCA--WENO) with
      \emph{per-equation} contact and acoustic dissipation tailored to
      Romenski-type mixtures, benchmarked against FORCE and against upwind
      HLLC/Roe solvers on identical meshes.
\item RICCA--RH: an RH-motivated dual-speed dissipation law
      $\alpha=\lvert V_n\rvert+Ba$ with $B=\kappa\sigma$ on contact equations and
      $B=\sigma$ on acoustic/slip equations, driven by a binary pressure jump
      and normalised RH residual; production ($\kappa=1$) and an oscillatory
      soft-$\sigma$ prototype are reported in 1D and 2D
      (Sections~\ref{sec:ricca-rh}, \ref{sec:1d-rh}, \ref{sec:2d-rh}).
\item A pragmatic multi-variable primitive \emph{combo} hybrid sensor that
      reduces residual ringing relative to a pressure-only blend on the
      reported suite, with modest 4--6\% roughness gains on the expansion tube
      and consistent improvement on a three-problem metric suite
      (Section~\ref{sec:sensor-discussion}); an explicit tuning guideline is
      provided, with honest acknowledgement that the sensor is empirical and
      that users may need to retune $(C,w)$ for other regimes.
\item A linearised dispersion/dissipation analysis of RICCA versus FORCE,
      together with a positivity/admissibility strategy (hybrid clipping, FORCE
      Lax--Friedrichs fallback, primitive guards) that keeps $\rho$, $\alpha$
      and $c$ in physical bounds under the reported CFLs.
\item A Metal GPU implementation delivering up to $3.9\times$ wall-clock
      speedup on completed 2D CPU/Metal comparisons (Apple M2 laptop;
      $3.93\times$ on the $101\times101$ expansion tube FORCE FO,
      $2.86\times$ on the coarse FFS), including RICCA--RH; RICCA--WENO5
      on the expansion tube is currently timed on Metal alone.
\end{enumerate}
Primitive WENO3/WENO5 with the combo shock hybrid (default in 1D and 2D) raises
spatial accuracy while controlling oscillations near discontinuities and
contacts.
The scheme is assessed on the complete one-dimensional gas--magma Riemann suite
of \cite{Zeidan2014} and the three two-dimensional gas--liquid benchmarks of
\cite{Zeidan2019}, with FORCE as the centred reference on identical meshes.
A smooth manufactured advection problem confirms that WENO5 recovers high-order
accuracy away from shocks, while discontinuous Riemann problems remain globally
first-order.

\paragraph{Positioning as a baseline methodology.}
We position the proposed algorithms as a centred, eigenstructure-free
baseline for Romenski-type mixtures: simple enough to reproduce and port to
GPU, but equipped with modern stabilisation (approximate face-normal
characteristic WENO, a pragmatic multi-variable hybrid sensor, and an
invariant-domain face limiter on $(\rho,\alpha,c)$).
The model and benchmark matrix follow Zeidan et~al.\ to enable direct
comparisons with FORCE and with upwind HLLC/Roe fluxes
(Sections~\ref{sec:hllc}--\ref{sec:1d-upwind}).
Scope is limited to the homogeneous system (no relaxation/source terms) and to
the reported Zeidan-style tests; in 2D the forward-facing step uses an
equal-EOS closure and therefore reduces to a single-velocity Euler-like
mixture ($c\equiv\alpha$, $\mathbf{u}_r=\mathbf{0}$).
RICCA--RH and the combo hybrid remain heuristic, suite-tuned constructions; we
report both the robust production choice and an under-dissipative soft-switch
variant to make the design trade-offs explicit.

\paragraph{Related work.}
The present work sits within a broader body of research on Riemann-solver-free and
high-order schemes for compressible multiphase flow. Within the same Romenski/SHTC
hyperbolic two-phase model family, exact and numerical (Godunov-type) solutions of
the associated Riemann problem have recently been derived \citep{ThienRomenskiDumbser2022},
providing a natural benchmark against which RICCA's centred, Riemann-solver-free
approach can be assessed. The idea of avoiding an explicit Riemann solver for
two-phase systems is not unique to RICCA: a KFVS-type central-upwind scheme for a
compressible two-phase model was developed with the same motivation of algorithmic
simplicity and robustness \citep{AhmedCentralUpwind2015}.

On the reconstruction side, several high-order WENO strategies for two-phase and
multicomponent flows have been proposed, including incremental-stencil WENO
\citep{WangIncrementalWENO2018}, ADER--WENO treatments of non-conservative
Baer--Nunziato-type systems \citep{DumbserWENOAMR2014}, and finite-volume
WENO/HLLC multicomponent schemes \citep{ColoniusCoralic2014}. Closest in spirit to
the positivity/robustness goals of RICCA--RH's dual-speed dissipation switch are
positivity- and boundedness-preserving high-order WCNS schemes for the five-equation
gas--liquid model \citep{WongPositivity2021}, and non-equilibrium relative-velocity
Riemann solvers with internal reconstruction (RSIR) \citep{CarmouzeRSIR2020}.
Relaxation-based alternatives to the mixture formulation used here are discussed in
\citep{PelantiRelaxation2022}.

On the computational side, open-source multiphase solvers such as MFC
\citep{BryngelsonMFC2021} have demonstrated GPU acceleration across multiple vendor
platforms via OpenACC, including recent cross-platform (AMD and NVIDIA) offloading
results \citep{WilfongOpenACC2024}. This contrasts with the single-platform (Apple
Metal) GPU implementation reported here, which we regard as a first demonstration of
feasibility rather than a claim of portable performance; extending RICCA--RH to
CUDA/HIP backends for direct cross-platform comparison is left to future work.

The paper is organised as follows.
Section~\ref{sec:ge} states the 1D/2D mixture equations.
Section~\ref{sec:nm} presents RICCA, the RICCA--RH dual-speed law
(Section~\ref{sec:ricca-rh}), HLLC and Roe upwind fluxes
(Sections~\ref{sec:hllc}--\ref{sec:roe}), the linearised spectral comparison
with FORCE (Section~\ref{sec:dispersion}), the positivity strategy
(Section~\ref{sec:positivity}), WENO hybrid sensors and time stepping.
Section~\ref{sec:1d} discusses the 1D gas--magma suite, the sensor study and
the RICCA--RH comparison (including the oscillatory prototype).
Section~\ref{sec:2d} presents the 2D gas--liquid benchmarks and the
RICCA--RH comparison.
Section~\ref{sec:conv} reports the smooth MMS convergence study and
discontinuous self-convergence.
Section~\ref{sec:perf} summarises the Metal GPU implementation.
Section~\ref{sec:disc} discusses the hybrid sensor design (scope, tuning and
robustness).
Conclusions and limitations are given in Section~\ref{sec:conc}.

%==============================================================================
\section{Governing equations}
\label{sec:ge}

We employ the thermodynamically compatible compressible mixture model of Romenski--type used by Zeidan and co-workers \cite{Romenski2007,Zeidan2014,Zeidan2019}.
Two phases share a hyperbolic, fully conservative flux; relative velocity is retained as an independent field.

\subsection{Mixture variables}

Subscripts $1$ and $2$ denote the condensed (liquid/magma) and gas phases.
With gas volume fraction $\alpha=\alpha_2$ ($\alpha_1=1-\alpha$),
\begin{align}
\rho &= \alpha\rho_2+(1-\alpha)\rho_1
&&\text{(mixture density)},\\
c &= \frac{\alpha\rho_2}{\rho}
&&\text{(gas mass fraction)},\\
\mathbf{u}
&=
\frac{(1-\alpha)\rho_1\mathbf{u}_1+\alpha\rho_2\mathbf{u}_2}{\rho}
&&\text{(mixture velocity)},\\
\mathbf{u}_r &= \mathbf{u}_2-\mathbf{u}_1
&&\text{(relative velocity)},\\
S &= \frac{(1-\alpha)\rho_1 S_1+\alpha\rho_2 S_2}{\rho}
&&\text{(mixture entropy)}.
\end{align}
Mixture pressure $P=P(\rho,c,\alpha,S)$ and mixture sound speed $a$ follow
from the dual stiffened/ideal EOS closures below.
For reproducibility we also make explicit the mixture specific internal energy
used to evaluate the thermodynamic potential $\Psi$.
Following \cite{Zeidan2014,Zeidan2019}, we take a single entropy field
($S_1=S_2=S$) and a mass-fraction mixture energy
\begin{equation}
 e(\rho,c,\alpha,S)=(1-c)\,e_1(\rho_1,S)+c\,e_2(\rho_2,S),
 \qquad
 \rho_1=\frac{\rho(1-c)}{1-\alpha},\quad
 \rho_2=\frac{\rho c}{\alpha}.
\end{equation}
The phase internal energies consistent with the stated pressure closures
$P_1=A_1\rho_1^{\gamma_1}e^{S/C_{1v}}-A_0$ and
$P_2=A_2\rho_2^{\gamma_2}e^{S/C_{2v}}$ are
\begin{equation}
 e_1(\rho_1,S)=\frac{A_1}{\gamma_1-1}\,\rho_1^{\gamma_1-1}e^{S/C_{1v}},
 \qquad
 e_2(\rho_2,S)=\frac{A_2}{\gamma_2-1}\,\rho_2^{\gamma_2-1}e^{S/C_{2v}},
\end{equation}
so that $P_1=(\gamma_1-1)\rho_1 e_1-A_0$ and $P_2=(\gamma_2-1)\rho_2 e_2$.
The relative-velocity chemical potential
\begin{equation}
\Psi
=
\frac{\partial e}{\partial c}\bigg\rvert_{\rho,\alpha,S}
\end{equation}
appears in the $\mathbf{u}_r$ flux.

\subsection{One-dimensional system}
\label{sec:ge-1d}

In one space dimension the conserved vector is
\begin{equation}
\mathbf{U}
=
\bigl(
\rho,\;
\rho u,\;
\rho\alpha,\;
\rho c,\;
u_r,\;
\rho S
\bigr)^{\!\mathsf{T}},
\label{eq:U1d}
\end{equation}
and the balance law (no relaxation sources in this paper) reads
\begin{equation}
\partial_t\mathbf{U}+\partial_x\mathbf{F}(\mathbf{U})=\mathbf{0},
\label{eq:cons1d}
\end{equation}
with flux
\begin{equation}
\mathbf{F}
=
\begin{pmatrix}
\rho u \\
\rho u^{2}+P+\rho\,c(1-c)\,u_r^{2} \\
\rho u\,\alpha \\
\rho u\,c+\rho\,c(1-c)\,u_r \\
u\,u_r+(1-2c)\tfrac12 u_r^{2}+\Psi \\
\rho u\,S
\end{pmatrix}.
\label{eq:F1d}
\end{equation}
\paragraph{Term-by-term interpretation.}
\begin{itemize}
\item $\rho u$: mixture mass flux.
\item $\rho u^{2}+P$: mixture momentum flux (advection + pressure);
$\rho\,c(1-c)\,u_r^{2}$ is the Reynolds-like stress from phase slip.
\item $\rho u\,\alpha$: advection of gas volume fraction.
\item $\rho u\,c$: advection of gas mass;
$\rho\,c(1-c)\,u_r$ is the relative-mass (drift) flux.
\item $u\,u_r+(1-2c)\tfrac12 u_r^{2}+\Psi$: evolution flux for relative
velocity (advective, quadratic slip, and thermodynamic $\Psi$ contributions).
\item $\rho u\,S$: advection of mixture entropy (isentropic in the absence of
sources; shocks numerically introduce entropy production through dissipation).
\end{itemize}

%\paragraph{Note on the forward-facing step (FFS).}
%Under the equal-EOS limit with $c\equiv\alpha$ and $\mathbf{u}_r=\mathbf{0}$,
%the relative-velocity and drift-flux terms vanish, reducing the system to a
%single-velocity Euler-like mixture (Section~\ref{sec:2d-ffs}).
%
\subsection{Two-dimensional system}
\label{sec:ge-2d}

In two dimensions,
\begin{equation}
\mathbf{U}
=
\bigl(
\rho,\;
\rho u,\;
\rho v,\;
\rho\alpha,\;
\rho c,\;
u_r,\;
v_r,\;
\rho S
\bigr)^{\!\mathsf{T}},
\label{eq:U2d}
\end{equation}
and
\begin{equation}
\partial_t\mathbf{U}
+
\partial_x\mathbf{F}_x(\mathbf{U})
+
\partial_y\mathbf{F}_y(\mathbf{U})
=
\mathbf{0}.
\label{eq:cons2d}
\end{equation}
With $\mathbf{u}=(u,v)$, $\mathbf{u}_r=(u_r,v_r)$, the Cartesian fluxes are
\begin{align}
\mathbf{F}_x
&=
\begin{pmatrix}
\rho u \\
\rho u^{2}+P+\rho\,c(1-c)\,u_r^{2} \\
\rho uv+\rho\,c(1-c)\,u_r v_r \\
\rho u\,\alpha \\
\rho u\,c+\rho\,c(1-c)\,u_r \\
u\,u_r+(1-2c)\tfrac12 u_r^{2}+\Psi \\
u\,v_r+v\,u_r+(1-2c)\,u_r v_r \\
\rho u\,S
\end{pmatrix},
&
\mathbf{F}_y
&=
\begin{pmatrix}
\rho v \\
\rho vu+\rho\,c(1-c)\,v_r u_r \\
\rho v^{2}+P+\rho\,c(1-c)\,v_r^{2} \\
\rho v\,\alpha \\
\rho v\,c+\rho\,c(1-c)\,v_r \\
v\,u_r+u\,v_r+(1-2c)\,u_r v_r \\
v\,v_r+(1-2c)\tfrac12 v_r^{2}+\Psi \\
\rho v\,S
\end{pmatrix}.
\label{eq:F2d}
\end{align}
On a face with outward unit normal $\mathbf{n}=(n_x,n_y)$ the normal flux
uses $V_n=\mathbf{u}\cdot\mathbf{n}$ and
$V_{rn}=\mathbf{u}_r\cdot\mathbf{n}$;
pressure contributes $P\mathbf{n}$ to momentum and $\Psi\mathbf{n}$ to the
relative-velocity equations.
\paragraph{Term-by-term interpretation.}
The meaning of each row mirrors the 1D case, extended by transverse
advection: mixture mass $(\rho V_n)$; momentum with pressure and
$\rho\,c(1-c)(\mathbf{u}_r\otimes\mathbf{u}_r)$ stresses; volume-fraction and
mass-fraction advection with drift flux
$\rho\,c(1-c)V_{rn}$; relative-velocity transport with quadratic slip and
$\Psi$; entropy advection $\rho S V_n$.
All 2D runs in this paper omit relative-velocity relaxation sources, as in
\cite{Zeidan2019}.

\subsection{Equations of state}

\paragraph{Gas--magma (1D suite).}
Following \cite{Zeidan2014},
\begin{equation}
\gamma_1=2.35,\quad
\gamma_2=1.43,\quad
A_2=8,\quad
A_0=10^5,\quad
A_1=1,\quad
C_{1v}=1815,\quad
C_{2v}=1039.86.
\end{equation}
Phase pressures are of stiffened/ideal form
$P_1=A_1\rho_1^{\gamma_1}\mathrm{e}^{S_1/C_{1v}}-A_0$,
$P_2=A_2\rho_2^{\gamma_2}\mathrm{e}^{S_2/C_{2v}}$,
assembled into the mixture $P(\rho,c,\alpha,S)$ used by the solver.

\paragraph{Gas--liquid equal EOS (2D CFD 2019 suite).}
\begin{equation}
\gamma_1=\gamma_2=1.4,\quad
A_1=A_2=1,\quad
A_0=0,\quad
C_{1v}=C_{2v}=1.
\end{equation}
For the forward-facing step the equal-EOS limit $c\equiv\alpha$,
$\mathbf{u}_r=\mathbf{0}$ reduces the system to a single-velocity Euler mixture
(no $\Psi$ / slip fluxes).

%==============================================================================
\section{Numerical method: RICCA, RICCA--RH, upwind solvers and WENO}
\label{sec:nm}

\subsection{Finite-volume stencils}
\label{sec:stencils}

\begin{figure}[htbp]
\centering
\begin{subfigure}[t]{\textwidth}
\centering
\begin{tikzpicture}[
  cell/.style={draw, minimum width=1.35cm, minimum height=0.85cm,
               fill=blue!8},
  flux/.style={-{Stealth[length=3mm]}, thick, red!70!black},
  lbl/.style={font=\footnotesize}
]
\node[cell] (im1) at (-2.6,0) {};
\node[cell] (i)   at (0,0) {};
\node[cell] (ip1) at (2.6,0) {};
\node[lbl, below=1pt] at (im1.south) {$i{-}1$};
\node[lbl, below=1pt] at (i.south) {$i$};
\node[lbl, below=1pt] at (ip1.south) {$i{+}1$};
\node[lbl] at (i.center) {$\mathbf{U}_i^n$};
\draw[flux] ($(im1.east)!0.5!(i.west)+(0,0.22)$) --
            ++(0.55,0) node[above,lbl] {$\mathbf{F}_{i-\frac12}$};
\draw[flux] ($(i.east)!0.5!(ip1.west)+(0,0.22)$) --
            ++(0.55,0) node[above,lbl] {$\mathbf{F}_{i+\frac12}$};
\end{tikzpicture}
\caption{Finite-volume update: $\mathbf{U}_i$ advances by the net flux through
the left and right faces.}
\label{fig:stencil-fvm}
\end{subfigure}
\begin{subfigure}[t]{\textwidth}
\centering
\begin{tikzpicture}[
  cell/.style={draw, minimum width=1.05cm, minimum height=0.75cm},
  active/.style={cell, fill=green!15},
  target/.style={cell, fill=orange!20, thick},
  lbl/.style={font=\footnotesize},
  brace/.style={decorate, decoration={brace, amplitude=5pt, mirror}}
]
\foreach \idx/\lab/\sty in {
  -2/{i{-}2}/active, -1/{i{-}1}/active, 0/{i}/target,
   1/{i{+}1}/active,  2/{i{+}2}/active}
{
  \node[\sty] (c\idx) at (1.2*\idx, 0) {};
  \node[lbl, below=1pt] at (c\idx.south) {$\lab$};
}
\draw[thick, red!70!black, dashed]
  ($(c0.east)+(0,-0.5)$) -- ++(0,1.35)
  node[above, lbl, black] {$i{+}\tfrac12$};
\draw[brace] (c-2.south west) ++(0,-0.28) --
  node[below=3pt,lbl] {5-cell WENO5 stencil}
  ($(c2.south east)+(0,-0.28)$);
\end{tikzpicture}
\caption{WENO5 reconstruction: five cells about~$i$ supply left-/right-biased
polynomials for the face $i{+}\tfrac12$.}
\label{fig:stencil-weno5}
\end{subfigure}
\begin{subfigure}[t]{\textwidth}
\centering
\begin{tikzpicture}[
  state/.style={draw, rounded corners=2pt, minimum width=1.55cm,
                minimum height=0.95cm, fill=blue!10},
  lbl/.style={font=\footnotesize},
  arr/.style={-{Stealth[length=2.5mm]}, thick}
]
\node[state] (L) at (-2.1,0) {$\mathbf{U}_L,\mathbf{F}_L$};
\node[state] (R) at ( 2.1,0) {$\mathbf{U}_R,\mathbf{F}_R$};
\draw[very thick, red!70!black] (0,-0.85) -- (0,0.85)
  node[above, lbl, black] {face};
\draw[arr, blue!60!black] (L.east) -- ++( 0.5,0);
\draw[arr, blue!60!black] (R.west) -- ++(-0.5,0);
\node[lbl, text width=5.2cm, align=center] at (0,-1.55)
  {$\mathbf{F}^{\mathrm{RICCA}}=\tfrac12(\mathbf{F}_L{+}\mathbf{F}_R)
   -\tfrac12\boldsymbol{\alpha}(\mathbf{U}_R{-}\mathbf{U}_L)$};
\node[lbl] at (0,-2.15)
  {$\boldsymbol{\alpha}$: contact / acoustic dual speed};
\end{tikzpicture}
\caption{RICCA face flux: left/right states give physical fluxes and
component-wise dual-speed dissipation $\boldsymbol{\alpha}$.}
\label{fig:stencil-ricca}
\end{subfigure}

\caption{Finite-volume building blocks used in this work:
(a)~cell update by numerical fluxes;
(b)~WENO5 face reconstruction stencil;
(c)~RICCA two-state flux with contact/acoustic dissipation.
First-order faces use neighbouring cell averages in place of~(b).}
\label{fig:stencils}
\end{figure}
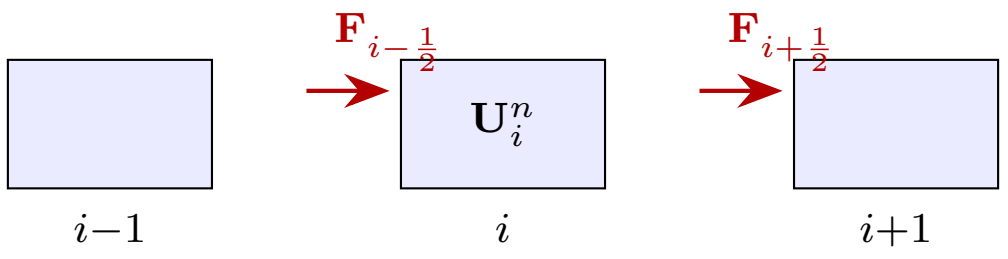
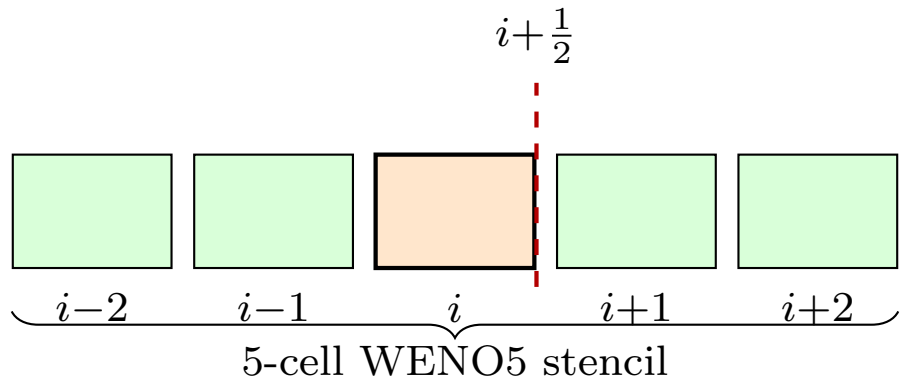
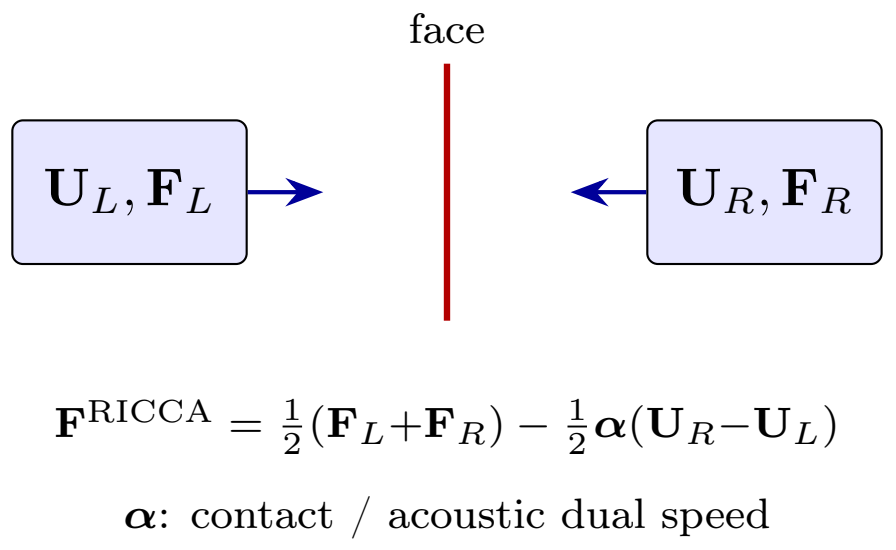

The method operates on a uniform Cartesian grid with cell averages
$\mathbf{U}_i$ and face fluxes
$\mathbf{F}^{\mathrm{num}}_{i\pm1/2}$.
Figure~\ref{fig:stencils} summarises the three building blocks used throughout:
the first-order finite-volume update, the five-cell WENO5 reconstruction about
a face, and the two-state RICCA flux with dual-speed dissipation.

\subsection{Finite-volume update}

Integrating \eqref{eq:cons1d}--\eqref{eq:cons2d} over a cell of measure $A_i$ gives
\begin{equation}
\mathbf{U}_i^{n+1}
=
\mathbf{U}_i^{n}
-
\frac{\Delta t}{A_i}
\sum_{f\in\partial i}
\mathbf{F}^{\mathrm{num}}_{f}\,
\mathrm{d}\ell_f.
\end{equation}
Face states are either first-order neighbours or WENO-reconstructed
primitives converted to conservatives.

\subsection{RICCA flux}
\label{sec:ricca}

RICCA (Riemann-invariant based contact-capturing algorithm)
\cite{KolluruRICCA,KolluruZeidan1D} is a centred scheme that avoids a full
eigenstructure decomposition.
On a face of length $\mathrm{d}\ell$ with left/right states
$(\mathbf{U}_L,\mathbf{F}_L)$ and $(\mathbf{U}_R,\mathbf{F}_R)$,
\begin{equation}
\mathbf{F}^{\mathrm{RICCA}}
=
\tfrac12\bigl(\mathbf{F}_L+\mathbf{F}_R\bigr)
-
\tfrac12\,\boldsymbol{\alpha}\,
(\mathbf{U}_R-\mathbf{U}_L)\,\mathrm{d}\ell.
\label{eq:ricca}
\end{equation}
The first term is the arithmetic average of the physical fluxes
\eqref{eq:F1d}--\eqref{eq:F2d}.
The second term is scalar (component-wise) artificial dissipation with
wave-speed coefficients $\boldsymbol{\alpha}$.

\paragraph{Dual wave-speed estimates.}
Two estimates are formed from the face-normal mixture velocity $V_n$ and
relative normal velocity $V_{rn}$:
\begin{align}
\alpha_{\mathrm{contact}}
&=
\max\bigl(|V_{n,L}|,|V_{n,R}|\bigr)
\quad\text{(augmented by a pressure jump when $|P_R-P_L|>0$)},
\\
\alpha_{\mathrm{acoustic}}
&=
\max\bigl(
|V_{n,L}|+a_L+c_L(1-c_L)|V_{rn,L}|,\;
|V_{n,R}|+a_R+c_R(1-c_R)|V_{rn,R}|
\bigr)
\end{align}
(again with a pressure-jump augmentation).
One then enforces
$\alpha_{\mathrm{acoustic}}\ge\alpha_{\mathrm{contact}}$.
\paragraph{Which equations use which speed.}
Contact-carrying equations---mixture density $\rho$, volume fraction
$\rho\alpha$ and entropy $\rho S$---use $\alpha_{\mathrm{contact}}$, so that
stationary contacts are not over-dissipated.
Acoustic/slip equations---momentum, gas mass $\rho c$ and relative velocity
$\mathbf{u}_r$---use $\alpha_{\mathrm{acoustic}}$.
\paragraph{RICCA pressure sensor (flux dissipation).}
If $|P_R-P_L|$ is negligible, only the convective speed
$\max(|V_{n,L}|,|V_{n,R}|)$ is retained; a genuine pressure jump adds an
acoustic contribution $\propto\sqrt{a}$.
This is distinct from the \emph{WENO hybrid sensors} of
Section~\ref{sec:weno}, which act on reconstructed face primitives rather
than on the intercell dissipation coefficient.

\subsection{RICCA--RH: Rankine--Hugoniot dual-speed dissipation}
\label{sec:ricca-rh}

Legacy RICCA selects between contact and acoustic speeds with a pressure
sensor (Section~\ref{sec:ricca}).
As a complementary, still heuristic design tuned to this mixture model,
\emph{RICCA--RH} keeps the same centred flux
structure~\eqref{eq:ricca} but replaces the per-face speed by the dual-speed
law
\begin{equation}
\alpha_m = \lvert V_n\rvert_{\max} + B_m\,a + \delta_{\mathrm{dens}},
\qquad
B_m =
\begin{cases}
\kappa\,\sigma & m\in\mathcal{C}\quad\text{(contact: $\rho$, $\rho\alpha$, $\rho S$)},\\
\sigma & m\in\mathcal{A}\quad\text{(acoustic/slip)},
\end{cases}
\label{eq:ricca-rh}
\end{equation}
where $\lvert V_n\rvert_{\max}=\max(\lvert V_{n,L}\rvert,\lvert V_{n,R}\rvert)$,
$a=\tfrac12(a_L+a_R)$ is a face mixture sound speed, and
$\kappa\in[0,1]$ is a contact factor (production: $\kappa=1$).
In 1D, $\mathcal{C}=\{0,2,5\}$ and $\mathcal{A}=\{1,3,4\}$; in 2D,
$\mathcal{C}=\{0,3,7\}$ and $\mathcal{A}=\{1,2,4,5,6\}$.

\paragraph{Shock switch.}
Define the normalised pressure and density jumps
\begin{equation}
\phi_P=\frac{\lvert P_R-P_L\rvert}{\tfrac12(\lvert P_L\rvert+\lvert P_R\rvert)+1},
\qquad
\phi_\rho=\frac{\lvert\rho_R-\rho_L\rvert}{\lvert\rho_{\mathrm{ref}}\rvert+1},
\end{equation}
and the Rankine--Hugoniot residual on flux densities
\begin{equation}
R=\bigl\lVert \mathbf{f}_R-\mathbf{f}_L-V_n(\mathbf{U}_R-\mathbf{U}_L)\bigr\rVert_2,
\quad
\phi_{\mathrm{RH}}=\frac{R}{\lVert\mathbf{f}_R-\mathbf{f}_L\rVert_2
+\lvert V_n\rvert\,\lVert\mathbf{U}_R-\mathbf{U}_L\rVert_2
+\lvert\rho_{\mathrm{ref}}\rvert a^2+1}.
\end{equation}
Production runs use a \emph{binary} switch
\begin{equation}
\sigma=
\begin{cases}
1 & \text{if }\phi_P>10^{-8}\text{ or }\phi_{\mathrm{RH}}>0.15,\\
0 & \text{otherwise}.
\end{cases}
\end{equation}
A density-jump safety floor
$\delta_{\mathrm{dens}}=0.15\,a$ when $\phi_\rho>10^{-4}$ and $\sigma<1$,
and an extra $0.25\,a$ plus a $c(1-c)\lvert V_{rn}\rvert$ slip term on the
relative-velocity equation(s) when $\sigma=1$, suppress residual ringing near
rarefaction edges and strong slip jumps.
In 2D the RH residual uses flux densities $\mathbf{F}/\mathrm{d}\ell$ so that
units match $\mathbf{U}$.

\paragraph{Design lesson.}
A continuous soft switch $\sigma=\mathrm{clip}(\max(\phi_P,\phi_{\mathrm{RH}}))$
with $\kappa\ll1$ was tried first and proved under-dissipative on the strong
gas--magma shocks (Section~\ref{sec:1d-rh}): contacts and slip waves rang
severely.
The binary switch with adequate contact $B$ ($\kappa=1$) restored robustness
while retaining a wave-speed structure close to legacy RICCA.
RICCA--RH is available in 1D and 2D (FO and WENO; Metal-resident path in 2D)
and is compared with legacy RICCA and FORCE in Sections~\ref{sec:1d-rh}
and~\ref{sec:2d-rh}.

\subsection{Linearised dispersion and dissipation}
\label{sec:dispersion}

\begin{figure}[htbp]
\centering
\includegraphics[width=0.95\textwidth]{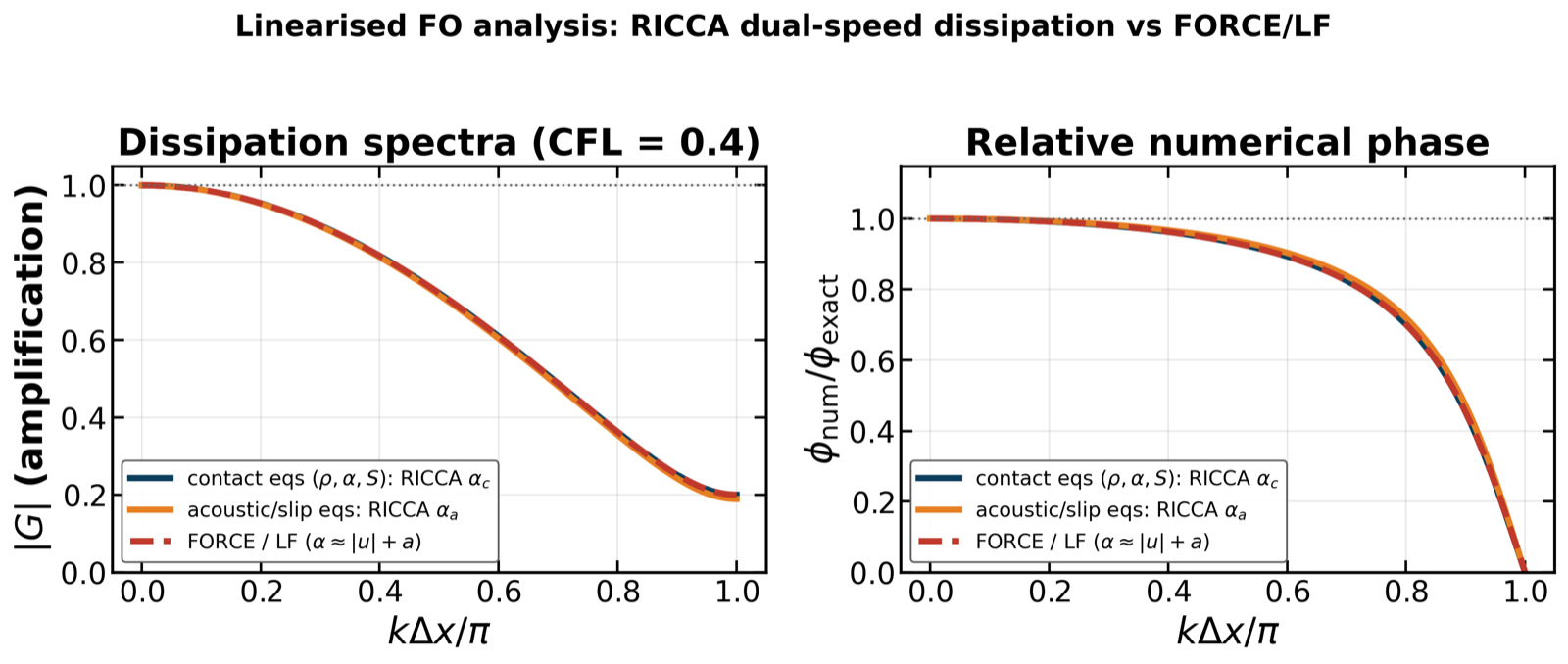}
\caption{Linearised FO analysis about a uniform mixture state:
amplification $\lvert G\rvert$ (left) and relative numerical phase (right)
versus $k\Delta x/\pi$ at CFL~$=0.4$.
RICCA contact dissipation is weaker than FORCE/LF; acoustic/slip dissipation
is comparable.}
\label{fig:dispersion}
\end{figure}

To support the claim that RICCA sharpens contacts while remaining robust on
acoustic/slip equations, we analyse a first-order centred flux of the form
\eqref{eq:ricca} on a uniform base state $(u_0,a_0,u_{r0})$ with Fourier modes
$\mathrm{e}^{\mathrm{i}j\kappa}$, $\kappa=k\Delta x$.
This linearised FO analysis is intentionally simplified: it does not replace a
nonlinear stability theory for WENO--RICCA, but it does quantify the intended
dual-speed mechanism in a controlled setting.
Contact-carrying equations use $\alpha_c=\lvert u_0\rvert$; acoustic/slip
equations use $\alpha_a=\lvert u_0\rvert+a_0+c(1-c)\lvert u_{r0}\rvert$; FORCE
is represented by a single Lax--Friedrichs speed
$\alpha_{\mathrm{LF}}=\lvert u_0\rvert+a_0$.
Forward-Euler amplification $G(\kappa)$ at CFL~$=0.4$ yields the spectra in
Fig.~\ref{fig:dispersion}.
High-wavenumber modes of the contact family are damped substantially less under
RICCA ($\alpha_c/\alpha_{\mathrm{LF}}\approx0.11$ for the representative state
$u_0=100$, $a_0=800$) than under a single acoustic LF speed, while the
acoustic/slip family tracks FORCE.
This is precisely the intended dual-speed mechanism: contacts are not
over-dissipated, yet acoustic jumps retain FORCE-like robustness.

\subsection{Positivity and admissibility strategy}
\label{sec:positivity}

Romenski mixture primitives require $\rho>0$ and
$(\alpha,c)\in(0,1)$.
The production codes enforce a layered admissibility strategy:
\begin{enumerate}
\item \textbf{CFL.} Explicit updates use CFL~$\lesssim0.4$ (WENO/SSP-RK) and
      CFL~$\lesssim0.5$ (FO); these are the ranges used throughout
      Sections~\ref{sec:1d}--\ref{sec:2d}.
\item \textbf{Characteristic WENO + combo hybrid.} Face states are reconstructed
      in an approximate face-normal acoustic characteristic basis
      (Section~\ref{sec:weno-char}), then blended toward neighbouring cell
      values by the combo sensor and clipped to the local $[\min,\max]$ of
      those cells, which prevents new extrema in $\rho$, $\alpha$,
      $c$ and $u_r$ at discontinuities.
\item \textbf{Invariant-domain (IDP) face limiter.} After hybridisation, a
      Zhang--Shu convex scaling
      $\mathbf{V}=(1-\theta)\mathbf{V}^{\mathrm{FO}}+\theta\mathbf{V}^{\mathrm{HO}}$
      chooses the largest $\theta\in[0,1]$ such that both face states remain in
      the admissible set
      $\mathcal{A}=\{\rho\ge\rho_{\min},\;\alpha,c\in[\varepsilon,1-\varepsilon]\}$
      \cite{ZhangShu2010}.
      The limiter is applied in 1D and 2D (CPU and Metal) and is active by
      default).
\item \textbf{Primitive guards.} After each update, $\alpha$ and $c$ are
      projected into $(10^{-10},1-10^{-10})$ before the equation of state;
      non-finite sound speeds abort the step.
\item \textbf{FORCE Lax--Friedrichs fallback.} If the Richtmyer intermediate
      state $\mathbf{U}^*$ violates $\rho>0$ or
      $(\alpha,c)\in(0,1)$, the FORCE face flux reverts to pure
      Lax--Friedrichs on that face
      (documented in the 1D \texttt{Force.cpp} path and the analogous 2D
      average-flux routine).
\end{enumerate}
Together these steps constitute a practical positivity-preserving /
invariant-domain strategy for the reported test matrix.
They make the scheme \emph{operationally} robust under the stated CFL and
reconstruction choices.
A complete proof that the nonlinear WENO--RICCA (or RICCA--RH) update maps
$\mathcal{A}$ into itself for arbitrary CFL remains open; the present claim is
therefore empirical robustness of the IDP-limited scheme on the Zeidan-style
suite.

\subsection{Primitive WENO reconstruction and hybrid sensors}
\label{sec:weno}

High-order face data are obtained by WENO-Z reconstruction
\cite{JiangShu1996,Borges2008} of
\begin{equation}
\mathbf{V}
=
\bigl[\rho,\,u,\,(v),\,\alpha,\,c,\,u_r,\,(v_r),\,S\bigr]
\end{equation}
(parenthetical components in 2D only), followed by
$\mathbf{V}\to\mathbf{U}$.
Both WENO3 and WENO5 are used on primitives
$(\rho,u,\alpha,c,u_r,S)$.
At each face the discontinuity indicators
\begin{align}
\phi_P
&=
\frac{|P_R-P_L|}{\tfrac12(|P_L|+|P_R|)+1},
&
\phi_\rho
&=
\frac{|\rho_R-\rho_L|}{\tfrac12(|\rho_L|+|\rho_R|)+1},
\\
\phi_S
&=
\frac{|S_R-S_L|}{\tfrac12(|S_L|+|S_R|)+1},
&
\phi_\alpha
&=
\frac{|\alpha_R-\alpha_L|}{\tfrac12(|\alpha_L|+|\alpha_R|)+10^{-3}},
\\
\phi_u
&=
\frac{|u_R-u_L|}{a_{\mathrm{ref}}},
&
\phi_{u_r}
&=
\frac{|u_{r,R}-u_{r,L}|}{a_{\mathrm{ref}}},
\end{align}
with $a_{\mathrm{ref}}=\tfrac12(a_L+a_R)+1$.
The \emph{original} (legacy) hybrid used the pressure indicator alone,
$\phi=\phi_P$.
That choice is retained in the comparisons below as the
``old sensor''.
The \emph{default production sensor} is the multi-variable combination
\begin{equation}
\phi
=
\max\!\bigl(
\phi_P,\phi_\rho,\phi_S,\phi_\alpha,\phi_u,1.5\,\phi_{u_r}
\bigr).
\end{equation}
Face primitives are then blended toward neighbouring cell values,
\begin{equation}
\mathbf{V}_{L,R}
\leftarrow
\theta\,\mathbf{V}_{L,R}^{\mathrm{WENO}}
+(1-\theta)\,\mathbf{V}_{L,R}^{\mathrm{FO}},
\qquad
\theta=\frac{1}{1+C\phi},
\end{equation}
with $C=50$ for $(\rho,\alpha,c,S)$, $C=120$ for $u$ and $C=200$ for
$u_r$, and clipped to the local min/max of the adjacent cells.
WENO time integration uses SSP Runge--Kutta (CFL$\approx 0.4$).
The same combo hybrid is used in two dimensions (CPU and Metal WENO paths),
with additional sensors on $v$ and $v_r$.
Relative-velocity jumps are measured on a local $|u_r|$ scale (not only the
mixture sound speed). On the 2D Riemann tubes, face states for
$u_r$, $v_r$ and entropy $S$ (which enters $\Psi$) are taken first-order,
while $\rho$, $(u,v)$, $\alpha$ and $c$ keep the combo-blended WENO faces,
and the RICCA acoustic viscosity on the slip equations is increased by a
factor $1.6$ under WENO. This cleans isolines in $\rho$ and $|V|$; RICCA
still yields a higher peak $|u_r|$ than FORCE because the centred FORCE flux
is more dissipative on the relative-velocity system.

\paragraph{Approximate characteristic WENO.}
\label{sec:weno-char}
Because a full $R\lvert\Lambda\rvert R^{-1}$ eigenstructure of the
Romenski Jacobian is not available, production WENO (default
\texttt{WENO\_CHAR=1}) reconstructs in an approximate face-normal acoustic
characteristic basis frozen at the arithmetic face average
$(\bar\rho,\bar a,\mathbf{n})$.
In 1D the characteristic fields are
\begin{equation}
w^\mp=\tfrac12\Bigl(\rho\,\bar a/\bar\rho\mp u\Bigr),\quad
w_\alpha=\alpha,\quad w_c=c,\quad w_{u_r}=u_r/\bar a,\quad w_S=S;
\end{equation}
WENO-Z is applied to $\mathbf{w}$, then
$\rho=\bar\rho/\bar a\,(w^-+w^+)$, $u=w^+-w^-$ recover the primitives.
In 2D the same construction uses the face-normal velocity
$u_n=\mathbf{u}\cdot\mathbf{n}$ and tangential $u_t$, together with
$(u_{rn},u_{rt})$ for the slip vector.
This is the standard pragmatic substitute for full characteristic WENO when
only mixture acoustics are known
\cite{JiangShu1996}, and it is implemented on both the CPU and Metal paths.
Disable with \texttt{WENO\_CHAR=0} to recover componentwise primitive WENO.

\begin{figure}[!t]
\centering
\includegraphics[width=0.92\textwidth]{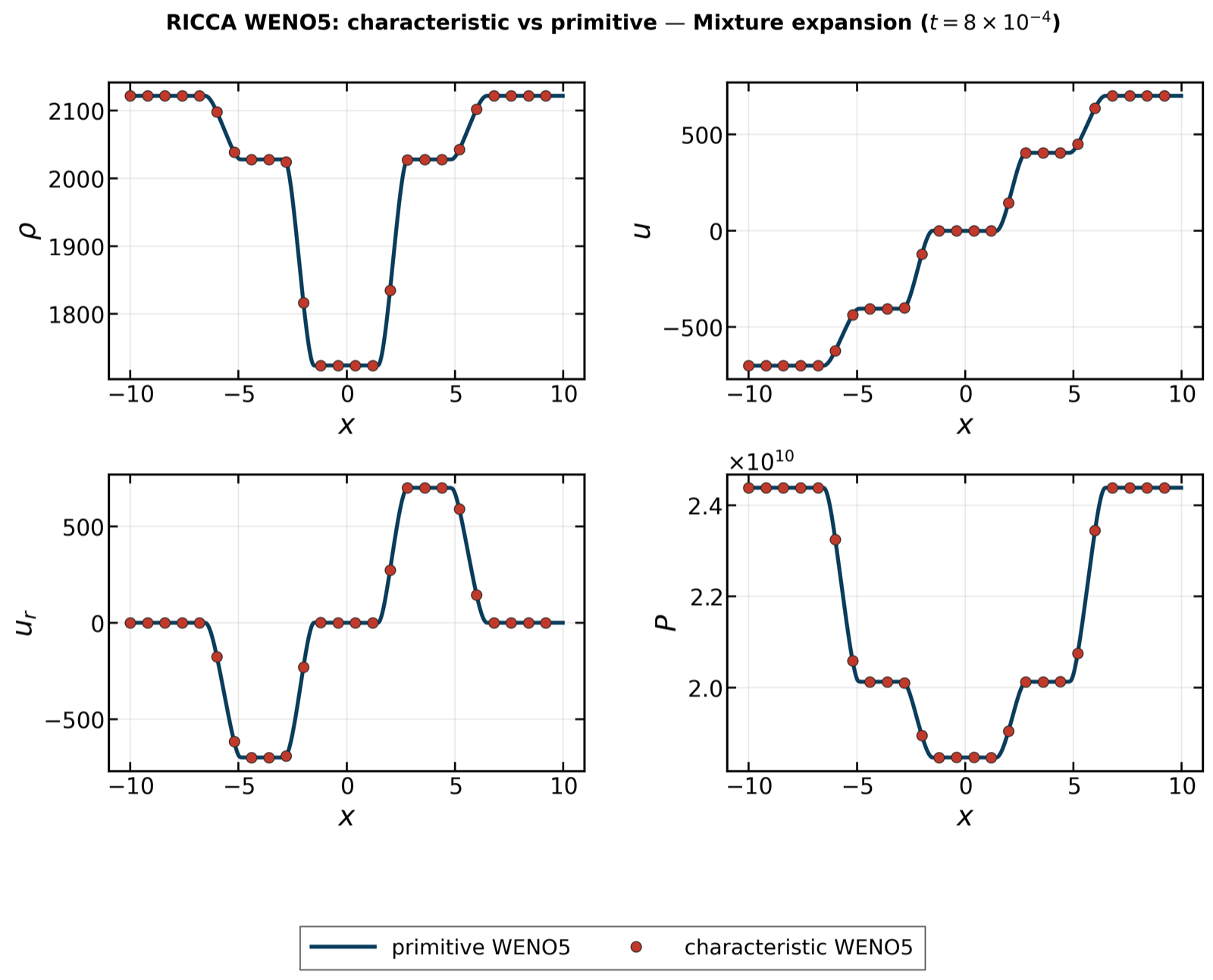}
\caption{1D mixture expansion: RICCA--WENO5 with approximate characteristic
reconstruction (markers) versus componentwise primitive WENO5 (line),
$N\approx1000$, $t=8\times10^{-4}$.}
\label{fig:char-vs-prim-1d}
\end{figure}

\begin{figure}[!t]
\centering
\includegraphics[width=0.92\textwidth]{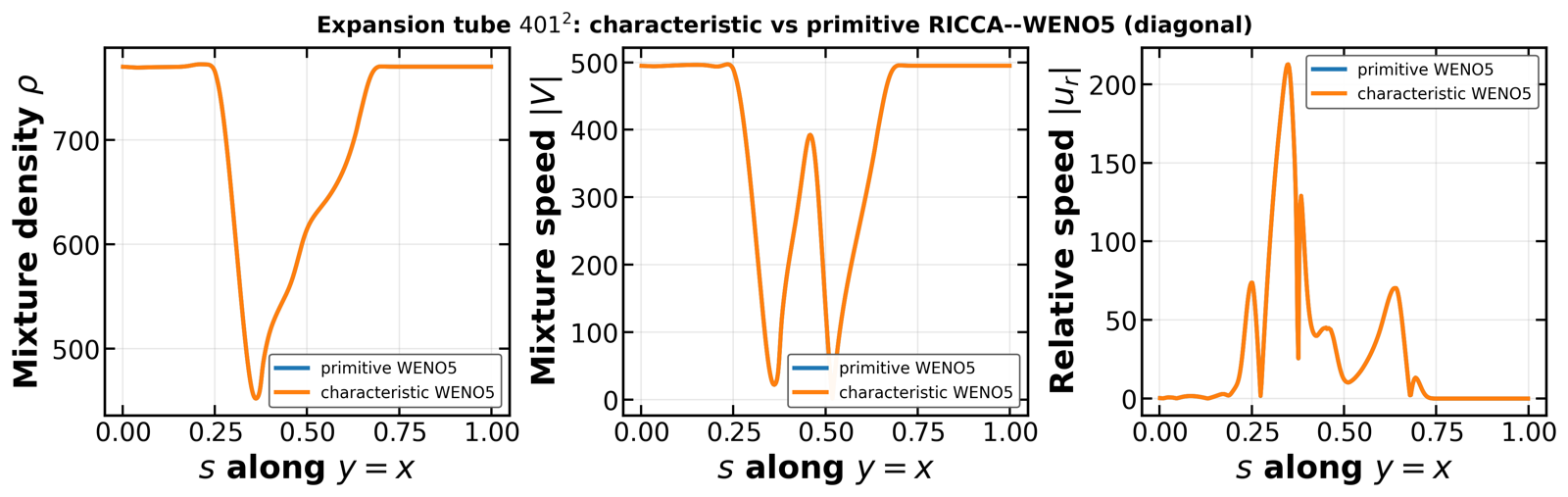}
\caption{2D expansion tube ($401^2$): diagonal $y=x$ cuts of
$\rho$, $|V|$ and $|u_r|$ for characteristic versus primitive RICCA--WENO5.}
\label{fig:char-vs-prim-2d}
\end{figure}

\paragraph{Characteristic versus primitive WENO.}
\label{sec:char-vs-prim}
Figures~\ref{fig:char-vs-prim-1d}--\ref{fig:char-vs-prim-2d} compare production
characteristic WENO5 (\texttt{WENO\_CHAR=1}) with componentwise primitive WENO5
(\texttt{WENO\_CHAR=0}) under the same combo hybrid and IDP settings.
On the 1D gas--magma suite ($N\approx1000$) the two reconstructions overlie to
plotting accuracy: roughness $\mathrm{std}(\Delta\cdot)$ changes by
$\lesssim0.1\%$ and pointwise differences remain localised near the waves
(Table~\ref{tab:char-vs-prim}).
On the 2D expansion tube ($401^2$) diagonal cuts and contour fields likewise
agree; peak $|u_r|$ differs by less than $0.1\%$.
Thus the approximate acoustic characteristic basis does not alter the
Zeidan-suite solutions relative to primitive WENO, while providing the
standard characteristic pathway for future full-eigenstructure upgrades.
We retain characteristic WENO as the production default.

\begin{table}[!t]
\centering\small
\caption{Characteristic vs primitive RICCA--WENO5 roughness
$\mathrm{std}(\Delta\cdot)$ on the 1D suite ($N\approx1000$).
Differences are negligible relative to FO/WENO and FORCE/RICCA gaps
reported elsewhere.}
\label{tab:char-vs-prim}
\begin{tabular}{@{}llcc@{}}
\toprule
Case & Field & primitive & characteristic \\
\midrule
Expansion & $\rho$ & $1.839$ & $1.838$ \\
Expansion & $u$ & $2.460$ & $2.458$ \\
Expansion & $u_r$ & $5.511$ & $5.508$ \\
HighV & $\rho$ & $7.394$ & $7.394$ \\
Collision & $\rho$ & $1.757$ & $1.757$ \\
\bottomrule
\end{tabular}
\end{table}

\paragraph{Invariant-domain face limiter.}
After characteristic reconstruction and combo hybridisation, the IDP scaling
of Section~\ref{sec:positivity} is applied (default \texttt{WENO\_IDP=1}).
On the 1D gas--magma suite and the 2D expansion/collision tubes the limiter
is typically inactive in smooth regions ($\theta=1$) and engages only near
strong contacts where high-order face $(\rho,\alpha,c)$ would leave
$\mathcal{A}$.

For the equal-EOS forward-facing step only
$\{\rho,u,v,\alpha,S\}$ are reconstructed and
$c=\alpha$, $u_r=v_r=0$ are enforced
(Section~\ref{sec:2d-ffs} discusses the single-velocity character of this case).

\begin{figure}[htbp]
\centering
\includegraphics[width=0.92\textwidth]{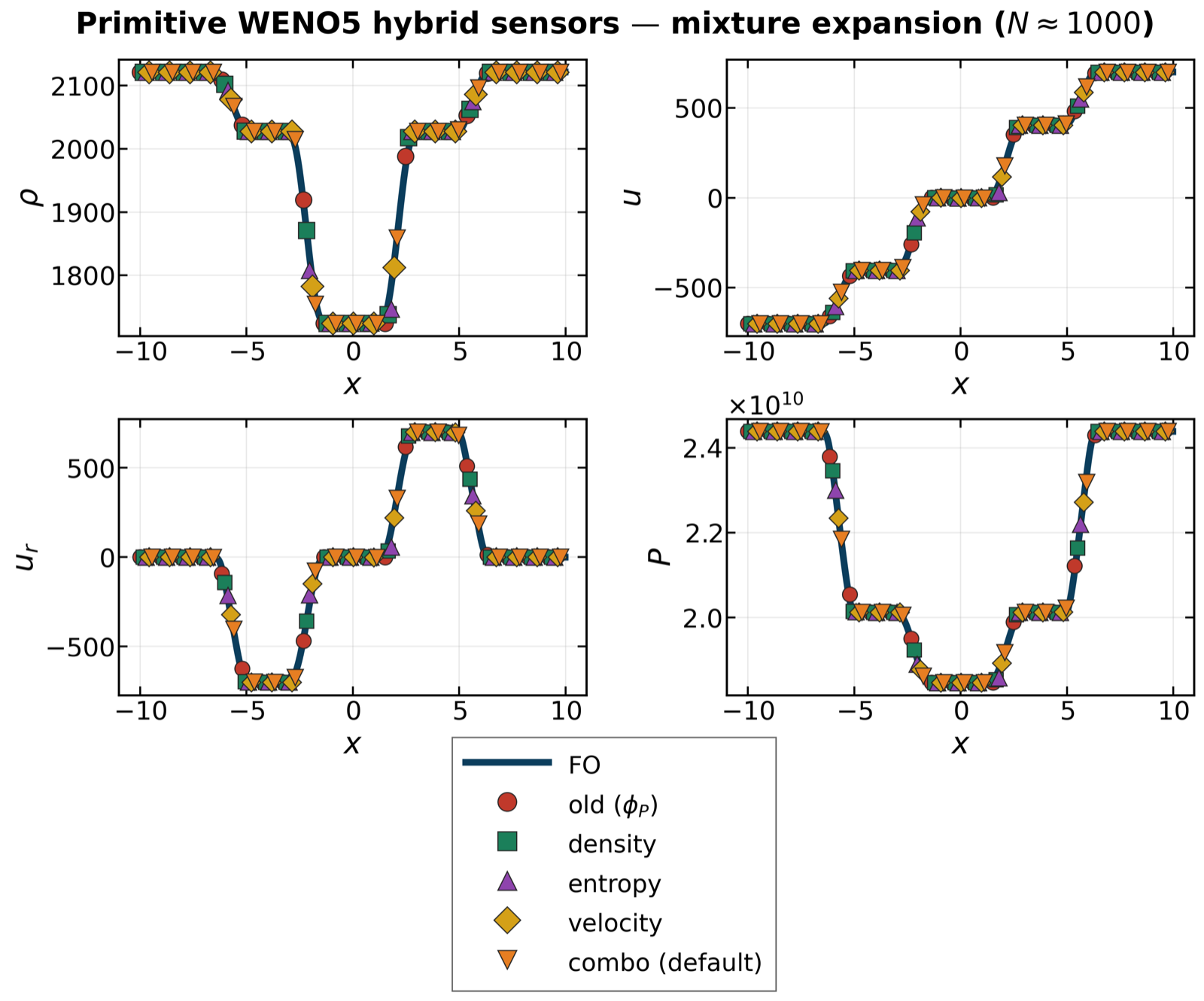}
\caption{1D mixture expansion, $N\approx1000$: primitive RICCA--WENO5 with
different hybrid sensors versus first-order RICCA.
The legacy pressure-only sensor is labelled ``old ($\phi_P$)'';
the combined ($P$/$\rho$/$S$/$\alpha$/$u$/$u_r$) sensor is the default.}
\label{fig:weno-sensor}
\end{figure}

\paragraph{Why a multi-variable primitive sensor.}
Characteristic WENO with a full eigenstructure remains the gold standard for
smoothness detection, but Romenski mixtures make that factorisation expensive
and awkward on GPU.
Entropy-only sensors miss contacts where $S$ is flat; pressure-only sensors
miss near-isobaric density/volume-fraction jumps
(Fig.~\ref{fig:weno-sensor}).
The combo sensor is deliberately cheap---normalised jumps of already available
primitives, one $\max$ reduction per face---and is offered as a
\emph{production heuristic} for this equation set, not as a replacement for
characteristic detectors
(Section~\ref{sec:sensor-discussion}).

\begin{figure}[htbp]
\centering
\includegraphics[width=0.95\textwidth]{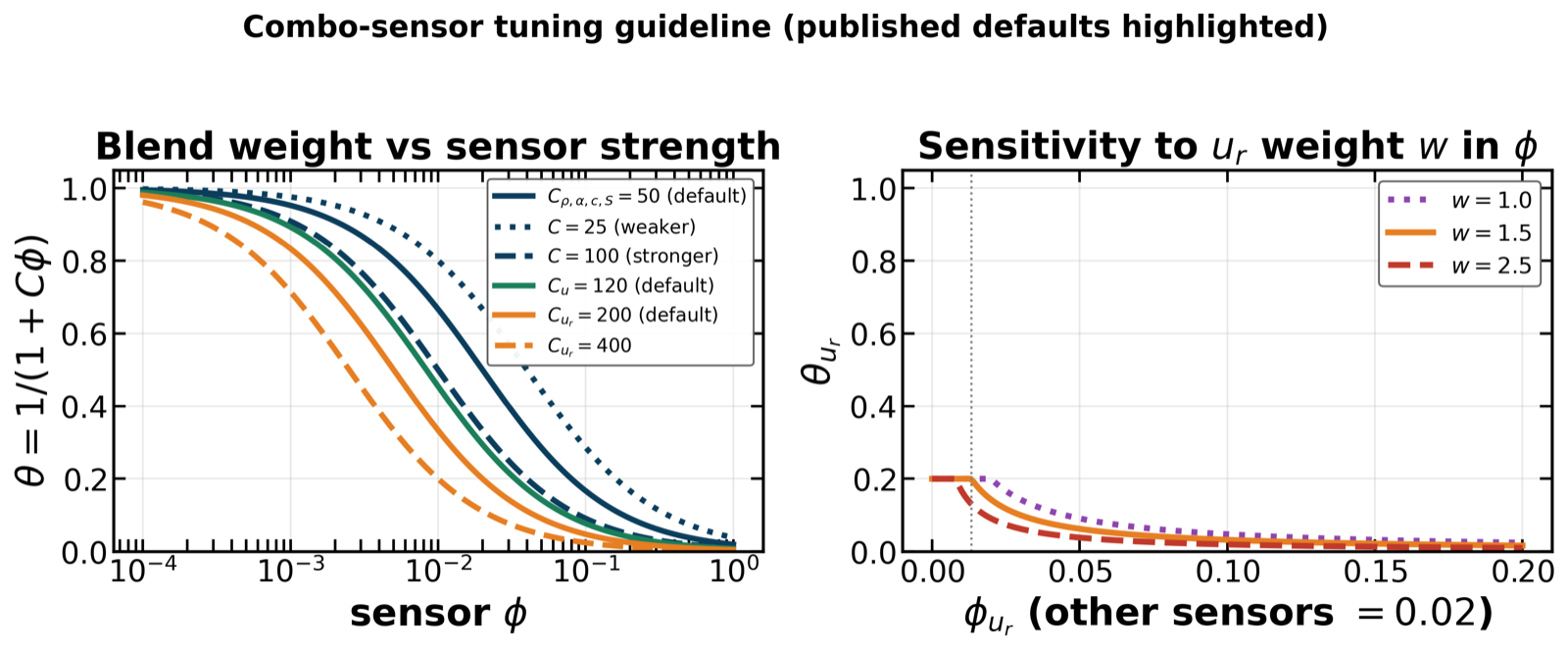}
\caption{Combo-sensor tuning: blend weight versus $\phi$ for the published
$C$ values (left), and sensitivity of $\theta_{u_r}$ to the weight $w$ on
$\phi_{u_r}$ (right).}
\label{fig:sensor-tuning}
\end{figure}

\paragraph{Tuning guideline.}
The constants $(C_{\mathrm{base}},C_u,C_{u_r},w)=(50,120,200,1.5)$ are
empirical: they were fixed by a short survey on the 1D expansion
($N\approx1000$) and then frozen for all other tests.
Figure~\ref{fig:sensor-tuning} shows $\theta=1/(1+C\phi)$ against $\phi$ for
the published $C$ values and neighbouring alternatives, and the effect of the
$u_r$ weight $w$.
Defaults keep smooth faces ($\phi\lesssim10^{-3}$) near $\theta\approx1$ and
trip FO blending once $\phi\gtrsim0.02$--$0.05$.
On the expansion tube, $\pm50\%$ perturbations of $C_{u_r}$ or $w$ leave the
Table~\ref{tab:sensor} ranking unchanged.
The 2D path uses stronger slip $C$ (up to $800$) because face-normal $u_r$
rings more on Cartesian Riemann tubes.
Transferability beyond the Zeidan-style suite is not claimed; retuning may be
needed for other Mach/slip/EOS regimes
(Section~\ref{sec:sensor-discussion}).

\begin{figure}[htbp]
\centering
\includegraphics[width=0.88\textwidth]{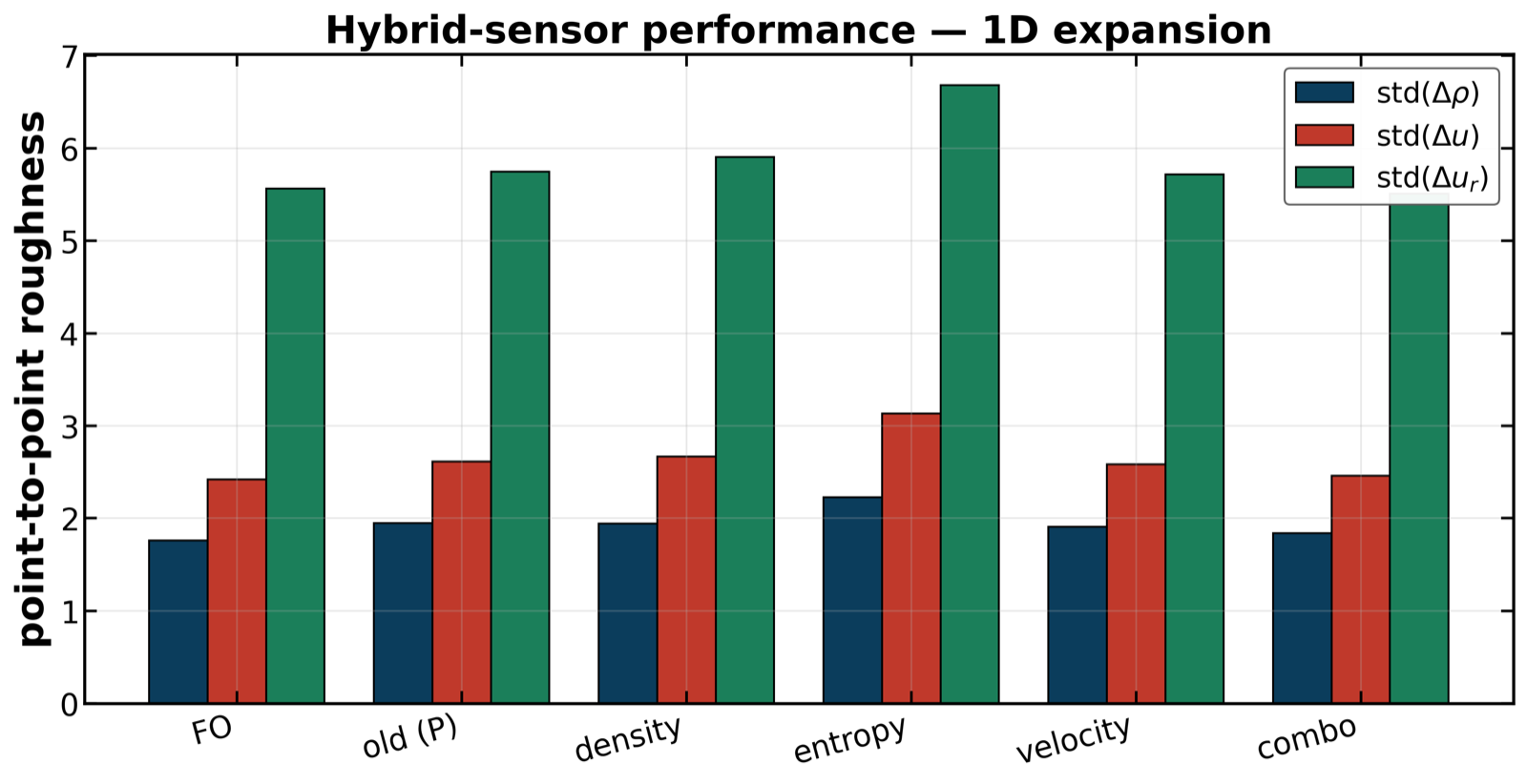}
\caption{Quantitative sensor performance (same case as
Fig.~\ref{fig:weno-sensor}): point-to-point roughness of $\rho$, $u$ and
$u_r$.}
\label{fig:weno-sensor-metrics}
\end{figure}

\begin{figure}[htbp]
\centering
\includegraphics[width=0.95\textwidth]{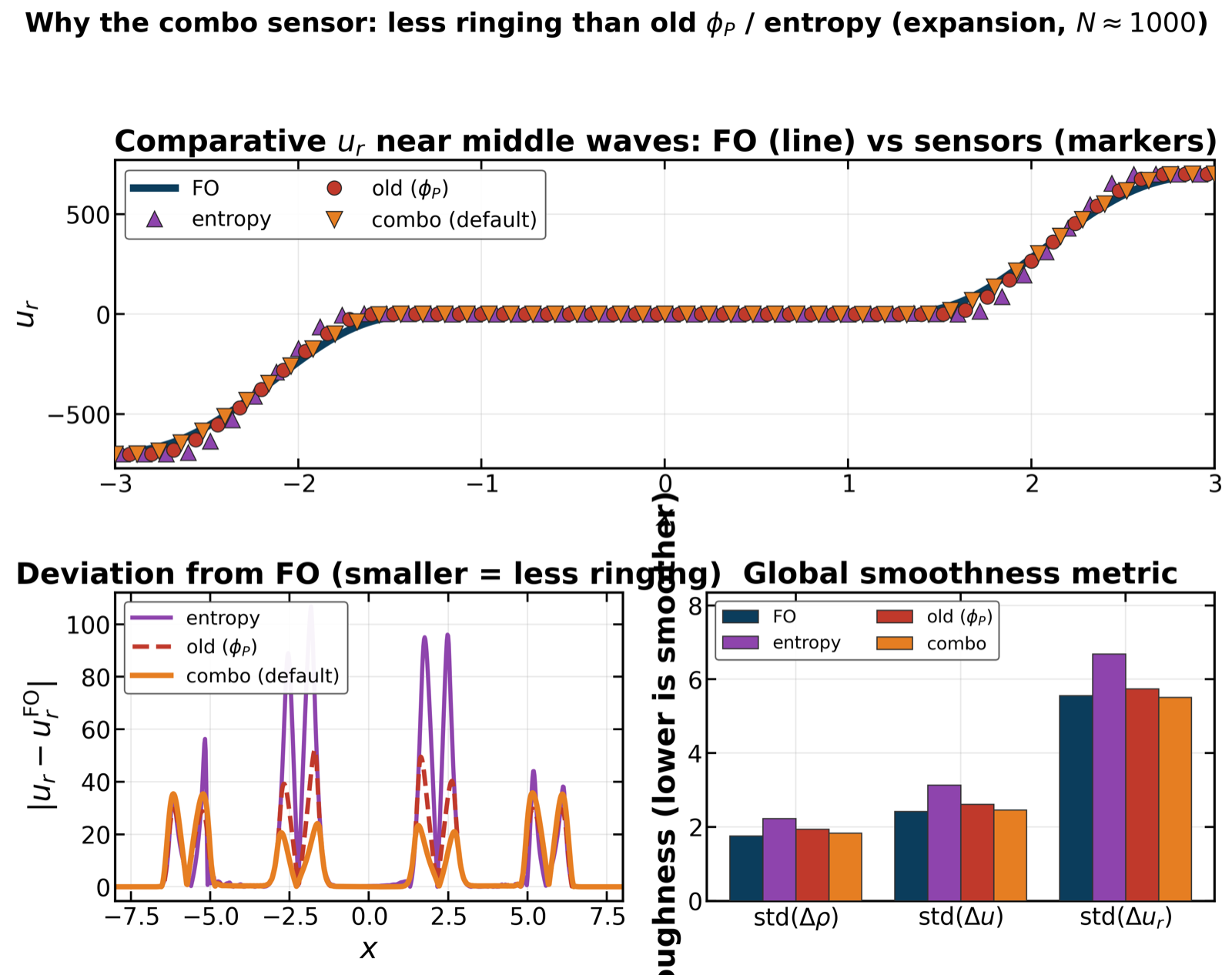}
\caption{Advantage of the combo sensor on the 1D expansion
($N\approx1000$).
Top: single comparative plot of $u_r$ about both middle waves
(FO as a thick line; entropy, old $\phi_P$ and combo as markers)---entropy
and old $\phi_P$ depart from FO, while combo tracks FO.
Bottom left: $|u_r-u_r^{\mathrm{FO}}|$ (combo stays lowest).
Bottom right: global roughness bars (combo closest to FO).}
\label{fig:weno-sensor-advantage}
\end{figure}

\paragraph{Performance of the hybrid sensors.}
On the 1D mixture expansion ($N\approx1000$) five choices of $\phi$ are
compared against first-order RICCA
(Figs.~\ref{fig:weno-sensor}--\ref{fig:weno-sensor-advantage},
Table~\ref{tab:sensor}).
The metrics are point-to-point roughness $\mathrm{std}(\Delta\cdot)$ for
$\rho$, $u$ and $u_r$, and the peak slip residual
$|u_r-u_r^{\mathrm{FO}}|$.
Here ``roughness'' is a sensor-quality proxy: it measures residual
point-to-point ringing of high-order solutions relative to the FO reference
(and is therefore distinct from the FO scheme-to-scheme dissipation
comparison in Table~\ref{tab:1d-rh-fo}).
Because FO is the most dissipative baseline, it is also the smoothest by this
metric, so it is expected that even a well-behaved WENO run can have
$\mathrm{std}(\Delta\rho)$ above the FO row in Table~\ref{tab:sensor}.
These metrics do not resolve long-time error growth or every non-monotone
interface detail.
Gains of the combo sensor over the legacy pressure-only blend are
\emph{modest} but consistent on this case: roughly $4$--$6\%$ lower
roughness, and $|u_r-u_r^{\mathrm{FO}}|$ reduced from $\approx55$ to
$\approx36$.

\begin{figure}[htbp]
\centering
\includegraphics[width=0.95\textwidth]{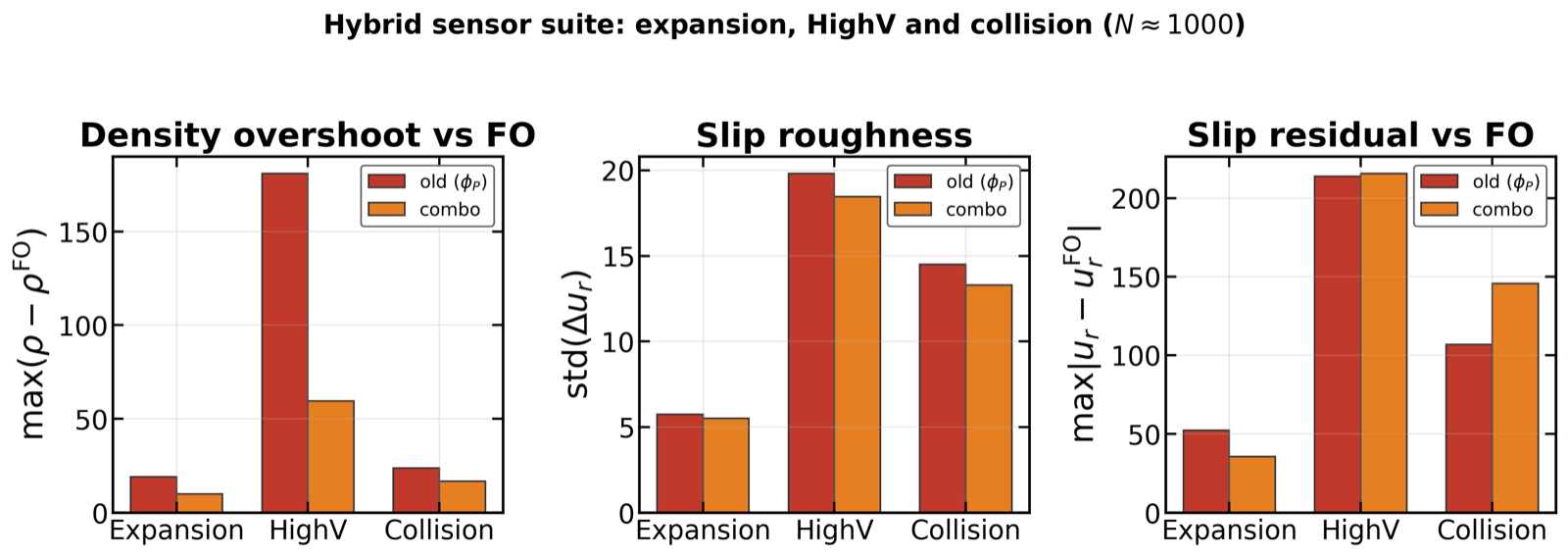}
\caption{Hybrid-sensor suite on expansion, HighV and collision
($N\approx1000$): density overshoot vs FO, slip roughness, and peak slip
residual. Combo consistently cuts overshoot and roughness; the peak
$|u_r-u_r^{\mathrm{FO}}|$ is problem-dependent.}
\label{fig:sensor-suite}
\end{figure}

The same quantitative metrics are repeated for the high-velocity and
collision problems (Fig.~\ref{fig:sensor-suite}).
Across all three Riemann problems the combo sensor reduces density overshoot
relative to FO and lowers slip roughness $\mathrm{std}(\Delta u_r)$.
The peak slip residual is case-dependent: it drops on the expansion, is
essentially unchanged on HighV, and can increase on the collision (where the
FO reference itself carries a strong $u_r$ spike).
The combo hybrid is therefore preferred for overshoot control and smoothness
on this suite, not as a universal minimiser of every slip residual.
\begin{table}[htbp]
\centering\small
\caption{1D expansion: hybrid-sensor roughness
$\mathrm{std}(\Delta\cdot)$ for RICCA--WENO5 versus FO
($N\approx1000$). Lower is smoother.
``Old'' denotes the legacy pressure-only sensor $\phi=\phi_P$.}
\label{tab:sensor}
\begin{tabular}{@{}lrrr@{}}
\toprule
Sensor $\phi$ & $\mathrm{std}(\Delta\rho)$ & $\mathrm{std}(\Delta u)$ & $\mathrm{std}(\Delta u_r)$ \\
\midrule
FO (reference) & $1.75$ & $2.41$ & $5.55$ \\
old (pressure $\phi_P$) & $1.99$ & $2.68$ & $5.85$ \\
density $\phi_\rho$ & $1.94$ & $2.67$ & $5.90$ \\
entropy $\phi_S$ & $2.23$ & $3.14$ & $6.68$ \\
velocity $\max(\phi_u,\phi_{u_r})$ & $1.91$ & $2.58$ & $5.72$ \\
\textbf{combo (default)} & $\mathbf{1.84}$ & $\mathbf{2.46}$ & $\mathbf{5.51}$ \\
\bottomrule
\end{tabular}
\end{table}

\subsection{FORCE centred flux}
\label{sec:force}

FORCE (First-Order Centred)
\cite{ToroFORCE,Zeidan2019} is retained as a centred reference on the same
meshes and the same left/right states as RICCA (cell averages at first order;
WENO face states when reconstruction is active).
Unlike RICCA, FORCE does not introduce a separate dual wave-speed /
pressure-sensor dissipation vector: the numerical flux is the arithmetic mean
of a Lax--Friedrichs (LF) flux and a Richtmyer (two-step Lax--Wendroff)
flux,
\begin{equation}
\mathbf{F}^{\mathrm{FORCE}}
=
\tfrac12\bigl(
\mathbf{F}^{\mathrm{LF}}
+
\mathbf{F}^{\mathrm{RI}}
\bigr).
\label{eq:force}
\end{equation}
If the Richtmyer intermediate state is unphysical
($\rho\le0$ or $(\alpha,c)\notin(0,1)$), the face reverts to pure LF
(Section~\ref{sec:positivity}).

\paragraph{Local face CFL.}
With left/right normal velocities $V_{n,L/R}$ and sound speeds $a_{L/R}$,
define
\begin{equation}
\lambda
=
\max\bigl(|V_{n,L}|+a_L,\;|V_{n,R}|+a_R\bigr),
\qquad
\Delta t_f
=
\mathrm{CFL}\,\frac{\Delta x_f}{\lambda},
\end{equation}
where $\Delta x_f$ is a local length scale (cell spacing $\Delta x$ in 1D;
area-to-edge length in 2D).
This $\Delta t_f$ is used only inside the FORCE face formula; the global
time step (below) remains separate.

\paragraph{Lax--Friedrichs branch.}
\begin{equation}
\mathbf{F}^{\mathrm{LF}}
=
\tfrac12\bigl(\mathbf{F}_L+\mathbf{F}_R\bigr)
+
\tfrac12\frac{\Delta x_f}{\Delta t_f}
\bigl(\mathbf{U}_L-\mathbf{U}_R\bigr)
\end{equation}
(with an analogous factor of face length $\mathrm{d}\ell$ when fluxes are
stored as integrated edge fluxes in 2D).
The first term averages the physical fluxes
\eqref{eq:F1d}--\eqref{eq:F2d}; the second is isotropic (component-wise)
numerical viscosity scaled by the local CFL speed
$\Delta x_f/\Delta t_f=\lambda/\mathrm{CFL}$.

\paragraph{Richtmyer branch.}
An intermediate conservative state is predicted by a centred half-step,
\begin{equation}
\mathbf{U}^{\mathrm{RI}}
=
\tfrac12\bigl(\mathbf{U}_L+\mathbf{U}_R\bigr)
+
\tfrac12\frac{\Delta t_f}{\Delta x_f}
\bigl(\mathbf{F}_L-\mathbf{F}_R\bigr),
\end{equation}
from which the physical flux $\mathbf{F}^{\mathrm{RI}}=\mathbf{F}(\mathbf{U}^{\mathrm{RI}})$
is evaluated (same mixture flux routines as for cell averages).
\paragraph{Robustness.}
If $\mathbf{U}^{\mathrm{RI}}$ is unphysical (non-positive density, volume or
mass fractions outside $(0,1)$, or non-finite components), the implementation
falls back to $\mathbf{F}^{\mathrm{FORCE}}=\mathbf{F}^{\mathrm{LF}}$ on that
face.
For the equal-EOS forward-facing step the intermediate state is projected onto
$c=\alpha$ and $\mathbf{u}_r=\mathbf{0}$ before forming $\mathbf{F}^{\mathrm{RI}}$.

\paragraph{Relation to RICCA.}
Both schemes are centred and Riemann-solver free.
RICCA modulates dissipation per equation with contact/acoustic speeds and a
pressure sensor; RICCA--RH uses the same equation split with the dual-speed law
\eqref{eq:ricca-rh}; FORCE applies a single LF--Richtmyer average to all
components.
Consequently FORCE is typically more dissipative at contacts, while remaining
a standard benchmark for the same WENO face data
\cite{Zeidan2019}.

\subsection{HLLC upwind flux}
\label{sec:hllc}

As an upwind reference we implement a mixture HLLC-family solver on the same
left/right states as RICCA/FORCE.
Wave bounds use Davis/Einfeldt-style estimates with slip augmentation,
\begin{equation}
S_L=\min(u_L-a_L^{\mathrm{eff}},u_R-a_R^{\mathrm{eff}}),\quad
S_R=\max(u_L+a_L^{\mathrm{eff}},u_R+a_R^{\mathrm{eff}}),
\end{equation}
where $a^{\mathrm{eff}}=a+|c(1-c)u_r|$, together with a contact speed $S^*$
from the classical pressure/velocity estimate written with mixture density.
Star states for $(\rho,\rho u,\rho\alpha,\rho c,\rho S)$ follow the standard
HLLC scaling; $u_r$ is treated as advected across the acoustic waves.
On the strong gas--magma Riemann problems the three-wave star states can drive
$(\alpha,c)$ out of $(0,1)$; the production path therefore uses the
\emph{Rusanov-stable} member of the same family,
\begin{equation}
\mathbf{F}^{\mathrm{HLLC}}
=
\tfrac12\bigl(\mathbf{F}_L+\mathbf{F}_R\bigr)
-
\tfrac12\,s_{\max}\,(\mathbf{U}_R-\mathbf{U}_L),
\label{eq:hllc-prod}
\end{equation}
with $s_{\max}=\max(|u_L|+a_L^{\mathrm{eff}},|u_R|+a_R^{\mathrm{eff}})$,
while the full three-wave formulas remain available via an optional strict mode
for development tests.
This is an honest engineering choice for stiff dual-EOS mixtures, not a claim
that classical perfect-gas HLLC applies unchanged.

\subsection{Roe upwind flux}
\label{sec:roe}

A Roe-type flux-difference splitting is built from $\sqrt{\rho}$-weighted
averages of $(u,a,c,u_r)$ and the same contact/acoustic equation split as
RICCA:
\begin{equation}
\mathbf{F}^{\mathrm{Roe}}
=
\tfrac12\bigl(\mathbf{F}_L+\mathbf{F}_R\bigr)
-
\tfrac12\,\lvert\hat{\boldsymbol{\Lambda}}\rvert\,
(\mathbf{U}_R-\mathbf{U}_L),
\label{eq:roe}
\end{equation}
where contact equations $(\rho,\rho\alpha,\rho S)$ use
$\lvert\hat\lambda\rvert=\lvert\hat u\rvert$ and acoustic/slip equations use
$\lvert\hat u\rvert+\hat a_{\mathrm{eff}}$, with a Harten entropy fix near
sonic points.
This is a field-by-field Roe linearisation using the analytical mixture wave
speeds at the Roe state; it is \emph{not} a full $R\lvert\Lambda\rvert R^{-1}$
decomposition of the complete six/eight-equation Jacobian (left/right
eigenvector matrices for this model remain future work).

\subsection{Time stepping}

A global CFL time step is used.
First-order runs use forward Euler (CFL$\approx 0.5$ in 1D;
CFL$=0.33$ for 2D Riemann problems and CFL$=0.90$ for the forward-facing
step \cite{Zeidan2019}).
WENO runs use SSP Runge--Kutta with CFL$\approx 0.4$ in 1D.

%==============================================================================
\section{One-dimensional gas--magma tests}
\label{sec:1d}

The 1D suite is that implemented in \texttt{1D\_MultiPhase\_FlowSV}
(Test cases~1--3), corresponding to the gas--magma Riemann problems of
\cite{Zeidan2014}.
The computational domain is $x\in[-10,10]$ with a diaphragm at $x=0$ and
extrapolated (transmissive) boundaries.
Representative grids range from $N=100$ to $N=6000$ (and $N=20000$ for
selected collision studies); the comparisons below use $N\approx1000$
RICCA, RICCA--RH and FORCE solutions (first-order and WENO5) in
\texttt{1D\_MultiPhase\_Solutions/\{RICCA,RICCA\_RH,FORCE\}}.
Section~\ref{sec:1d-rh} focuses on the three-way flux comparison and the
under-dissipative RICCA--RH prototype.

\subsection{Mixture expansion (Test case~1)}
\label{sec:1d-exp}

\begin{figure}[htbp]
\centering
\includegraphics[width=0.92\textwidth]{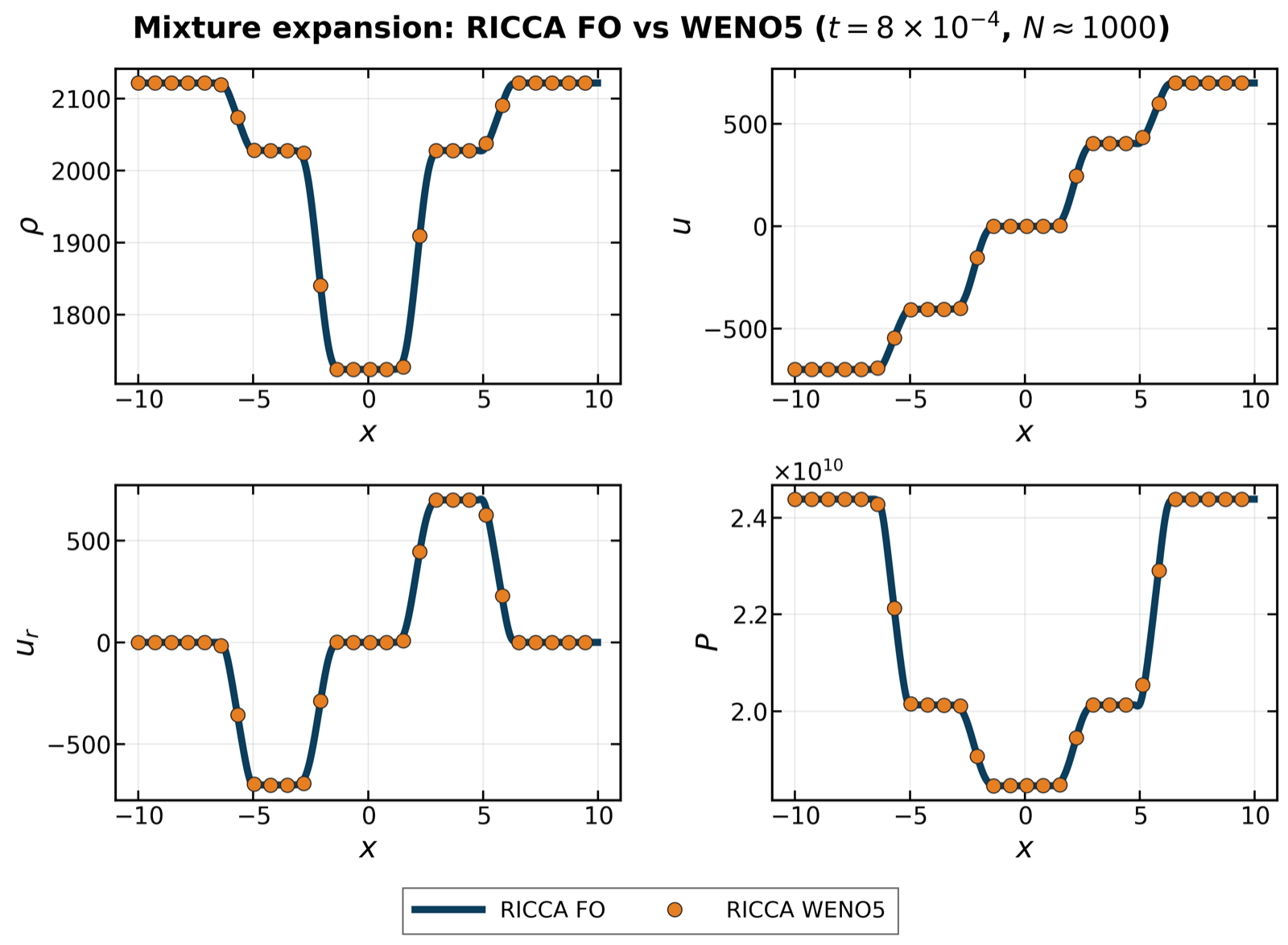}
\caption{1D gas--magma mixture expansion (Test case~1), $t=0.0008$:
overlapped RICCA FO (line) versus RICCA--WENO5 (markers),
$\rho$, $u$, $u_r$, $P$ ($N\approx1000$).}
\label{fig:1d-exp}
\end{figure}

End time $t=8\times10^{-4}$.
Left and right states (identical densities/entropies; opposing velocities):
\begin{center}
\small
\begin{tabular}{@{}lcc@{}}
\toprule
 & Left ($x\le 0$) & Right ($x>0$) \\
\midrule
$(\rho_1,\rho_2)$ & $(3800,1562)$ & $(3800,1562)$ \\
$(u_1,u_2)$ & $(-700,-700)$ & $(+700,+700)$ \\
$\alpha$ & $0.75$ & $0.75$ \\
$(S_1,S_2)$ & $(7556.5895,12658.879)$ & same \\
\bottomrule
\end{tabular}
\end{center}
Figure~\ref{fig:1d-exp} compares first-order RICCA with RICCA--WENO.
WENO sharpens the rarefaction heads and the middle-wave structure while
preserving the overall wave pattern of the expansion tube.
(Sensor variants for this case appear in Section~\ref{sec:weno};
RICCA--FORCE overlays in Section~\ref{sec:1d-force}.)

\subsection{High-velocity shock tube (Test case~2)}
\label{sec:1d-highv}

\begin{figure}[htbp]
\centering
\includegraphics[width=0.92\textwidth]{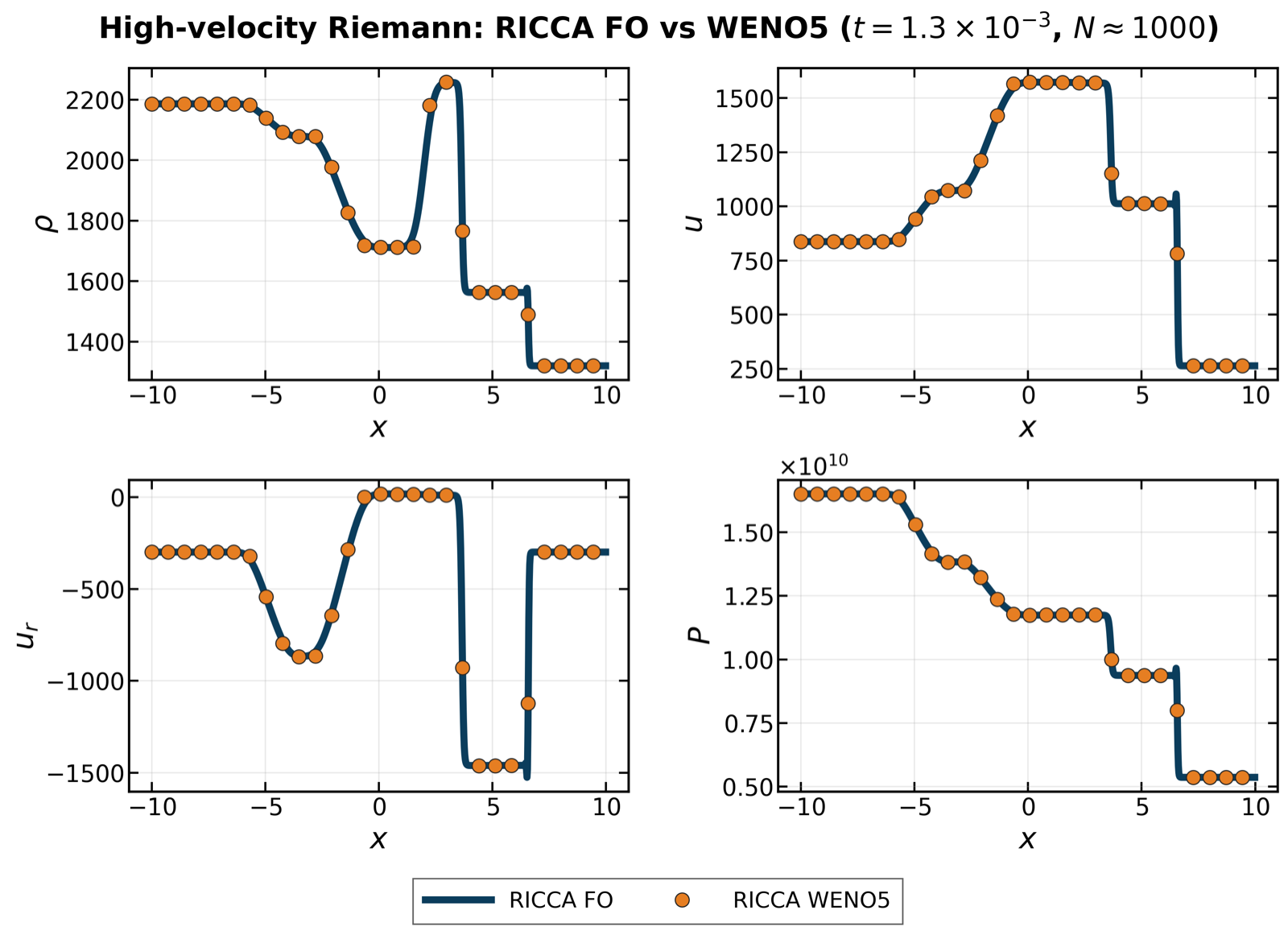}
\caption{1D gas--magma high-velocity Riemann problem (Test case~2),
$t=0.0013$: overlapped RICCA FO (line) versus RICCA--WENO5 (markers).}
\label{fig:1d-highv}
\end{figure}

End time $t=1.3\times10^{-3}$, $\alpha=0.6$:
\begin{center}
\small
\begin{tabular}{@{}lcc@{}}
\toprule
 & Left & Right \\
\midrule
$(\rho_1,\rho_2)$ & $(2500,1977)$ & $(1800,1000)$ \\
$(u_1,u_2)$ & $(1000,700)$ & $(400,100)$ \\
\bottomrule
\end{tabular}
\end{center}
Figure~\ref{fig:1d-highv} shows RICCA and RICCA--WENO.
The high-velocity left state produces strong shocks; WENO improves shock
resolution without destroying the contact.
RICCA--FORCE overlays for this case are in Section~\ref{sec:1d-force}.

\subsection{Mixture collision (Test case~3)}
\label{sec:1d-col}

\begin{figure}[htbp]
\centering
\includegraphics[width=0.92\textwidth]{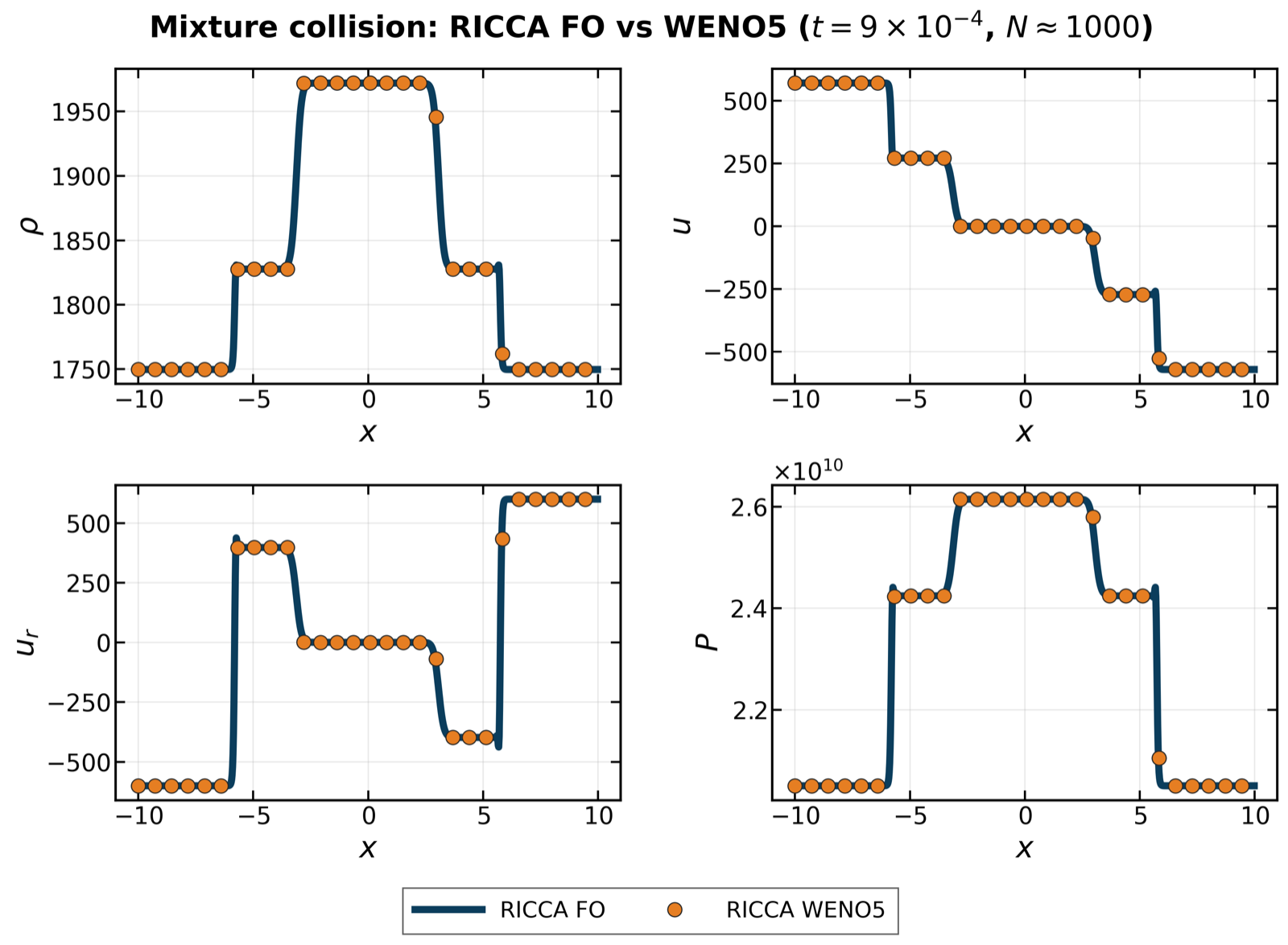}
\caption{1D gas--magma mixture collision (Test case~3), $t=0.0009$:
overlapped RICCA FO (line) versus RICCA--WENO5 (markers).}
\label{fig:1d-col}
\end{figure}

\begin{figure}[htbp]
\centering
\includegraphics[width=0.92\textwidth]{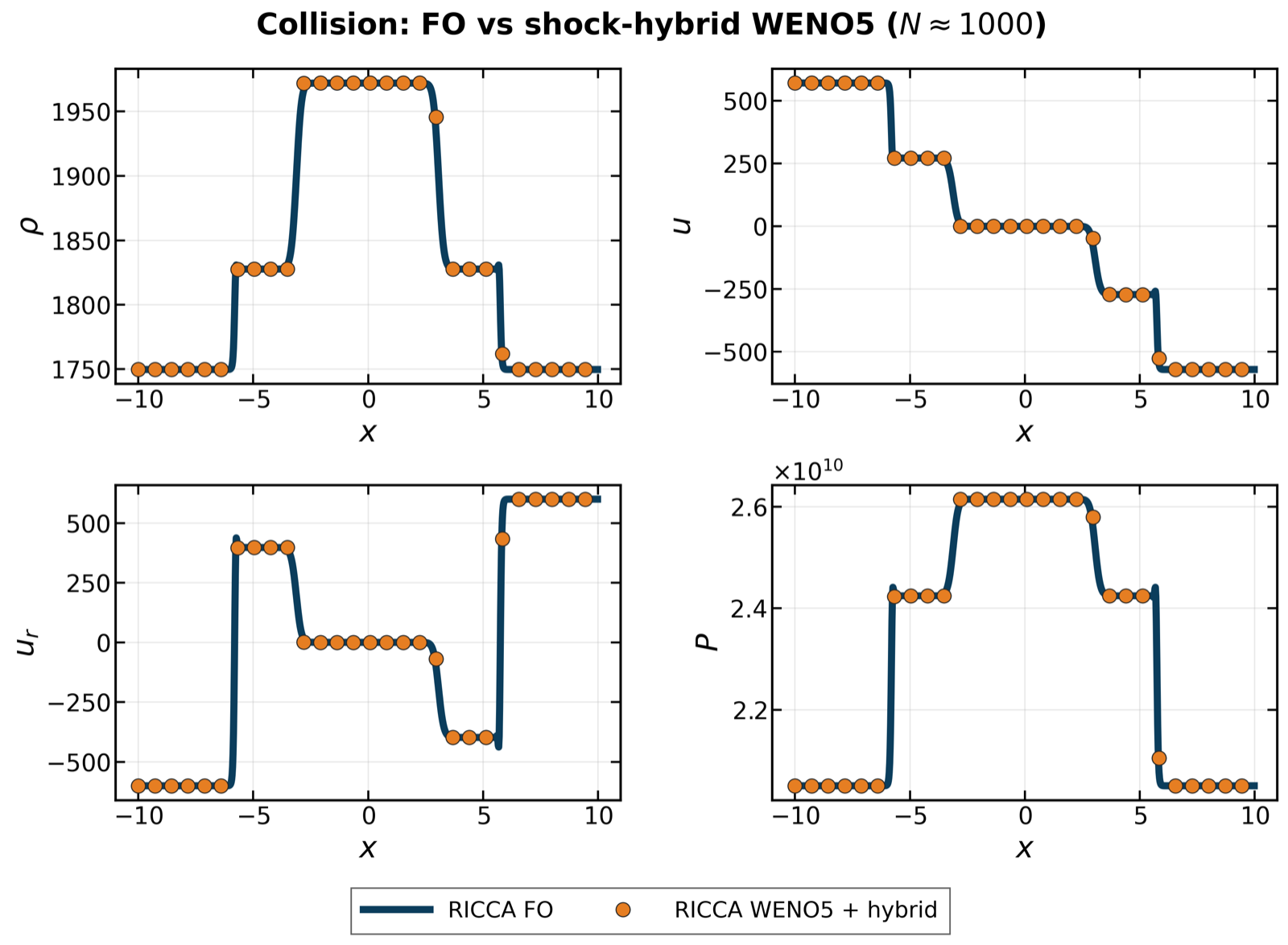}
\caption{1D collision ($N\approx1000$): overlapped FO (line) versus
shock-hybrid WENO5 (markers); same test as Fig.~\ref{fig:1d-col}.}
\label{fig:1d-col-weno}
\end{figure}

End time $t=9\times10^{-4}$, $\alpha=0.8$:
\begin{center}
\small
\begin{tabular}{@{}lcc@{}}
\toprule
 & Left & Right \\
\midrule
$(\rho_1,\rho_2)$ & $(2500,1562)$ & $(2500,1562)$ \\
$(u_1,u_2)$ & $(+1000,+400)$ & $(-1000,-400)$ \\
\bottomrule
\end{tabular}
\end{center}
Colliding streams generate strong shocks about a trivial contact.
Figure~\ref{fig:1d-col} compares RICCA and RICCA--WENO;
Fig.~\ref{fig:1d-col-weno} isolates the effect of the shock-hybrid limiter
on the same collision problem (not a new Riemann case).

\subsection{WENO with the old pressure sensor}
\label{sec:1d-old-sensor}

\begin{figure}[htbp]
\centering
\includegraphics[width=0.95\textwidth]{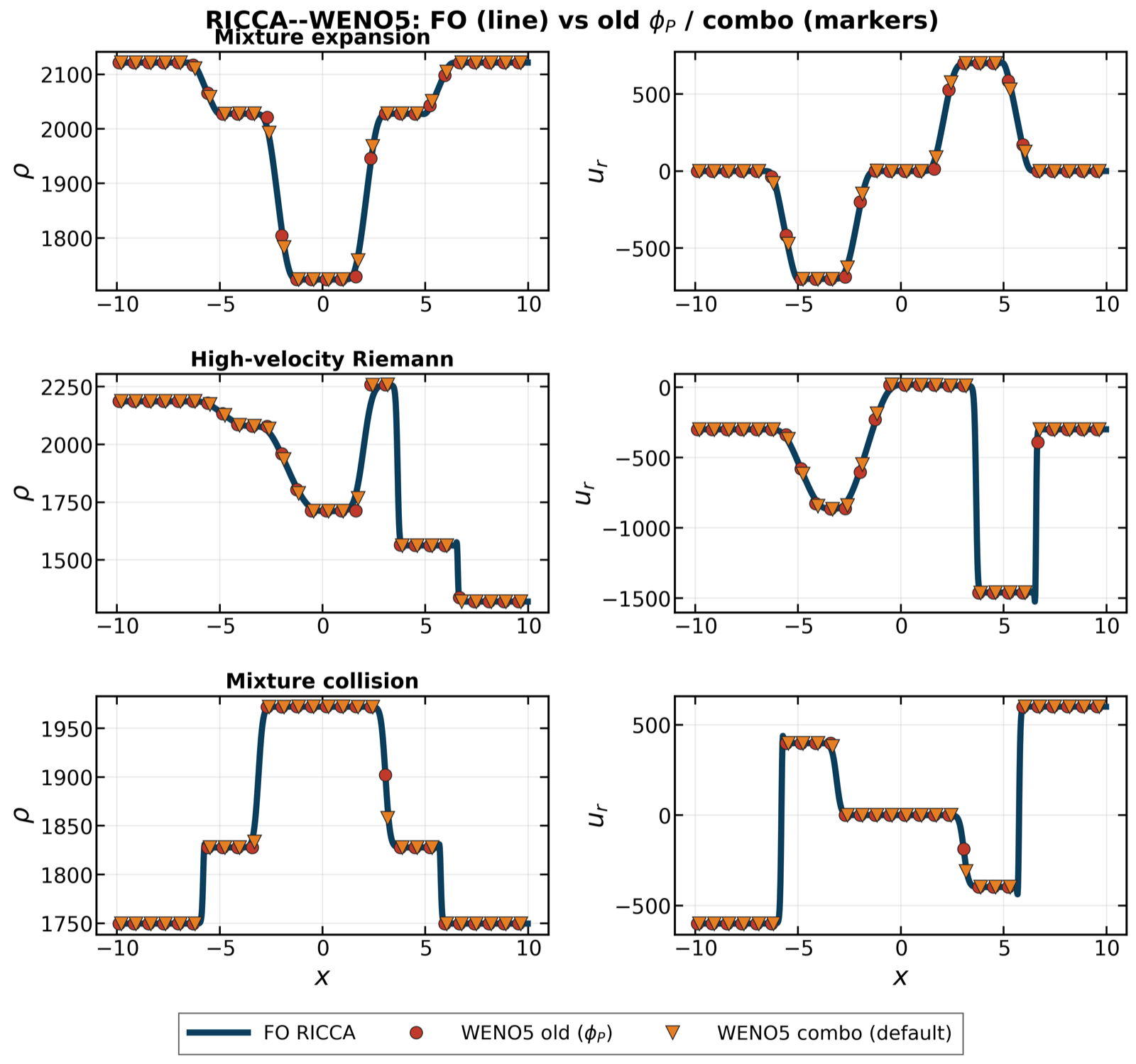}
\caption{1D overlapped comparison: FO RICCA (line) versus WENO5 with old
$\phi_P$ and combo sensors (markers), $N\approx1000$.
Left: $\rho$; right: $u_r$ (one row per Riemann problem).}
\label{fig:1d-old-combo}
\end{figure}

Unless noted otherwise, RICCA--WENO5 profiles use the default combo sensor.
Figure~\ref{fig:1d-old-combo} collects, in one place, the comparison of
first-order RICCA, WENO5 with the \emph{old} pressure-only sensor
$\phi=\phi_P$, and WENO5 with the combo sensor for all three 1D tests
($N\approx1000$; $\rho$ and $u_r$).
The expansion A/B study of Section~\ref{sec:weno} already details the
sensor metrics; Fig.~\ref{fig:sensor-suite} extends them quantitatively to
HighV and collision.
The overlays in Fig.~\ref{fig:1d-old-combo} are qualitative confirmation:
the old sensor leaves more ringing in $\rho$ and $u_r$, while the combo
sensor tracks FO more closely without erasing WENO sharpness.
As noted in the sensor suite (Fig.~\ref{fig:sensor-suite}), not every single
metric improves on every case---in particular the peak
$|u_r-u_r^{\mathrm{FO}}|$ residual is case-dependent; it can increase on the
collision where the FO reference itself carries a strong $u_r$ spike.

\subsection{RICCA versus FORCE in 1D}
\label{sec:1d-force}

\begin{figure}[htbp]
\centering
\includegraphics[width=0.92\textwidth]{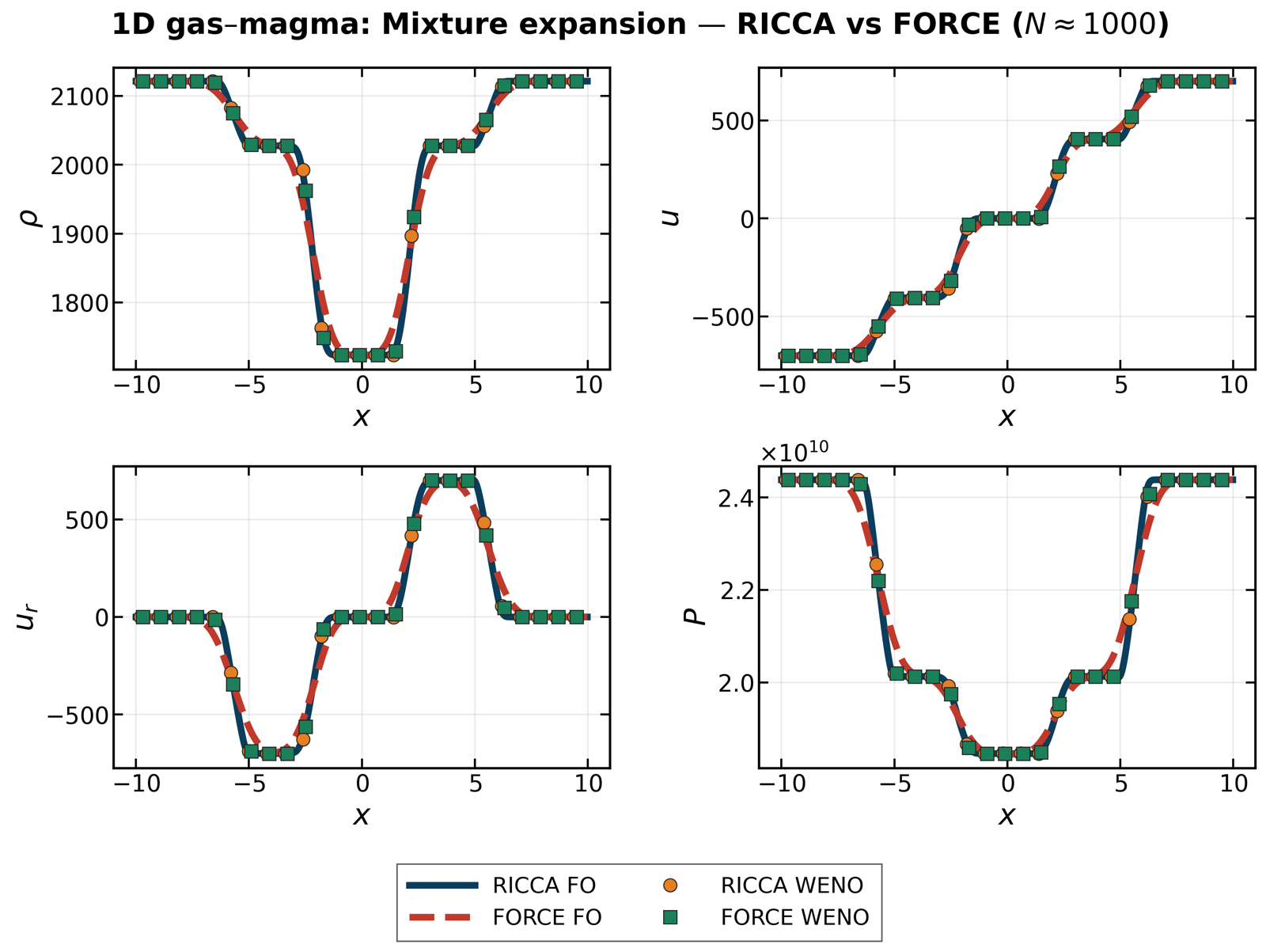}
\caption{1D mixture expansion: overlapped RICCA/FORCE comparison
(FO as lines, WENO5 as markers), $\rho$, $u$, $u_r$, $P$.}
\label{fig:1d-rf-exp}
\end{figure}

\begin{figure}[htbp]
\centering
\includegraphics[width=0.92\textwidth]{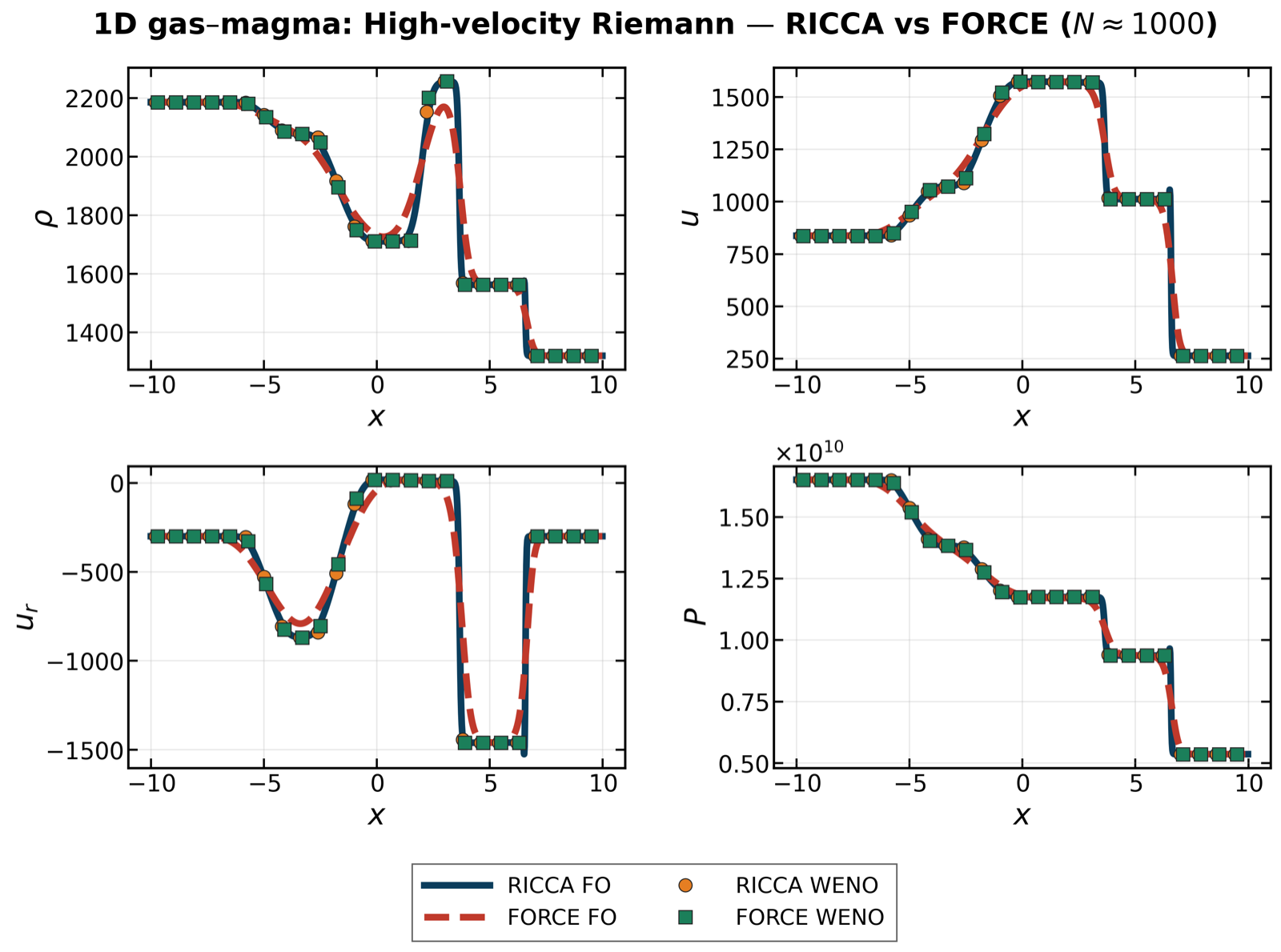}
\caption{1D high-velocity Riemann: overlapped RICCA/FORCE
(FO lines, WENO5 markers).}
\label{fig:1d-rf-highv}
\end{figure}

\begin{figure}[htbp]
\centering
\includegraphics[width=0.92\textwidth]{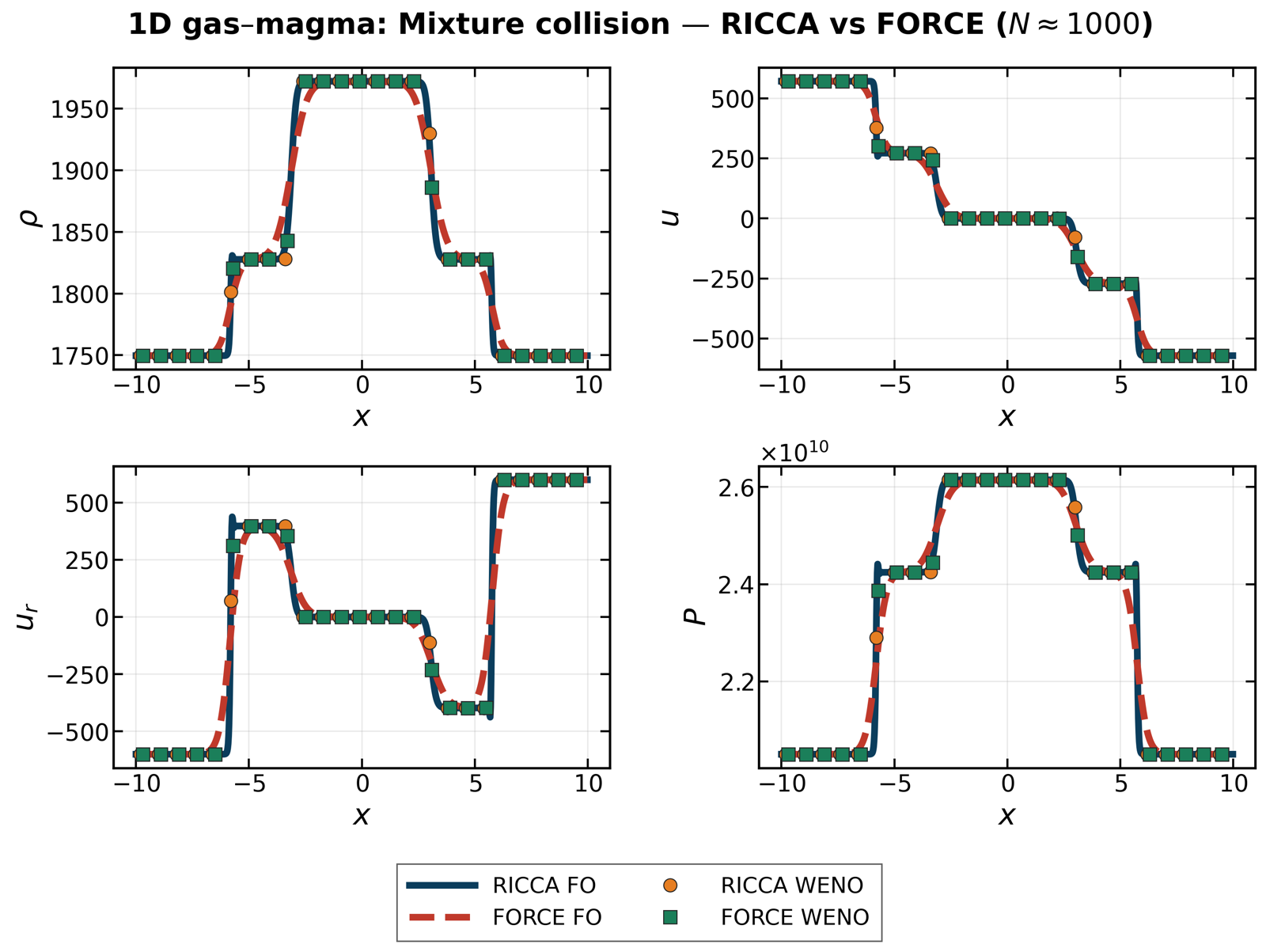}
\caption{1D mixture collision: overlapped RICCA/FORCE
(FO lines, WENO5 markers).}
\label{fig:1d-rf-col}
\end{figure}

Having shown RICCA FO/WENO for each Riemann problem above, we now compare
the two centred fluxes on the same $N\approx1000$ mesh
(Figs.~\ref{fig:1d-rf-exp}--\ref{fig:1d-rf-col}).
FORCE (Section~\ref{sec:force}) recovers the same wave pattern as RICCA;
RICCA is typically less diffusive at contacts, while FORCE remains a robust
centred baseline.
WENO5 sharpens shocks and contacts for both schemes.

\subsection{RICCA--RH versus RICCA and FORCE}
\label{sec:1d-rh}

\begin{figure}[htbp]
\centering
\includegraphics[width=0.95\textwidth]{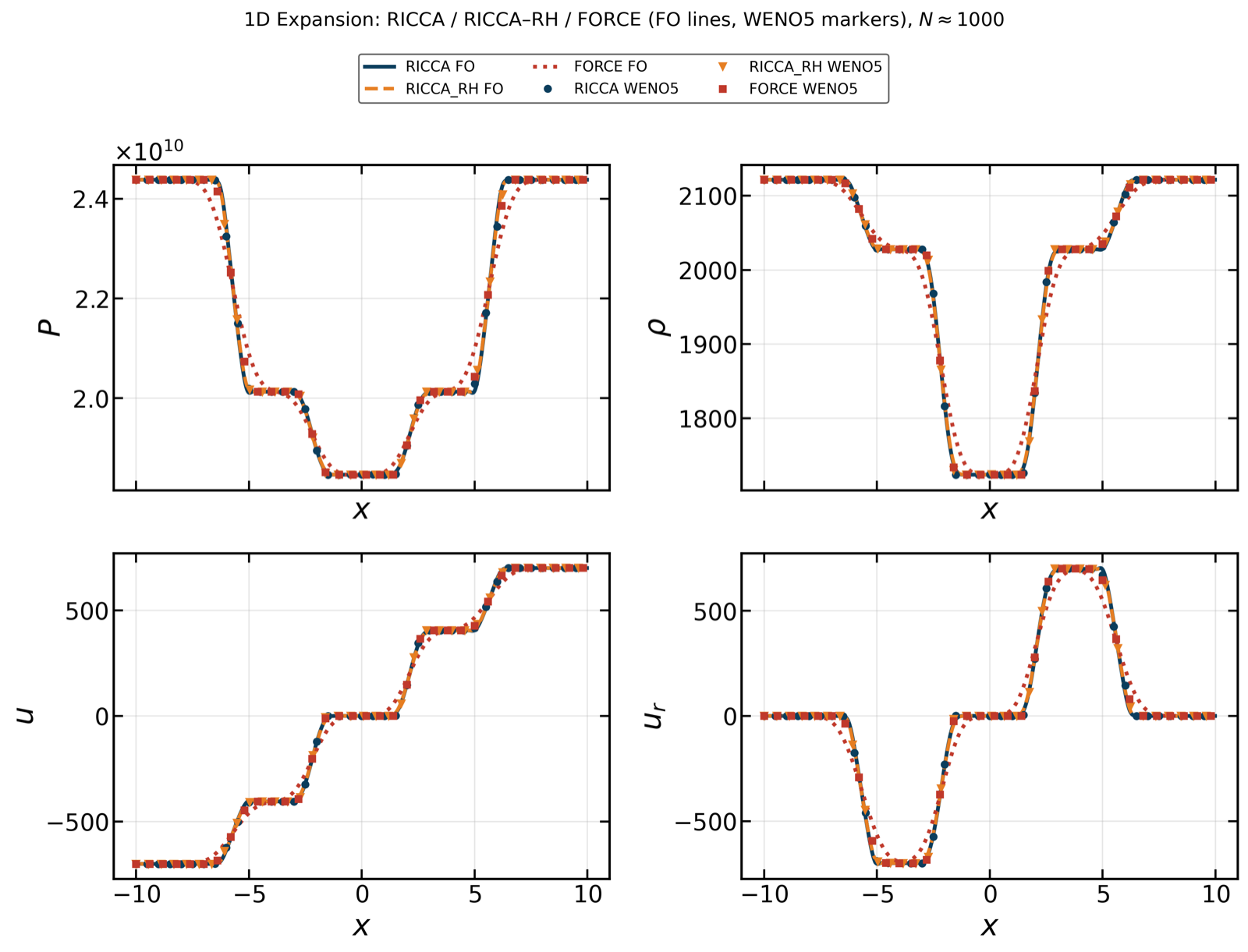}
\caption{1D mixture expansion: RICCA / RICCA--RH / FORCE
(FO lines, WENO5 markers), $N\approx1000$.}
\label{fig:1d-rh-exp}
\end{figure}

\begin{figure}[htbp]
\centering
\includegraphics[width=0.95\textwidth]{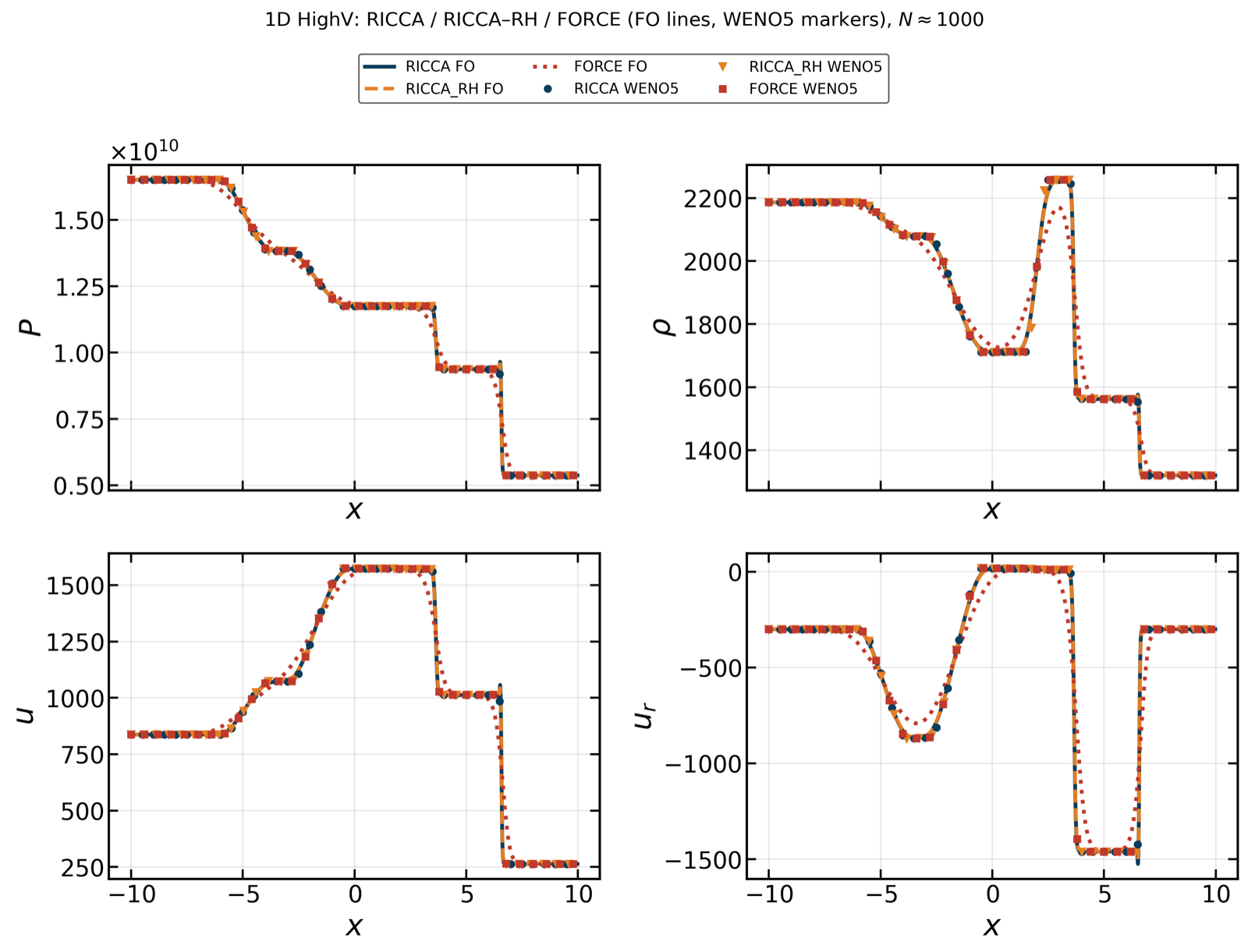}
\caption{1D high-velocity Riemann: RICCA / RICCA--RH / FORCE
(FO lines, WENO5 markers).}
\label{fig:1d-rh-highv}
\end{figure}

\begin{figure}[htbp]
\centering
\includegraphics[width=0.95\textwidth]{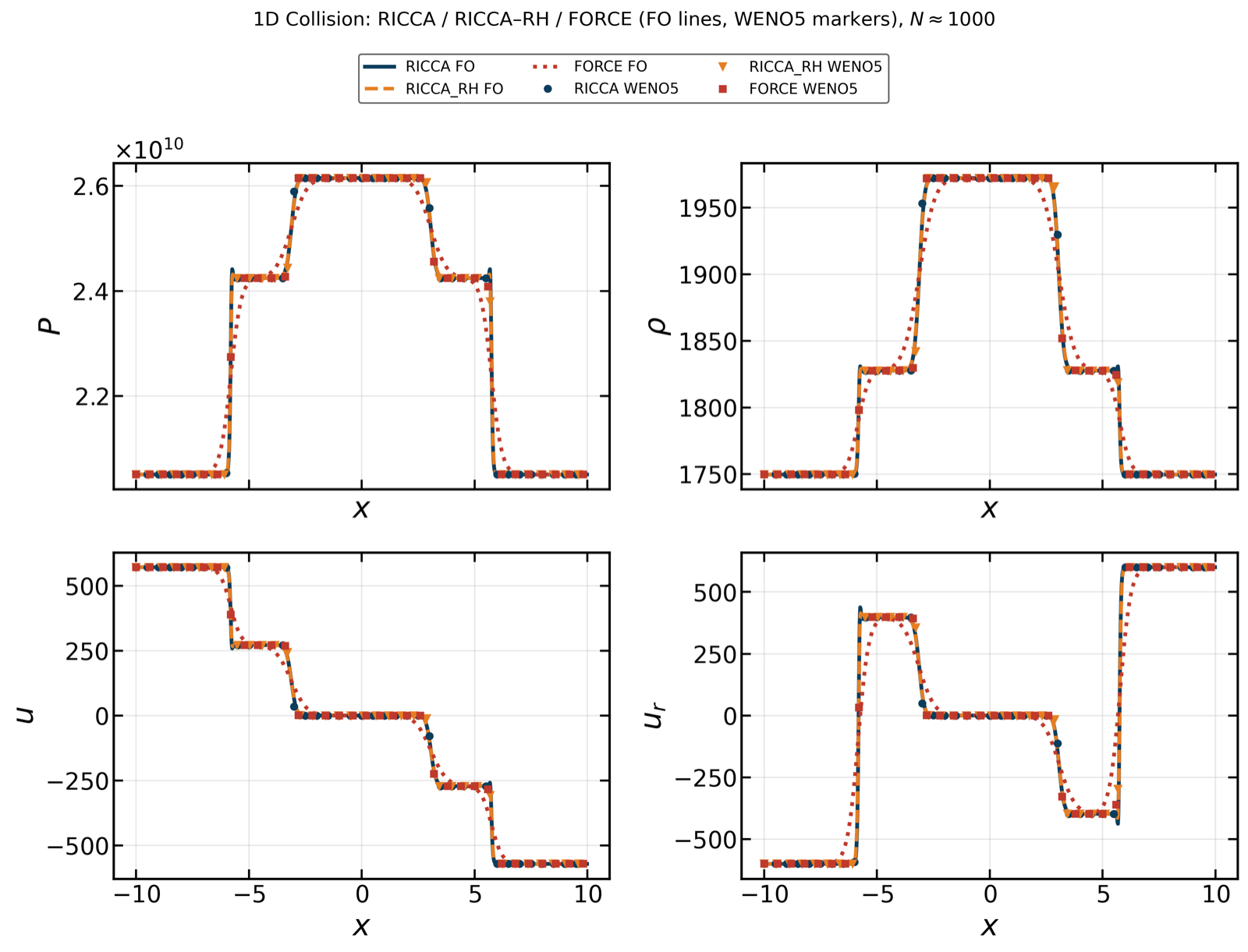}
\caption{1D mixture collision: RICCA / RICCA--RH / FORCE
(FO lines, WENO5 markers).}
\label{fig:1d-rh-col}
\end{figure}

We now include the RH-motivated dual-speed flux of Section~\ref{sec:ricca-rh}
on the same $N\approx1000$ meshes.
Figures~\ref{fig:1d-rh-exp}--\ref{fig:1d-rh-col} overlay RICCA, RICCA--RH
and FORCE (FO lines, WENO5 markers).
Production RICCA--RH ($\kappa=1$, binary pressure/RH switch) tracks legacy
RICCA closely and remains sharper than FORCE FO at contacts; WENO5 solutions
of all three fluxes collapse onto essentially the same profiles.

\begin{figure}[htbp]
\centering
\includegraphics[width=0.85\textwidth]{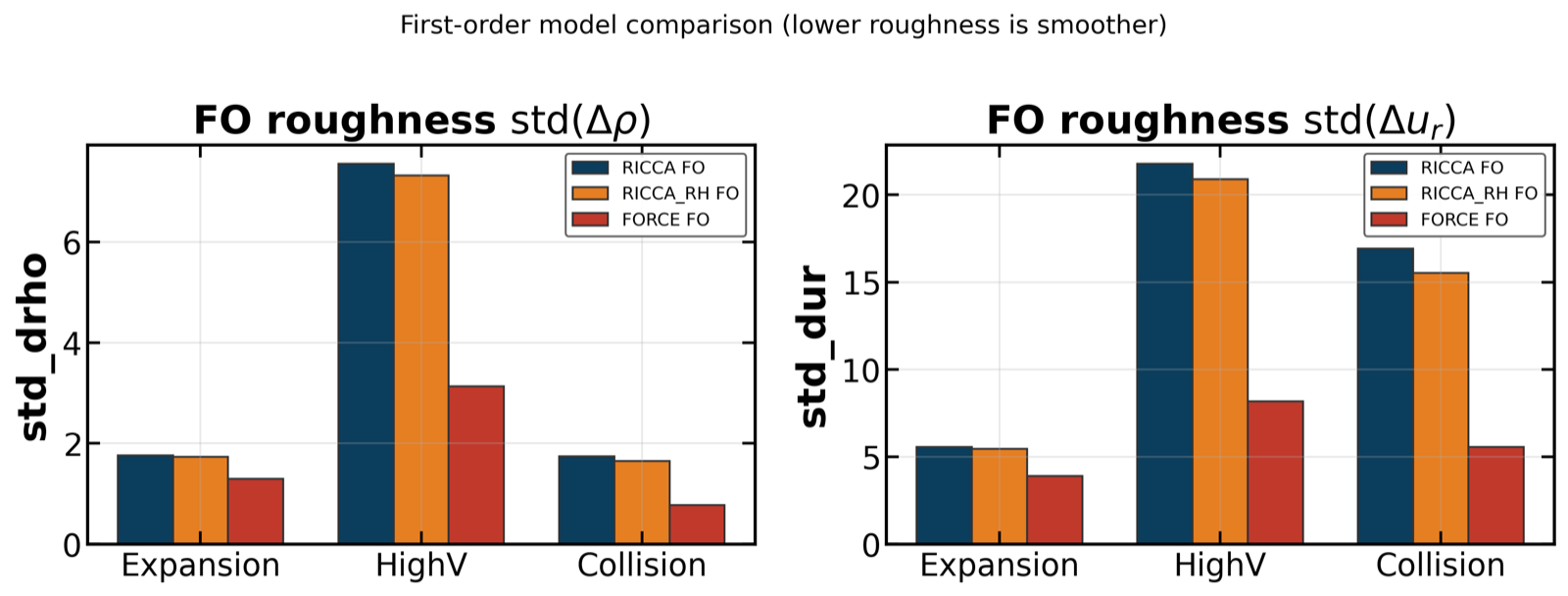}
\caption{FO density-roughness comparison of RICCA, RICCA--RH and FORCE on the
three 1D gas--magma problems ($N\approx1000$).}
\label{fig:1d-rh-fo-met}
\end{figure}

Table~\ref{tab:1d-rh-fo} and Fig.~\ref{fig:1d-rh-fo-met} summarise FO density
roughness $\mathrm{std}(\Delta\rho)$ and peak slip $\max\lvert u_r\rvert$.
RICCA--RH is slightly smoother than legacy RICCA on all three problems
(e.g.\ expansion $\mathrm{std}(\Delta\rho)$: $1.758\to1.728$; collision:
$1.739\to1.647$); FORCE FO remains the most dissipative, as expected from the
single-speed LF mechanism.

\begin{table}[htbp]
\centering
\caption{First-order density roughness and peak relative velocity on
$N\approx1000$ meshes (lower $\mathrm{std}(\Delta\rho)$ is smoother).}
\label{tab:1d-rh-fo}
\begin{tabular}{@{}llrrr@{}}
\toprule
Problem & Scheme (FO) & $\mathrm{std}(\Delta\rho)$ & $\max|\Delta\rho|$ & $\max|u_r|$ \\
\midrule
Expansion & RICCA & 1.758 & 6.050 & 703.8 \\
          & RICCA--RH & 1.728 & 5.870 & 700.0 \\
          & FORCE & 1.292 & 3.460 & 693.8 \\
\midrule
HighV & RICCA & 7.550 & 83.350 & 1524.3 \\
      & RICCA--RH & 7.321 & 78.560 & 1503.6 \\
      & FORCE & 3.133 & 15.400 & 1459.3 \\
\midrule
Collision & RICCA & 1.739 & 14.990 & 600.0 \\
          & RICCA--RH & 1.647 & 13.090 & 600.0 \\
          & FORCE & 0.775 & 2.110 & 600.0 \\
\bottomrule
\end{tabular}

\end{table}

\begin{figure}[htbp]
\centering
\includegraphics[width=0.95\textwidth]{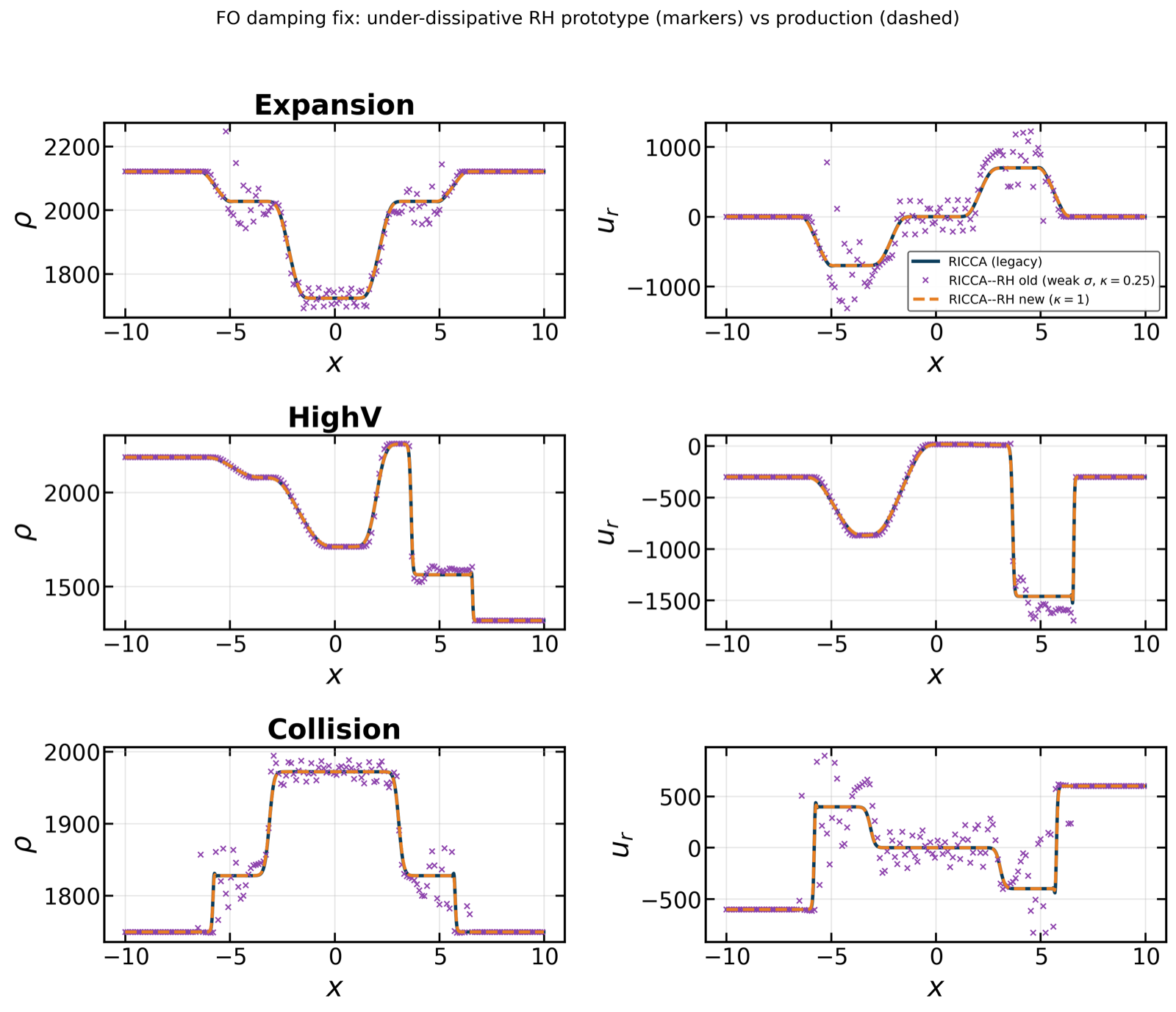}
\caption{FO damping lesson: under-dissipative RICCA--RH prototype
(soft $\sigma$, $\kappa=0.25$; markers) versus production RICCA--RH
($\kappa=1$; dashed) and legacy RICCA (solid).
Rows: expansion, HighV, collision; columns: $\rho$ and $u_r$.}
\label{fig:1d-rh-old}
\end{figure}

\begin{figure}[htbp]
\centering
\includegraphics[width=0.85\textwidth]{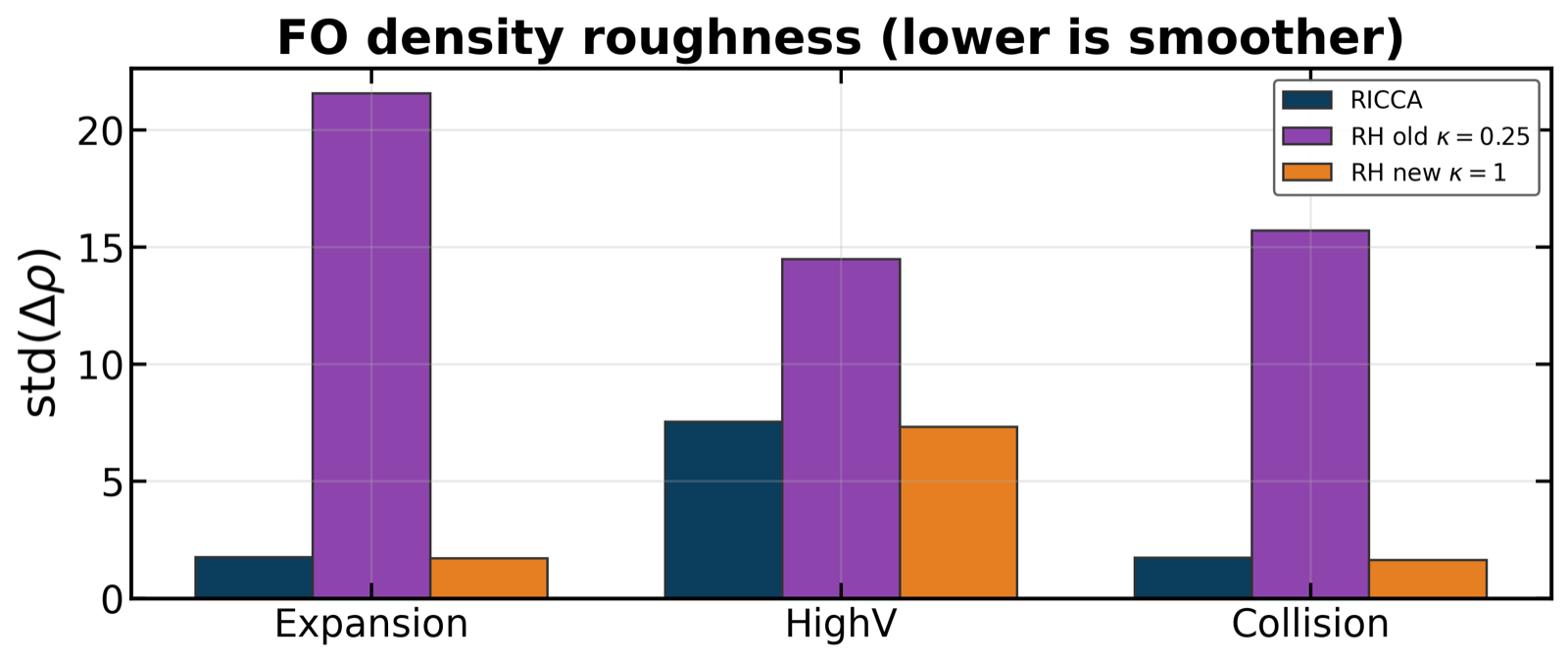}
\caption{FO density roughness $\mathrm{std}(\Delta\rho)$ for legacy RICCA,
the oscillatory RICCA--RH prototype and production RICCA--RH.}
\label{fig:1d-rh-old-metrics}
\end{figure}

\paragraph{Under-dissipative prototype (cautionary result).}
An early continuous soft-$\sigma$ prototype with contact factor
$\kappa=0.25$ under-dissipated contacts and relative-velocity jumps, producing
severe ringing---especially on the collision problem
(Fig.~\ref{fig:1d-rh-old}).
On the expansion tube the FO density roughness rose from
$\mathrm{std}(\Delta\rho)\approx1.73$ (production) to $\approx21.6$
(prototype); peak $|u_r|$ roughly doubled.
Raising $\kappa$ to~$1$ and replacing the soft switch by a binary
pressure/RH trip removes the oscillations and recovers a solution
indistinguishable from legacy RICCA at plotting resolution
(Fig.~\ref{fig:1d-rh-old-metrics}).
We report this deliberately: soft continuous $\sigma\in[0,1]$ alone was
\emph{not} sufficient on these strong gas--magma shocks; adequate contact
$B$ was required for robustness.

\subsection{Comparison with HLLC and Roe}
\label{sec:1d-upwind}

\begin{figure}[htbp]
\centering
\includegraphics[width=\textwidth]{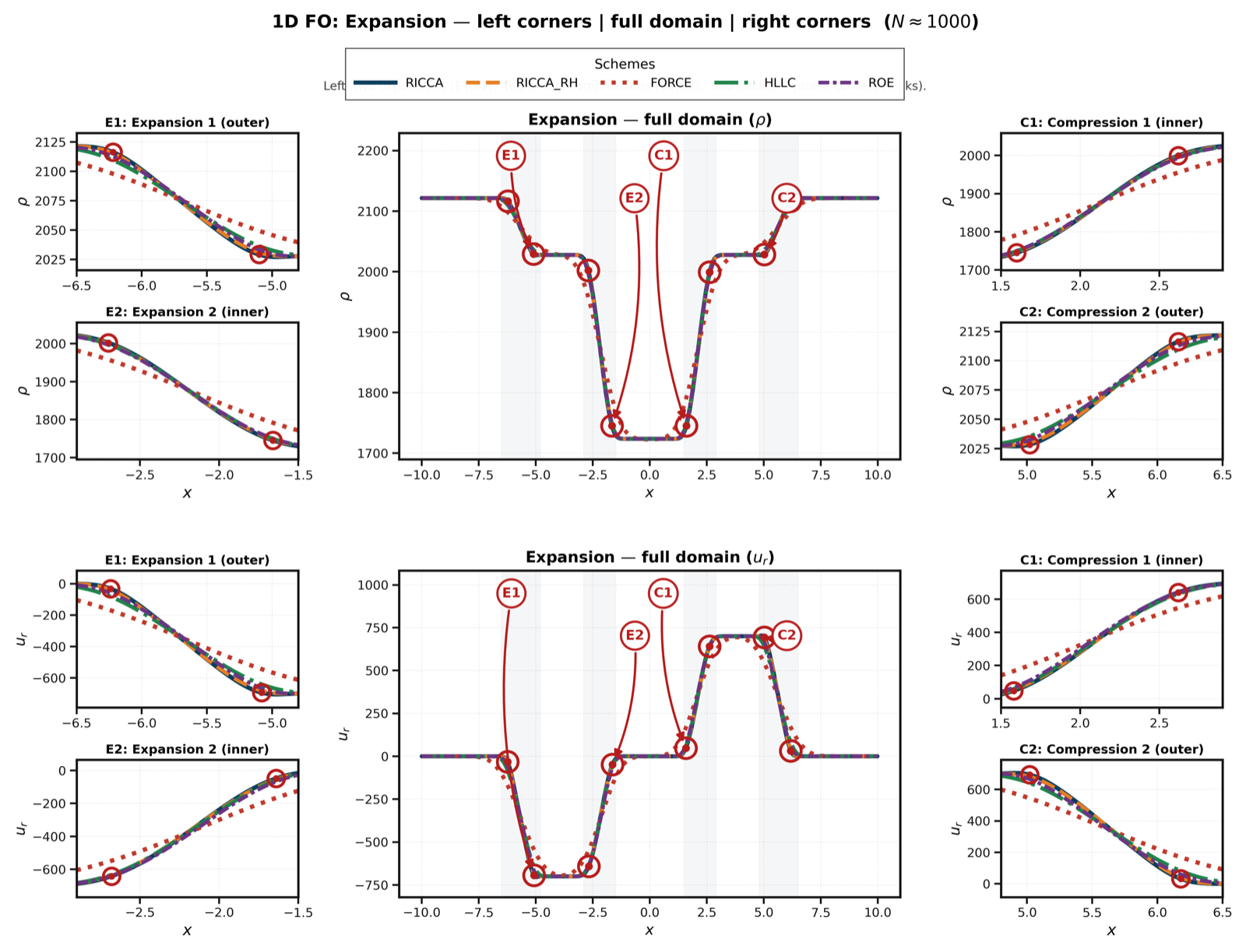}
\caption{1D FO expansion: RICCA, RICCA--RH, FORCE, HLLC and Roe
($N\approx1000$).
Top: $\rho$; bottom: $u_r$.
Each row is two left-corner zooms $|$ full domain $|$ two right-corner zooms.
Left = expansions E1/E2; right = compressions C1/C2.
Red circles mark profile corners (kinks).}
\label{fig:1d-upwind-exp}
\end{figure}

\begin{figure}[htbp]
\centering
\includegraphics[width=\textwidth]{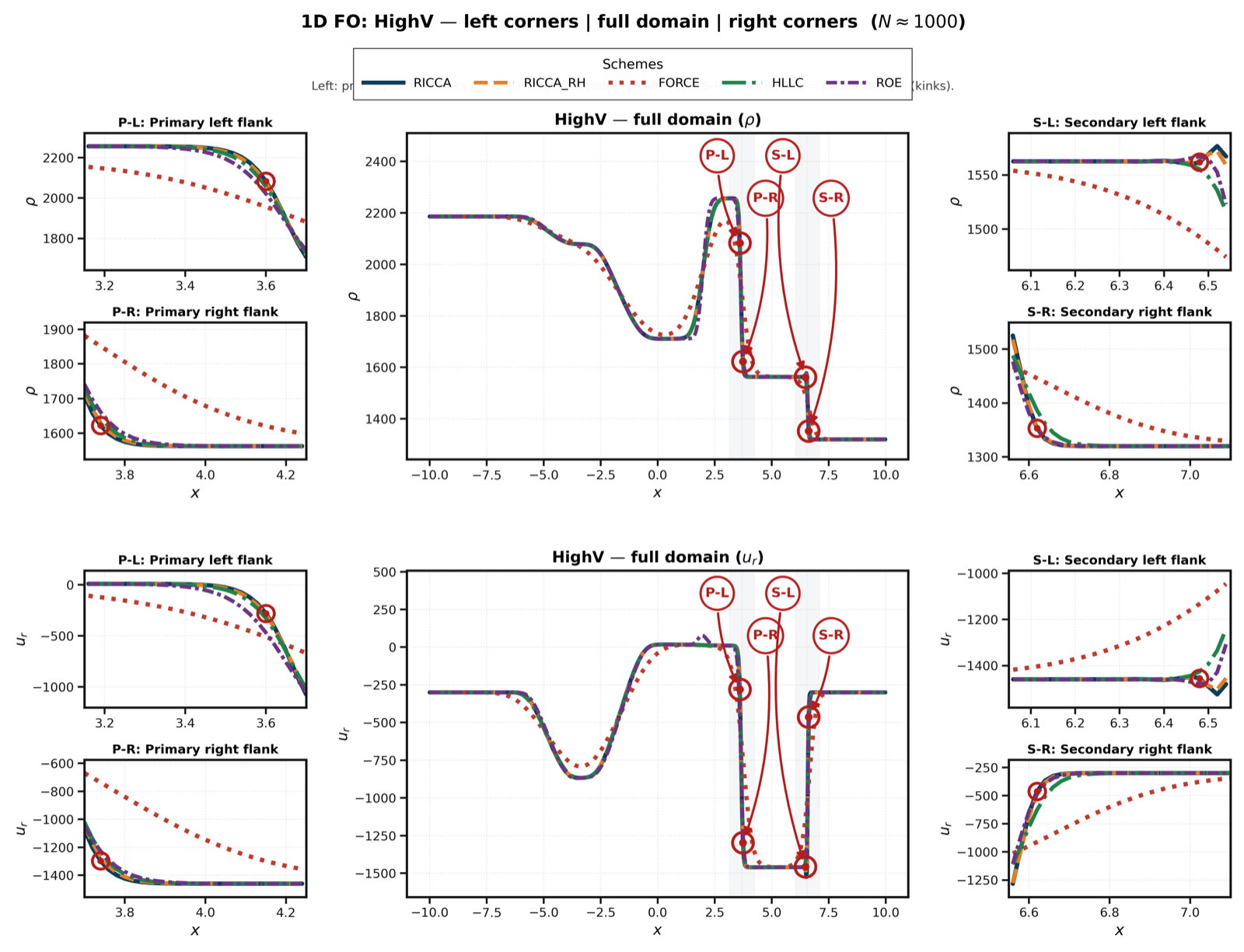}
\caption{Same layout as Fig.~\ref{fig:1d-upwind-exp} for the high-velocity
Riemann problem.
Left: primary-shock flanks (P-L, P-R); right: secondary flanks (S-L, S-R).}
\label{fig:1d-upwind-highv}
\end{figure}

\begin{figure}[htbp]
\centering
\includegraphics[width=\textwidth]{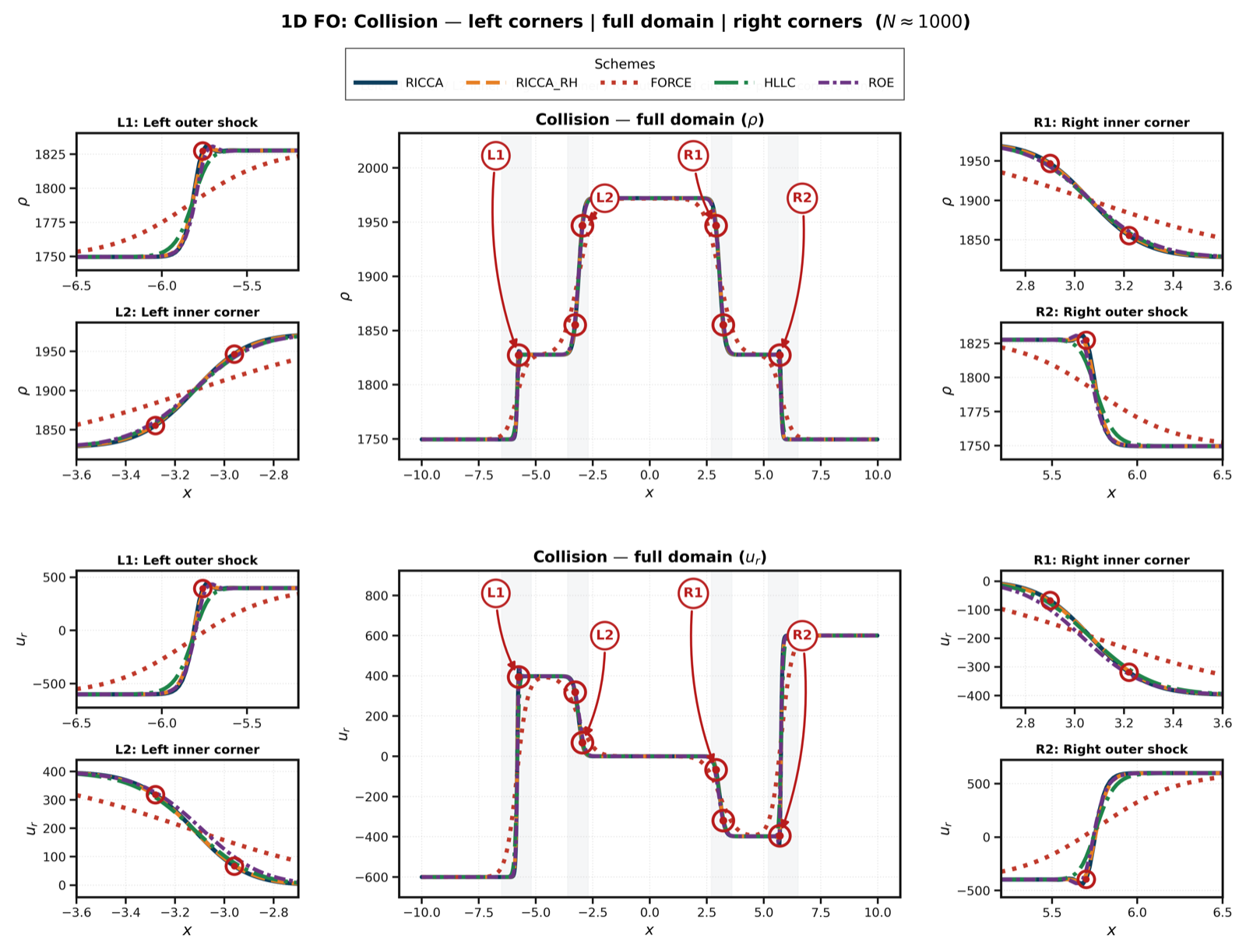}
\caption{Same layout as Fig.~\ref{fig:1d-upwind-exp} for the mixture collision.
Left: L1 outer / L2 inner; right: R1 inner / R2 outer.}
\label{fig:1d-upwind-col}
\end{figure}

\begin{figure}[htbp]
\centering
\includegraphics[width=0.85\textwidth]{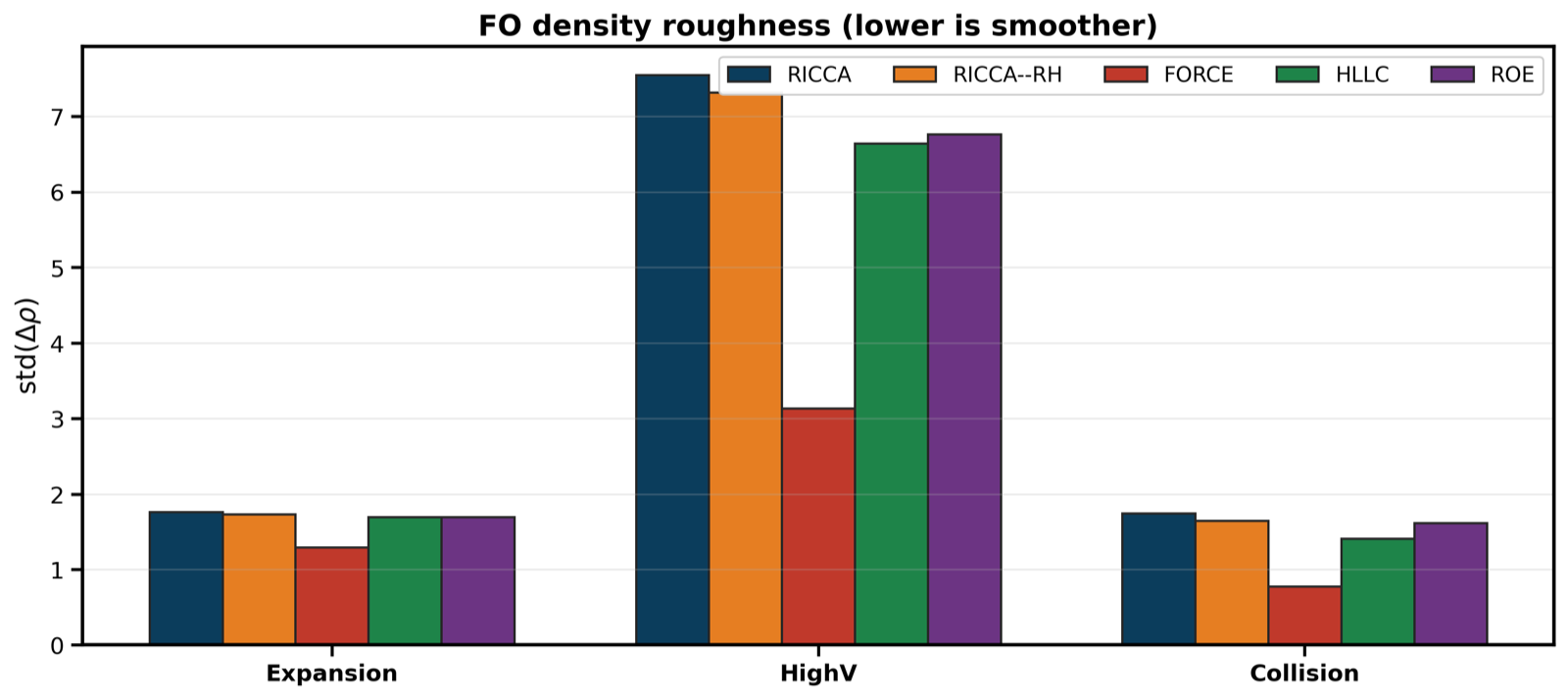}
\caption{FO density roughness $\mathrm{std}(\Delta\rho)$ for the five fluxes
on the three gas--magma problems.}
\label{fig:1d-upwind-met}
\end{figure}

Figures~\ref{fig:1d-upwind-exp}--\ref{fig:1d-upwind-met} and
Table~\ref{tab:1d-upwind} place RICCA and RICCA--RH against FORCE and the
upwind HLLC/Roe solvers of Sections~\ref{sec:hllc}--\ref{sec:roe} on the same
$N\approx1000$ FO meshes.
HLLC (Rusanov-stable production path) and Roe track the RICCA family closely;
FORCE remains the most dissipative centred reference.
On HighV and collision, HLLC/Roe reduce FO density roughness relative to
legacy RICCA while remaining sharper than FORCE, consistent with their
upwind/acoustic dissipation.

Figures~\ref{fig:1d-upwind-exp}--\ref{fig:1d-upwind-col} show each 1D case
as a separate full-width diagram so panels remain print-readable without
heavy zooming.
For both $\rho$ and $u_r$, a centre full-domain profile is flanked by
\emph{two} left and \emph{two} right corner zooms outside the main axes.
On expansion (Fig.~\ref{fig:1d-upwind-exp}) the left pair are the two
expansion corners (E1 outer, E2 inner) and the right pair the two compression
corners (C1 inner, C2 outer); HighV and collision
(Figs.~\ref{fig:1d-upwind-highv}--\ref{fig:1d-upwind-col}) use the same
layout for their wave corners.
Red circles mark profile \emph{corners} (curvature kinks), not mid-slope
points.
At full-domain scale the five fluxes share the same wave pattern; the corner
panels discriminate them locally: FORCE rounds every corner, while RICCA,
RICCA--RH, HLLC and Roe remain tightly clustered and sharp.
On the collision slip corners, HLLC and Roe show a mild overshoot that is
absent (or smaller) in RICCA/RICCA--RH.
Global metrics are in Fig.~\ref{fig:1d-upwind-met} and Table~\ref{tab:1d-upwind}.

\begin{table}[htbp]
\centering
\caption{First-order density roughness and peak relative velocity
($N\approx1000$).}
\label{tab:1d-upwind}
\begin{tabular}{@{}llrr@{}}
\toprule
Problem & Scheme (FO) & $\mathrm{std}(\Delta\rho)$ & $\max|u_r|$ \\
\midrule
Expansion & RICCA & 1.758 & 703.8 \\
 & RICCA--RH & 1.728 & 700.0 \\
 & FORCE & 1.292 & 693.8 \\
 & HLLC & 1.694 & 700.3 \\
 & ROE & 1.692 & 700.4 \\
\midrule
HighV & RICCA & 7.550 & 1524.3 \\
 & RICCA--RH & 7.321 & 1503.6 \\
 & FORCE & 3.133 & 1459.3 \\
 & HLLC & 6.645 & 1459.9 \\
 & ROE & 6.763 & 1479.4 \\
\midrule
Collision & RICCA & 1.739 & 600.0 \\
 & RICCA--RH & 1.647 & 600.0 \\
 & FORCE & 0.775 & 600.0 \\
 & HLLC & 1.407 & 600.0 \\
 & ROE & 1.617 & 600.0 \\
\bottomrule
\end{tabular}

\end{table}

%==============================================================================
\section{Two-dimensional gas--liquid benchmarks}
\label{sec:2d}

The 2D tests follow \cite{Zeidan2019} with equal-EOS gas--liquid closure
(Section~\ref{sec:ge}).
RICCA--WENO with the \emph{combo} hybrid sensor
(Section~\ref{sec:weno}) is the primary scheme; FORCE FO/WENO uses the same
reconstruction and hybrid on the same meshes.
Production RICCA--RH (Section~\ref{sec:ricca-rh}) is assessed on the same
grids with the same WENO5 hybrid (Section~\ref{sec:2d-rh}).
Source terms are omitted; global CFL matches the reference.
All WENO solutions reported below were recomputed with the combo limiter.
We emphasise the scope of the suite: the expansion and collision tubes exercise
nontrivial relative velocity, whereas the Mach~3 forward-facing step
(Section~\ref{sec:2d-ffs}) is effectively single-velocity and should be read as
a shock-capturing benchmark rather than a full two-velocity validation.

\subsection{Expansion tube}
\label{sec:2d-exp}

\begin{figure}[htbp]
\centering
\includegraphics[width=\textwidth]{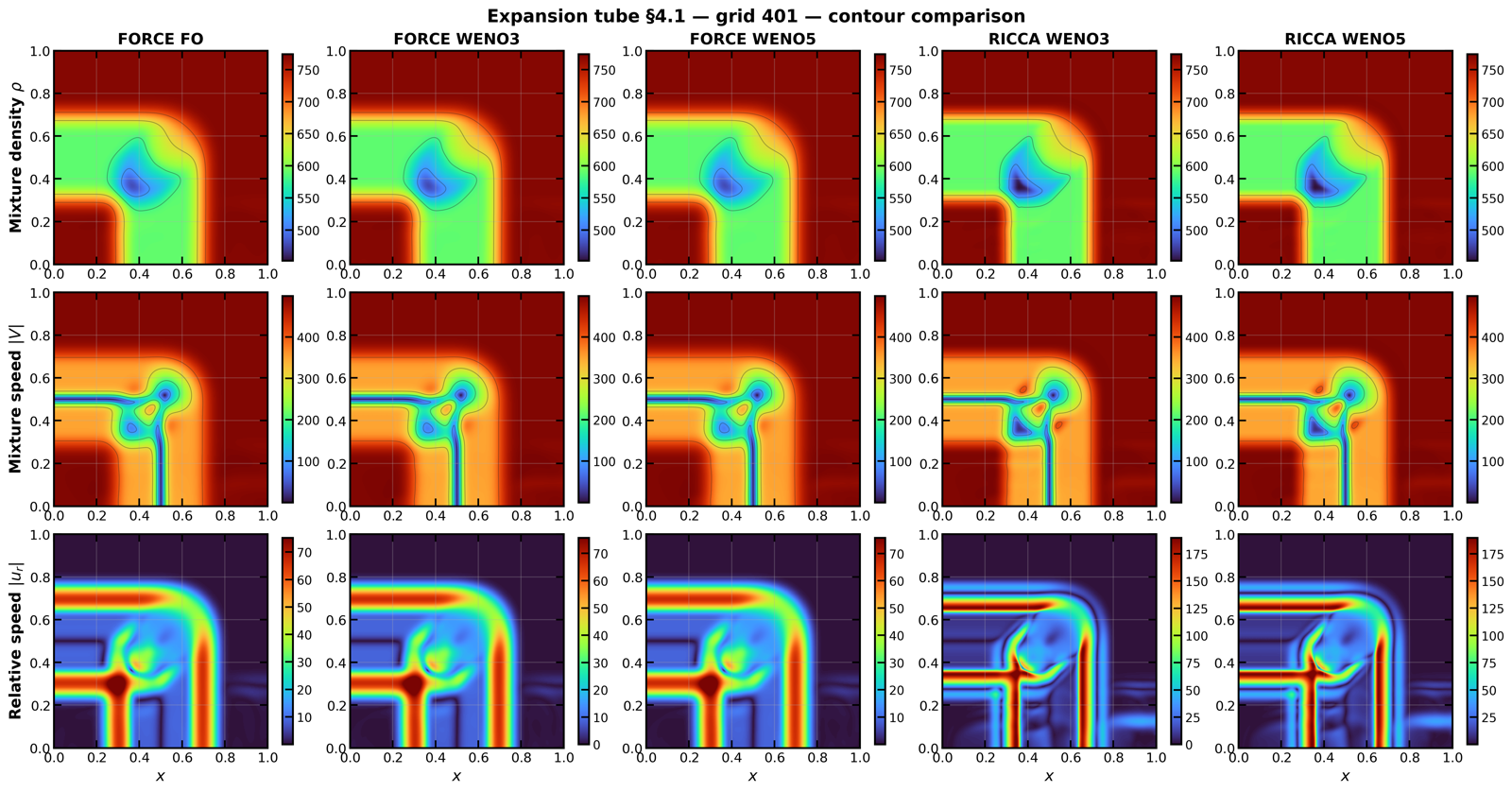}
\caption{2D expansion tube: $\rho$, $|V|$, $|u_r|$ at $t=1.3\times10^{-4}$
on $400\times400$ (FORCE FO/WENO3/WENO5 and RICCA WENO3/WENO5;
filled + line contours).}
\label{fig:exp401c}
\end{figure}

\begin{figure}[htbp]
\centering
\includegraphics[width=\textwidth]{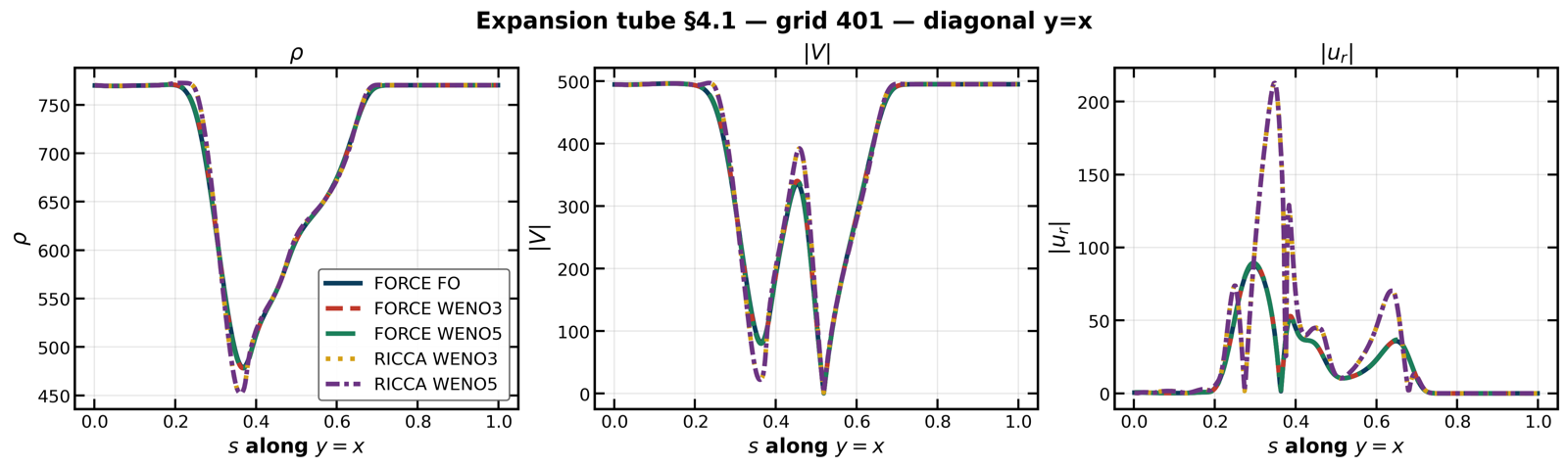}
\caption{2D expansion tube: diagonal cuts $y=x$.}
\label{fig:exp401l}
\end{figure}

Domain $[0,1]^2$, $t=1.3\times10^{-4}$, $P=1\,\mathrm{gPa}$, $\alpha=0.9$,
$\rho_1=1225.8912$, $\rho_2=719.6856$;
lower-left quarter velocities $(-350,-350)$, elsewhere $(+350,+350)$.
Figures~\ref{fig:exp401c}--\ref{fig:exp401l} show $400\times400$ solutions.
RICCA--WENO recovers the rarefaction--contact pattern of
\cite{Zeidan2019}; WENO sharpens contacts relative to FORCE FO.

\subsection{Mixture collision}
\label{sec:2d-col}

\begin{figure}[htbp]
\centering
\includegraphics[width=\textwidth]{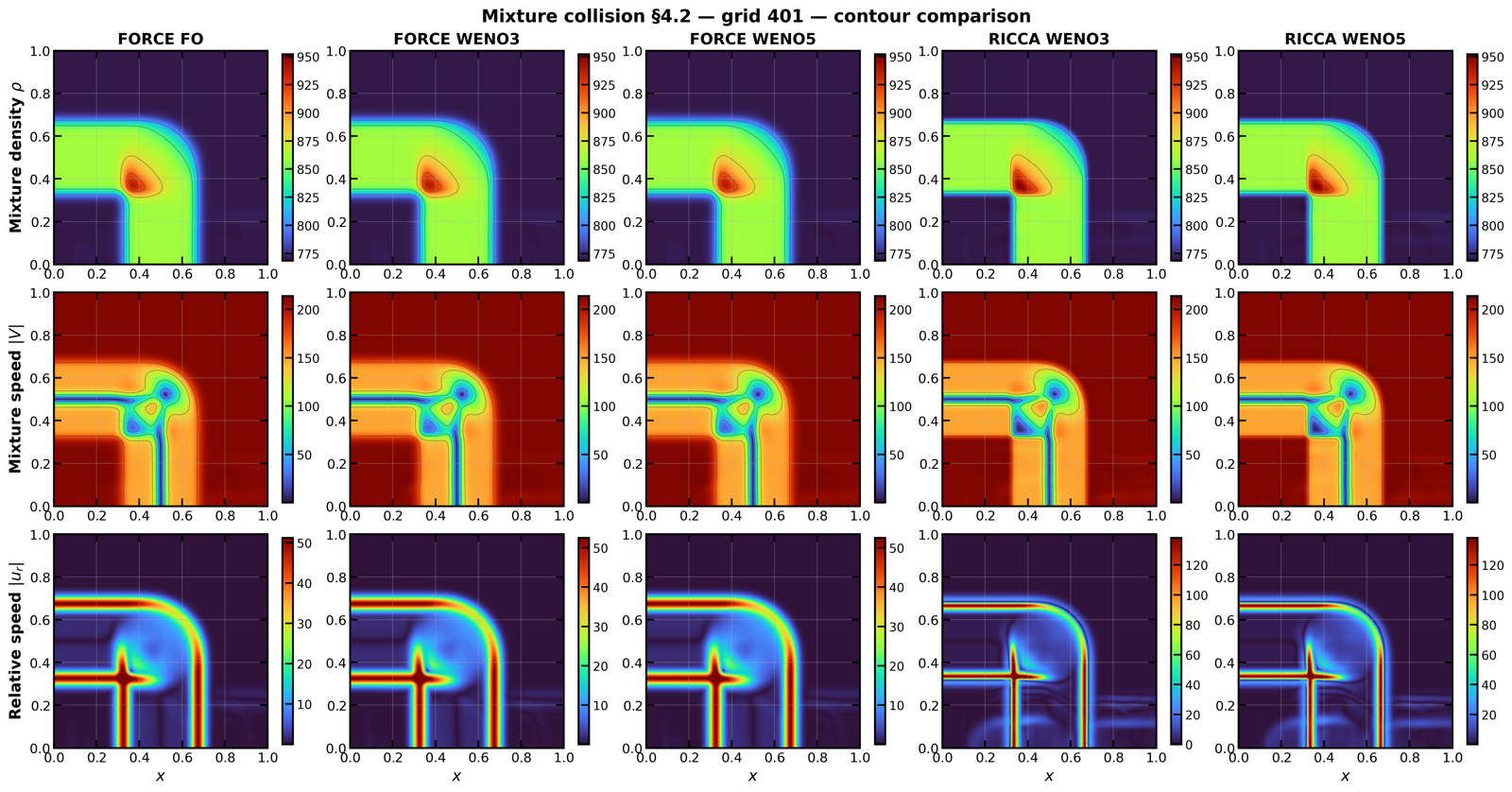}
\caption{2D mixture collision on $400\times400$ cells.}
\label{fig:col401c}
\end{figure}

\begin{figure}[htbp]
\centering
\includegraphics[width=\textwidth]{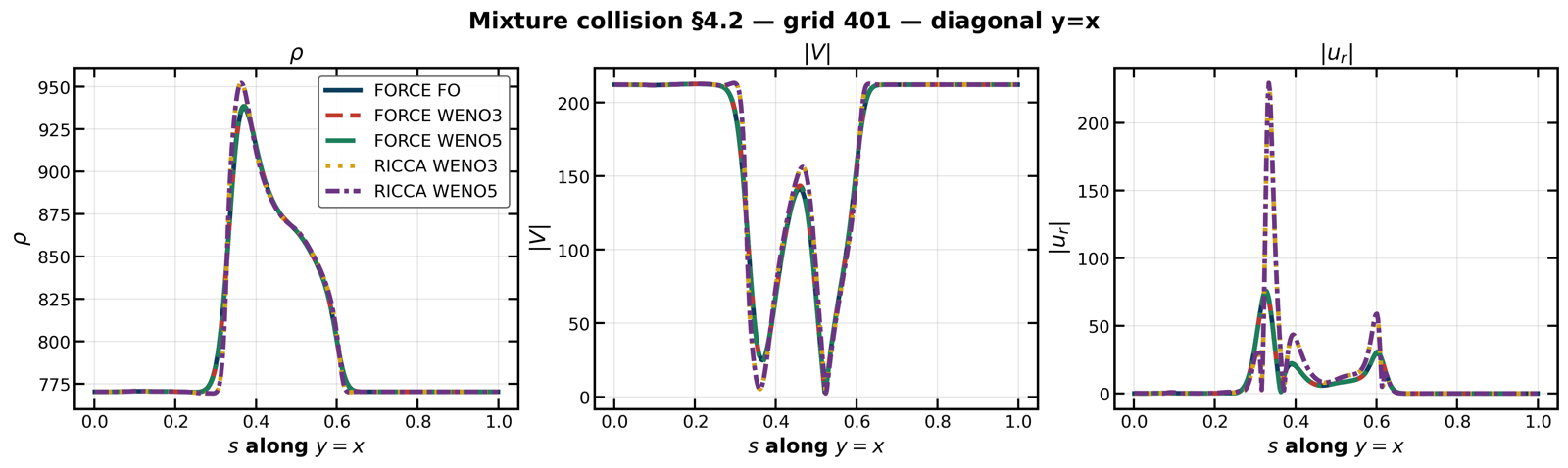}
\caption{2D mixture collision: diagonal cuts $y=x$.}
\label{fig:col401l}
\end{figure}

Same domain/EOS/time; velocities $\pm150$ with opposite orientation to the
expansion tube.
Figures~\ref{fig:col401c}--\ref{fig:col401l} show colliding shocks about a
trivial contact; RICCA--WENO matches FORCE on the same mesh.

\subsection{RICCA--RH in two dimensions}
\label{sec:2d-rh}

\begin{figure}[htbp]
\centering
\includegraphics[width=0.95\textwidth]{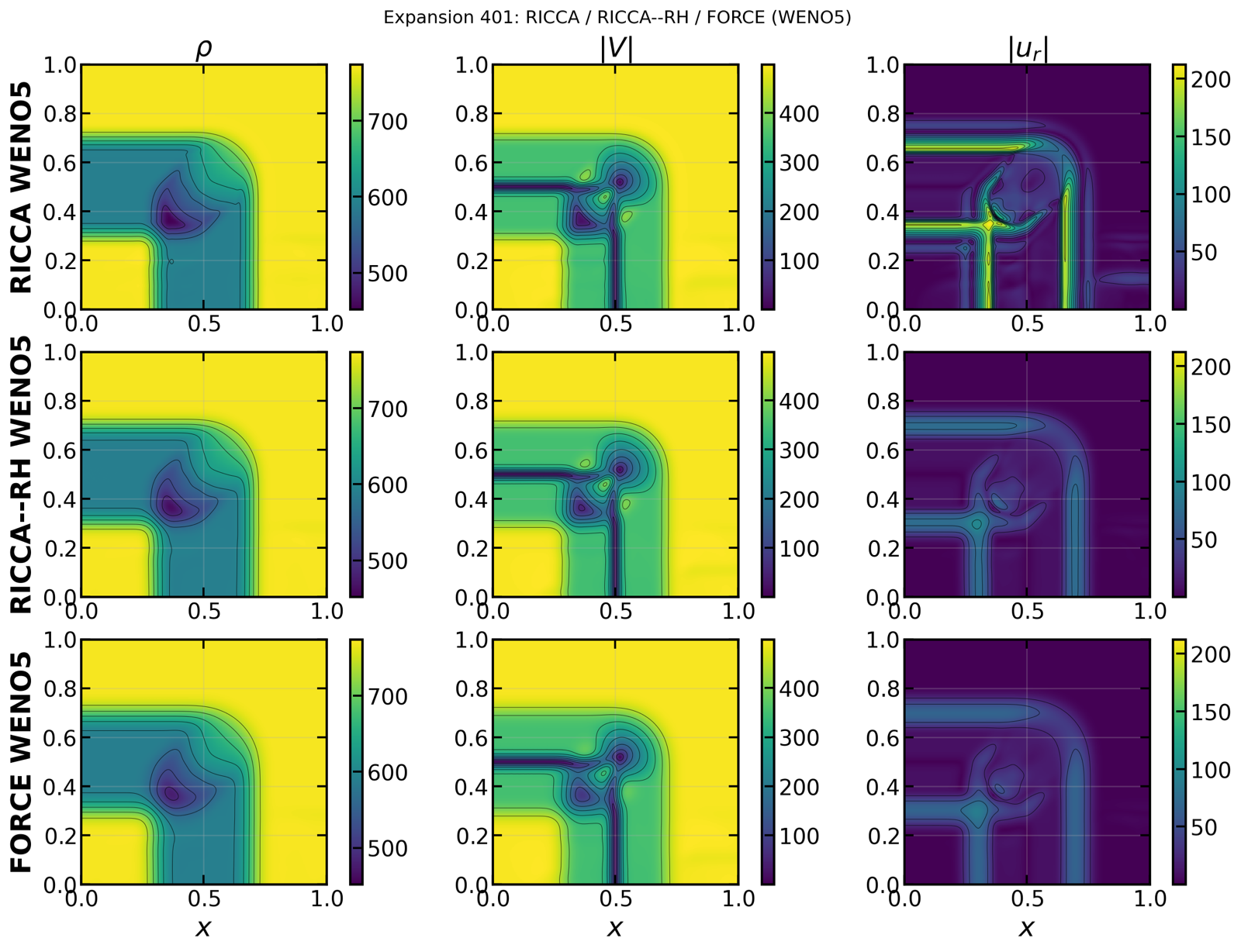}
\caption{2D expansion tube, $401^2$, WENO5: $\rho$, $|V|$, $|u_r|$ for
RICCA / RICCA--RH / FORCE.}
\label{fig:2d-rh-exp}
\end{figure}

\begin{figure}[htbp]
\centering
\includegraphics[width=0.95\textwidth]{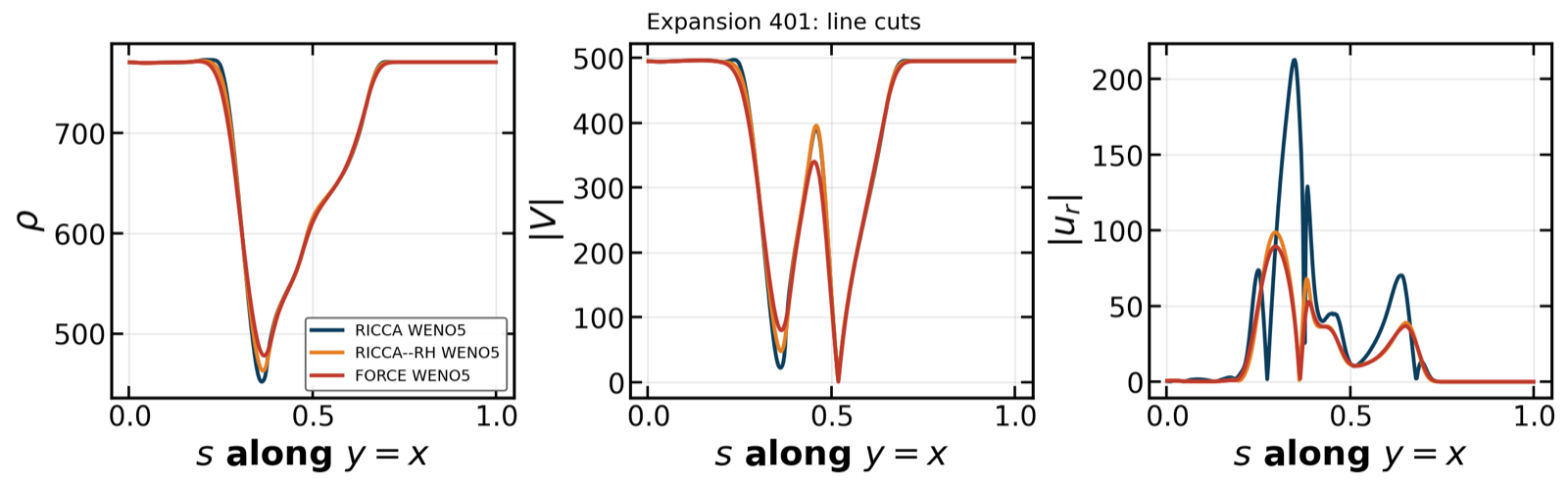}
\caption{2D expansion tube: diagonal cuts $y=x$ (WENO5).}
\label{fig:2d-rh-exp-l}
\end{figure}

\begin{figure}[htbp]
\centering
\includegraphics[width=0.95\textwidth]{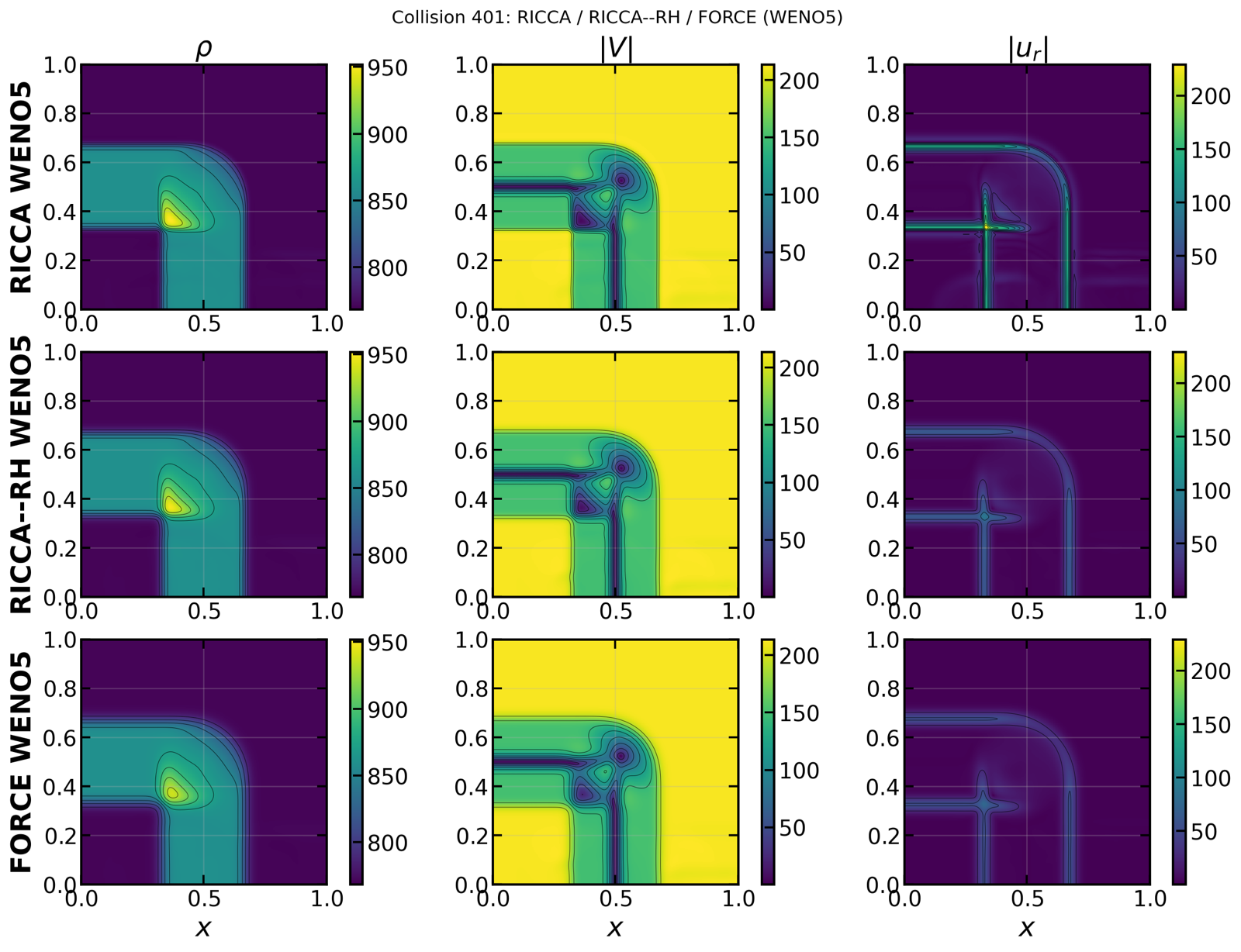}
\caption{2D mixture collision, $401^2$, WENO5: RICCA / RICCA--RH / FORCE.}
\label{fig:2d-rh-col}
\end{figure}

\begin{figure}[htbp]
\centering
\includegraphics[width=0.95\textwidth]{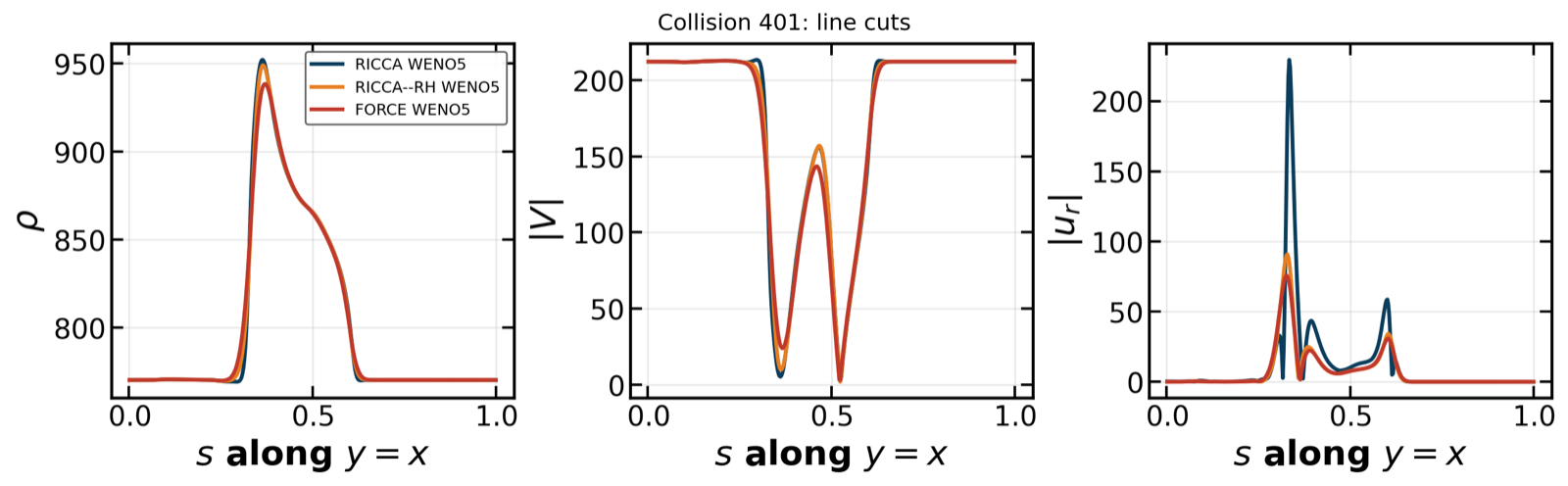}
\caption{2D mixture collision: diagonal cuts $y=x$ (WENO5).}
\label{fig:2d-rh-col-l}
\end{figure}

\begin{figure}[htbp]
\centering
\includegraphics[width=0.95\textwidth]{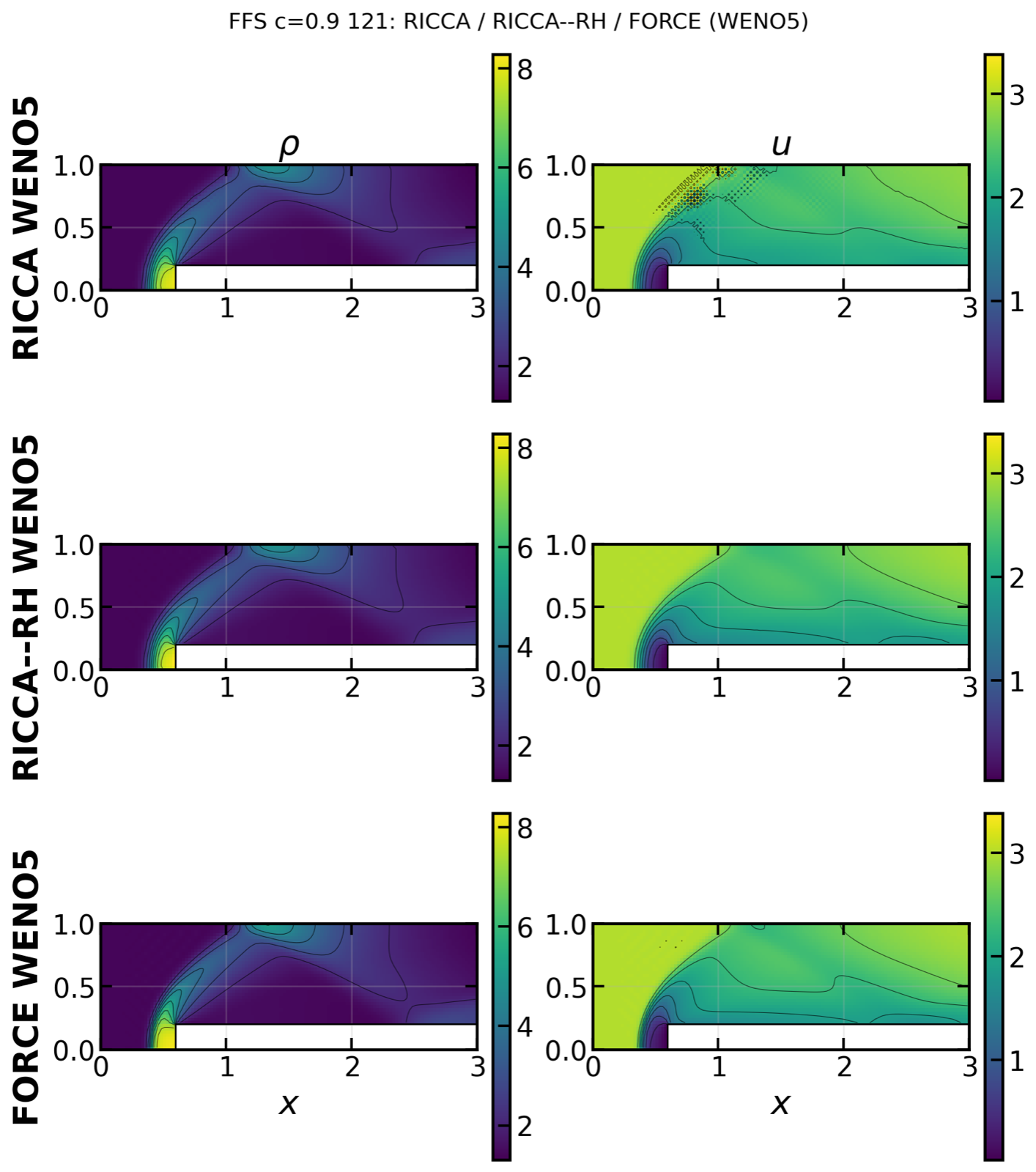}
\caption{FFS, $c=0.9$, $121\times41$, WENO5: RICCA / RICCA--RH / FORCE.}
\label{fig:2d-rh-ffs}
\end{figure}

\begin{figure}[htbp]
\centering
\includegraphics[width=0.95\textwidth]{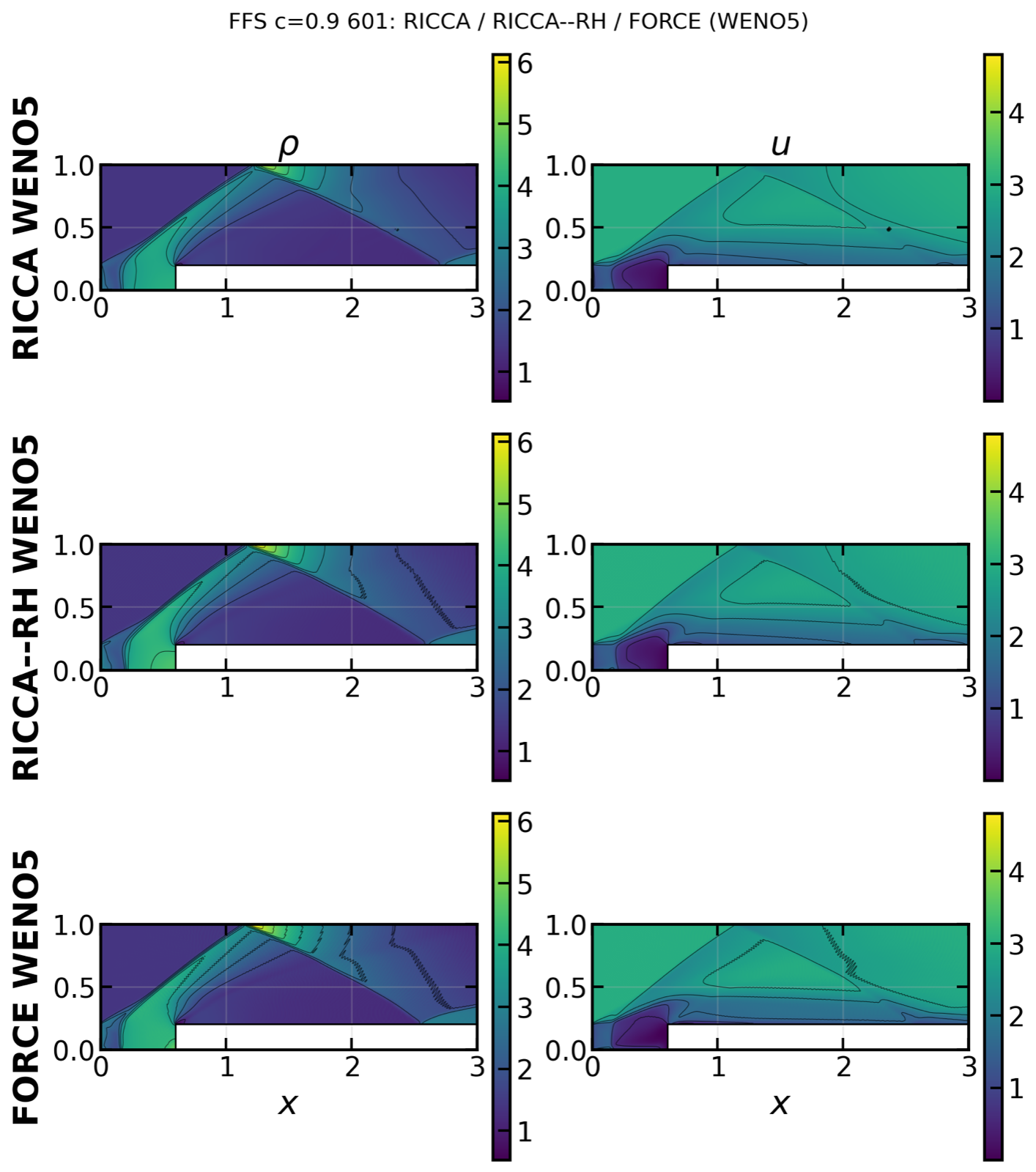}
\caption{FFS, $c=0.9$, $601\times201$, WENO5: RICCA / RICCA--RH / FORCE.}
\label{fig:2d-rh-ffs601}
\end{figure}

The production RICCA--RH law of Section~\ref{sec:ricca-rh} was ported to the
2D solver (CPU and Metal-resident face kernels) and run with the same combo
WENO5 hybrid on the CFD2019 meshes.
Figures~\ref{fig:2d-rh-exp}--\ref{fig:2d-rh-ffs601} compare RICCA, RICCA--RH
and FORCE WENO5; diagonal (Riemann) or $y\approx0.5$ (FFS) cuts confirm that
RICCA--RH tracks legacy RICCA and sits between RICCA and the more dissipative
FORCE baseline.
On the $401^2$ expansion tube the diagonal density roughness falls from
$2.96$ (RICCA) to $2.76$ (RICCA--RH) versus $2.48$ (FORCE); on the collision
tube the corresponding values are $2.49$, $2.18$ and $1.86$
(Table~\ref{tab:2d-rh}).
Wave patterns (rarefaction--contact in the expansion tube; colliding shocks in
the collision tube; bow and reflected shocks on the FFS) match the RICCA/FORCE
reference figures of Sections~\ref{sec:2d-exp}--\ref{sec:2d-ffs} at plotting
resolution.

\begin{figure}[htbp]
\centering
\includegraphics[width=0.95\textwidth]{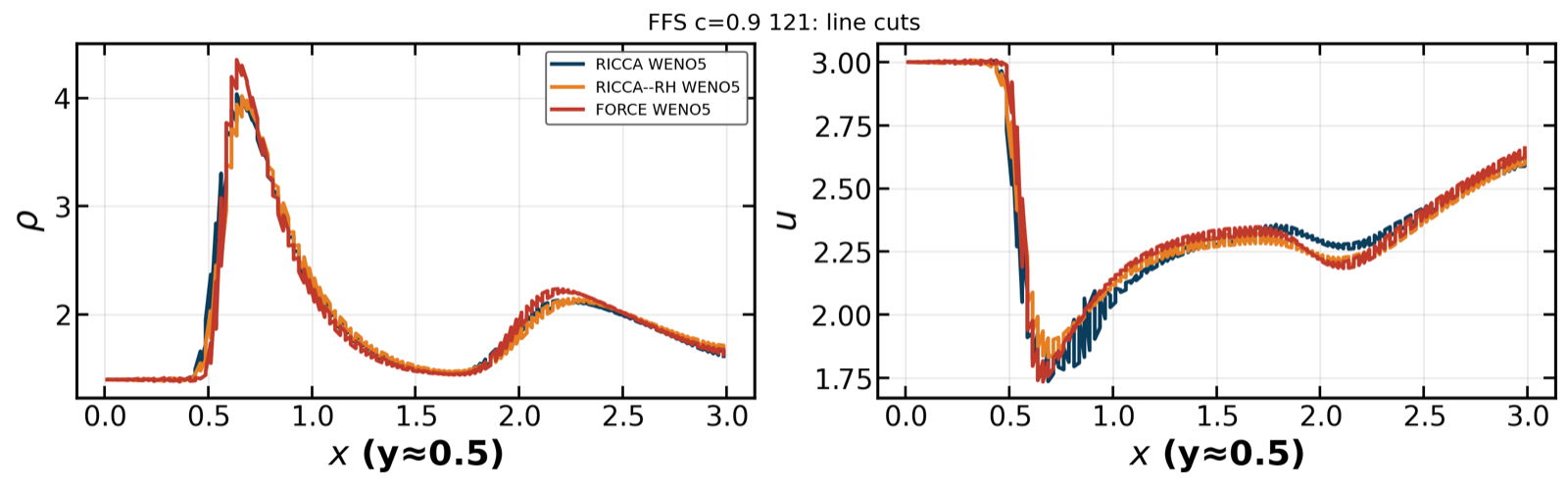}
\caption{FFS, $c=0.9$, $121\times41$: horizontal cuts $y\approx0.5$ (WENO5).}
\label{fig:2d-rh-ffs-l}
\end{figure}

\begin{table}[htbp]
\centering
\caption{2D WENO5 cut roughness of density (diagonal for Riemann; $y\approx0.5$
for FFS).}
\label{tab:2d-rh}
\begin{tabular}{@{}llrr@{}}
\toprule
Problem & Scheme (WENO5) & $\mathrm{std}(\Delta\rho)_{\mathrm{cut}}$ & $\rho$ range \\
\midrule
Expansion $401^2$ & RICCA & 2.96 & $[452,774]$ \\
 & RICCA--RH & 2.76 & $[463,772]$ \\
 & FORCE & 2.48 & $[478,772]$ \\
\midrule
Collision $401^2$ & RICCA & 2.49 & $[769,952]$ \\
 & RICCA--RH & 2.18 & $[770,949]$ \\
 & FORCE & 1.86 & $[770,939]$ \\
\midrule
FFS $c{=}0.9$, $121\times41$ & RICCA & 0.121 & $[1.31,8.06]$ \\
 & RICCA--RH & 0.116 & $[1.38,8.29]$ \\
 & FORCE & 0.145 & $[1.36,8.17]$ \\
\bottomrule
\end{tabular}

\end{table}

\subsection{Mach~3 forward-facing step}
\label{sec:2d-ffs}

\begin{figure}[htbp]
\centering
\includegraphics[width=\textwidth]{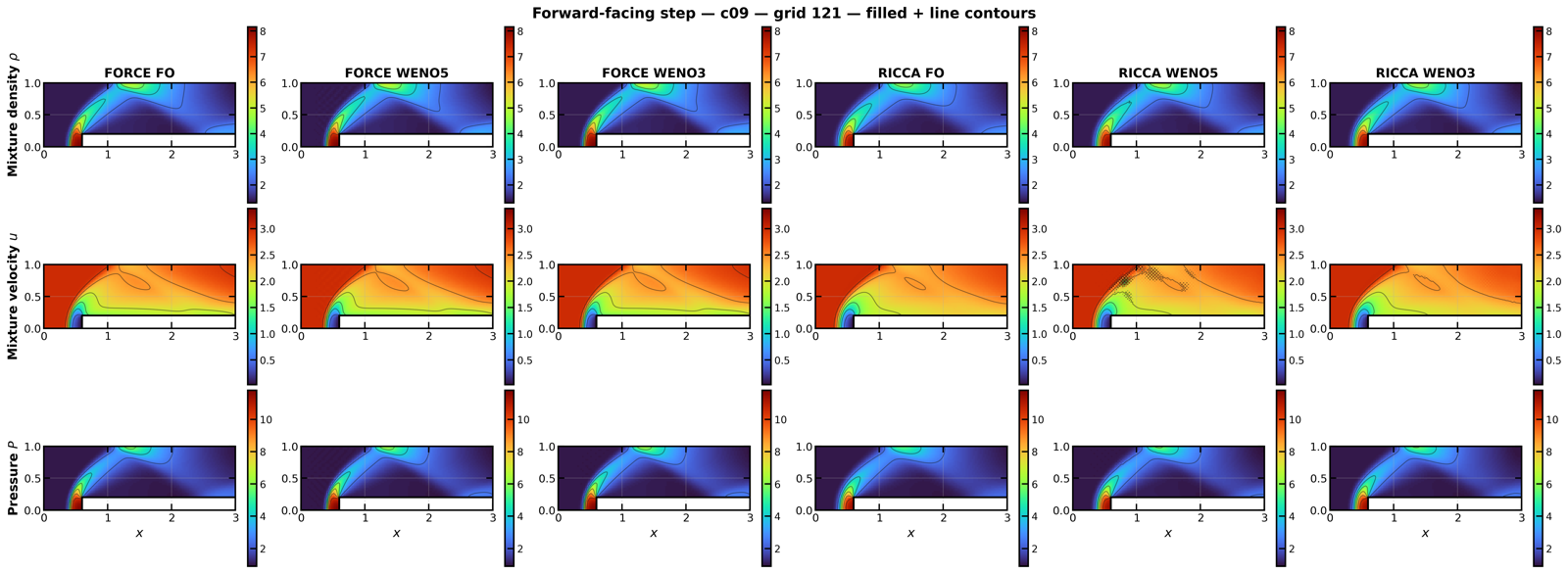}
\caption{FFS, $c=0.9$, $121\times41$: $\rho$, $u$, $P$
(RICCA and FORCE; filled + line contours).}
\label{fig:ffs121}
\end{figure}

\begin{figure}[htbp]
\centering
\includegraphics[width=\textwidth]{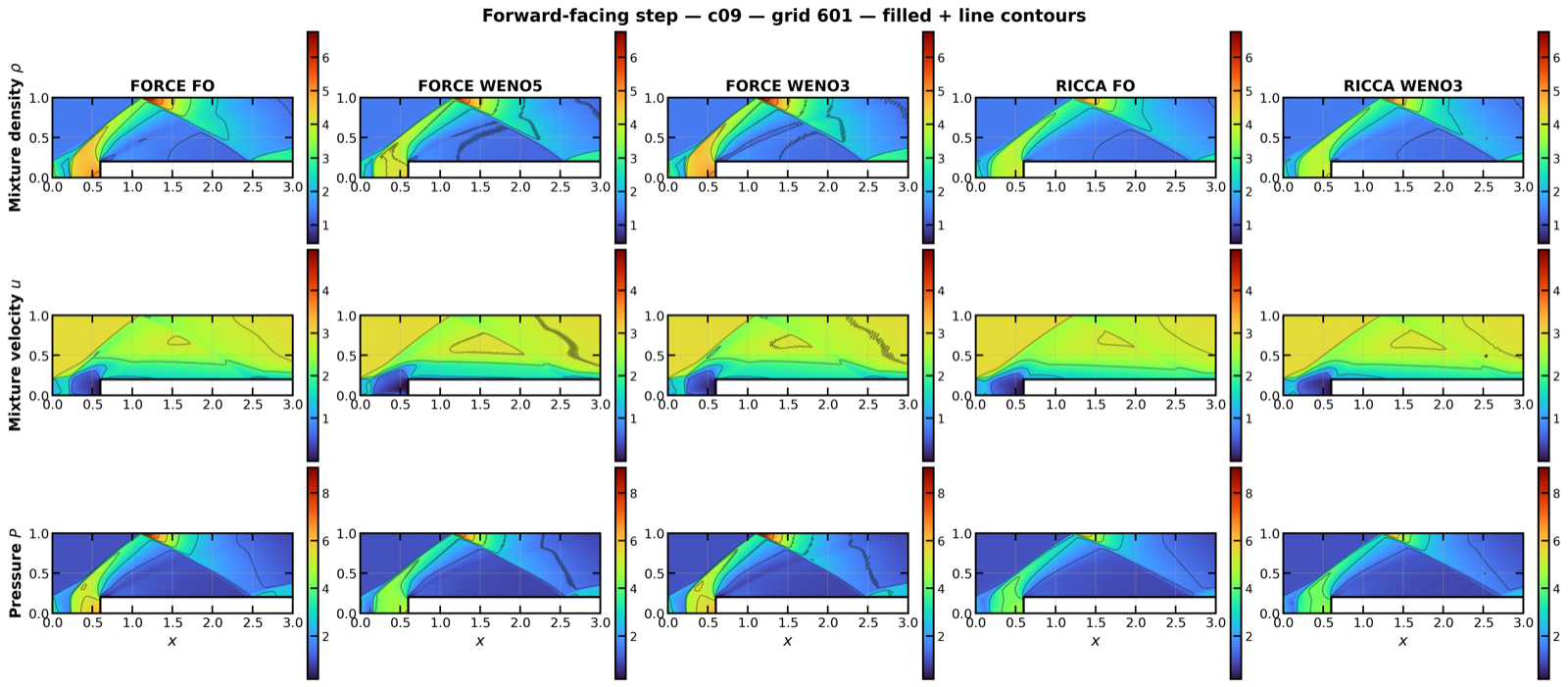}
\caption{FFS, $c=0.9$, $601\times201$.}
\label{fig:ffs601}
\end{figure}

\begin{figure}[htbp]
\centering
\includegraphics[width=\textwidth]{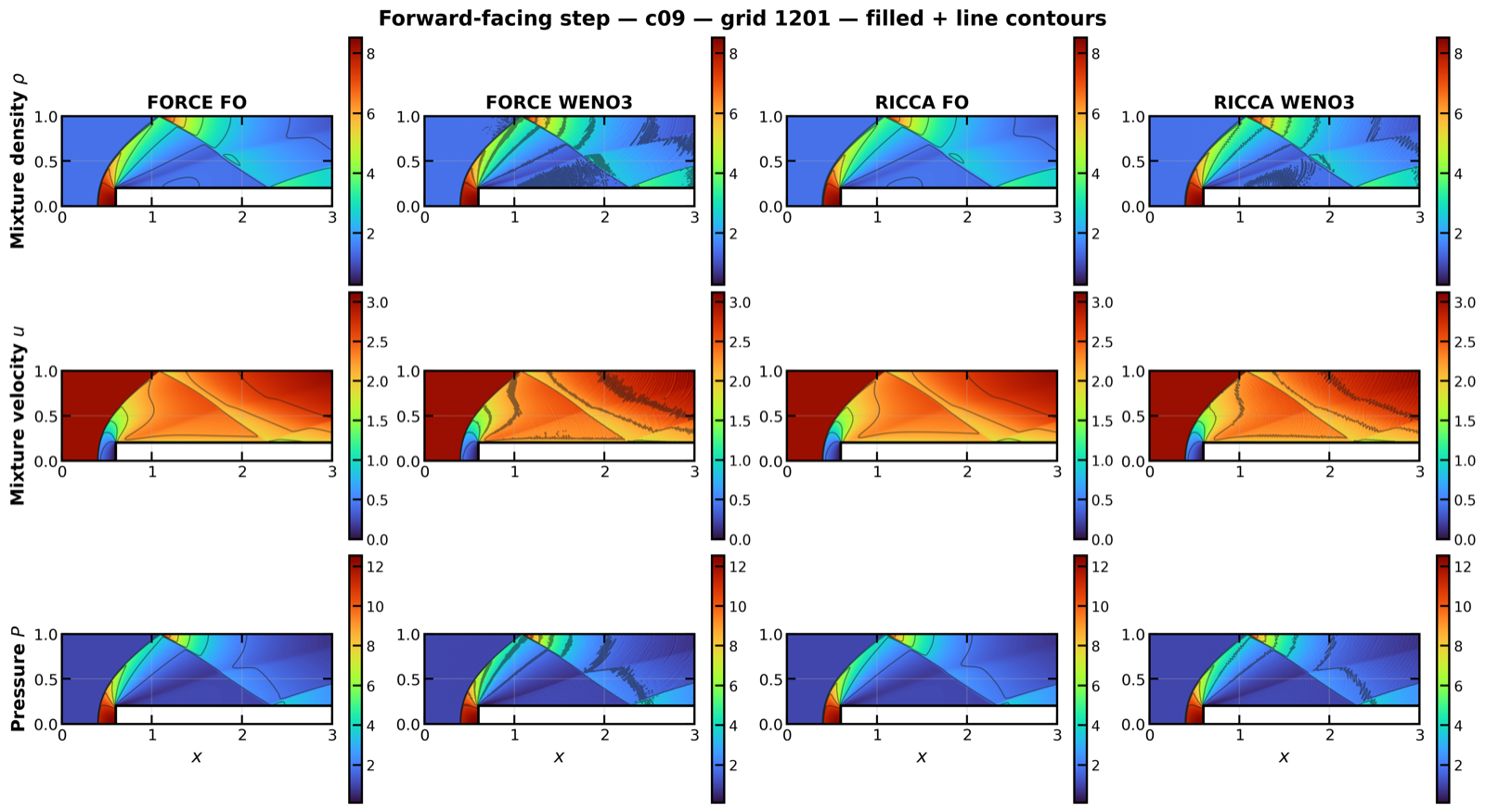}
\caption{FFS, $c=0.9$, $1201\times401$: FORCE FO vs RICCA FO.}
\label{fig:ffs1201}
\end{figure}

\begin{figure}[htbp]
\centering
\includegraphics[width=0.95\textwidth]{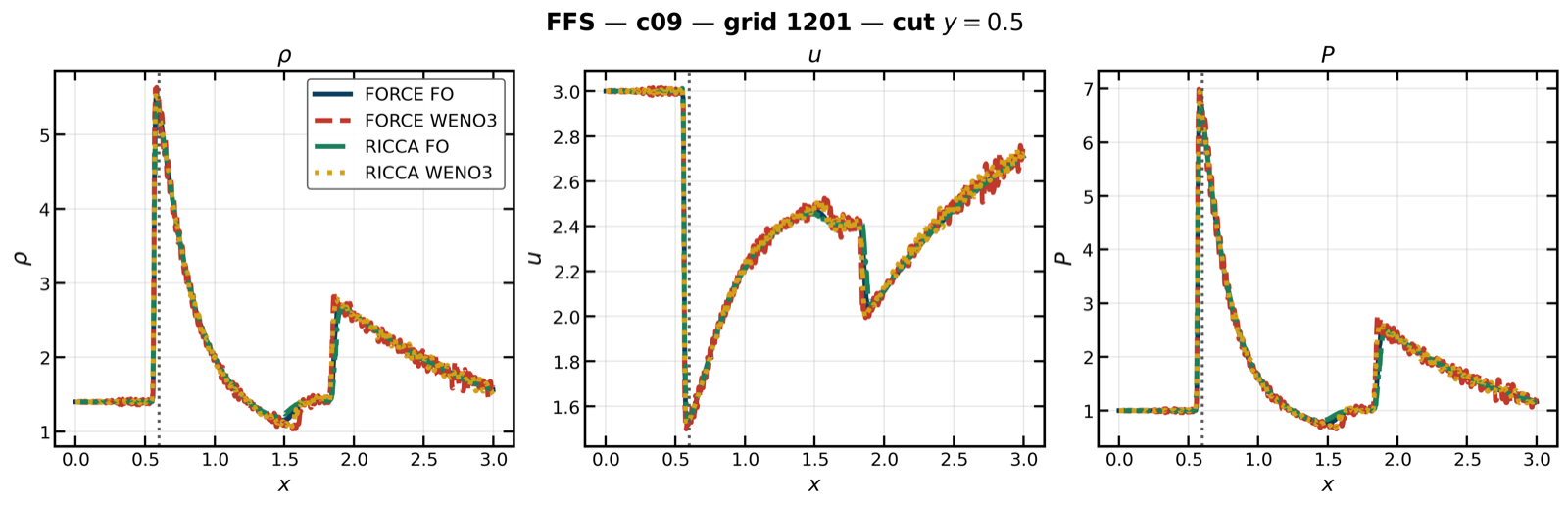}
\caption{FFS, $c=0.9$, $1201\times401$: cut $y=0.5$.}
\label{fig:ffs1201l}
\end{figure}

Domain $[0,3]\times[0,1]$, step height $0.2$ at $x=0.6$,
$(\rho,u,v)=(1.4,3,0)$, $P=1$, $t=2$, $c\in\{0.7,0.9\}$.
Under the equal-EOS closure of \cite{Zeidan2019} this configuration enforces
$c\equiv\alpha$ and $\mathbf{u}_r=\mathbf{0}$ throughout, so it is effectively
a \emph{single-velocity} mixture test of the shock-capturing machinery rather
than a full two-velocity problem.
We retain it because it is the third benchmark of the reference paper and
stresses strong shocks on fine grids; genuinely two-velocity 2D behaviour is
carried by the expansion and collision tubes
(Sections~\ref{sec:2d-exp}--\ref{sec:2d-col}), where $|u_r|$ is nontrivial.
Conclusions about relative-velocity capturing should therefore \emph{not} be
drawn from the FFS alone.
A future FFS variant with prescribed inlet slip would strengthen the
two-phase narrative further.
Figures~\ref{fig:ffs121}--\ref{fig:ffs1201l} show $c=0.9$ for RICCA and FORCE.
On the paper grid $1201\times401$, first-order RICCA and FORCE are nearly
indistinguishable; WENO on coarser meshes sharpens the bow and reflected
shocks.

%==============================================================================
\section{Order of accuracy and grid independence}
\label{sec:conv}

\subsection{Smooth manufactured solution}
\label{sec:mms}

\begin{figure}[htbp]
\centering
\includegraphics[width=0.95\textwidth]{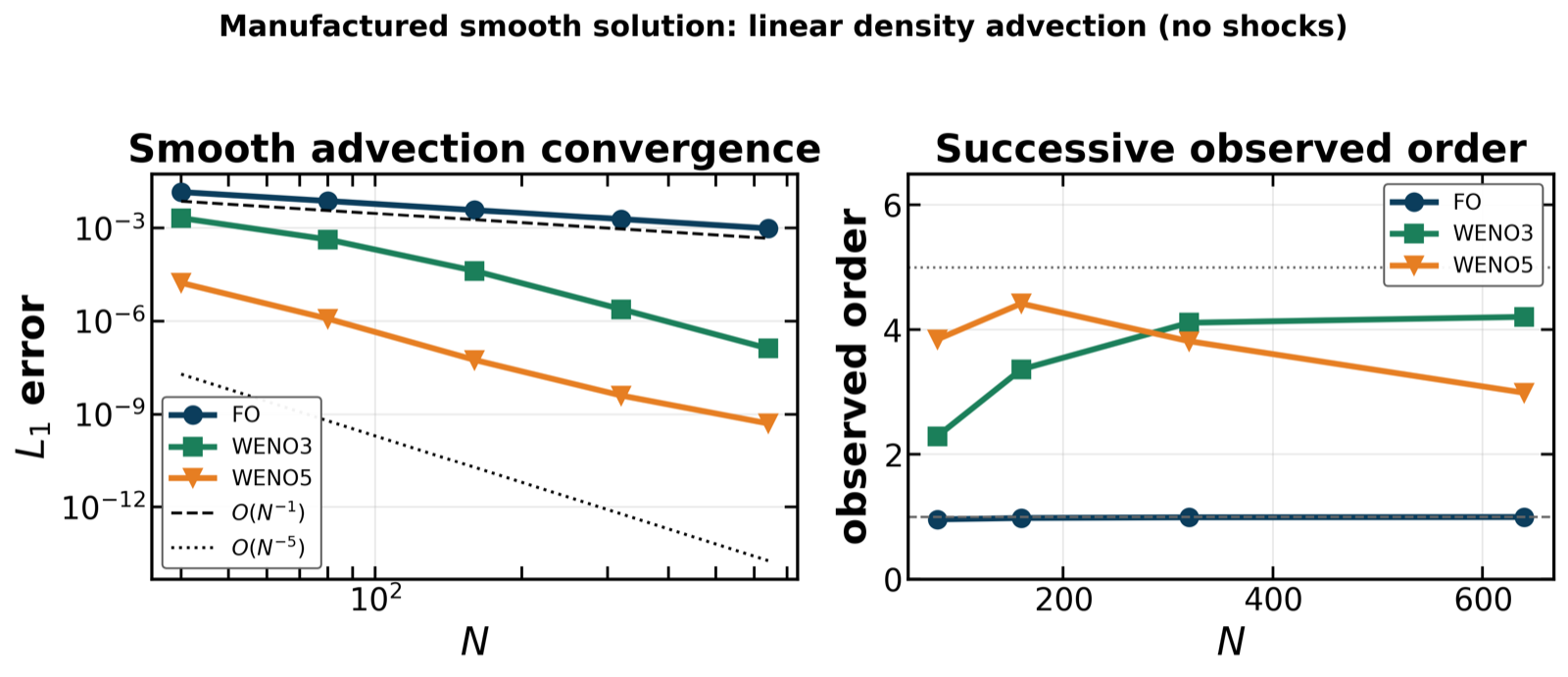}
\caption{Smooth manufactured density wave: $L_1$ error and observed order
for FO, WENO3 and WENO5 (SSP-RK3, CFL~$=0.4$).}
\label{fig:mms}
\end{figure}

To separate design-order behaviour from shock limiting we integrate the linear
advection equation $\partial_t\rho+U_0\partial_x\rho=0$ with manufactured
solution $\rho=1+0.2\sin\bigl(2\pi(x-U_0 t)\bigr)$ ($U_0=1$, $t=0.25$,
periodic, CFL~$=0.4$, SSP-RK3) using the same FO / WENO3 / WENO5 face
reconstructions as in the mixture solver.
Table~\ref{tab:mms-smooth} and Fig.~\ref{fig:mms} show that FO remains
first-order, while WENO5 recovers observed orders $p\approx3.8$--$4.4$ on
moderate grids (round-off and RK3 eventually limit the finest levels).
This confirms that the reconstruction itself is high-order on smooth data;
global first-order rates on Riemann problems are therefore due to
discontinuities, not a defective WENO implementation.

% Auto-generated — smooth linear advection MMS
\begin{table}[htbp]\centering\small
\caption{Smooth manufactured density wave ($\rho=1+0.2\sin(2\pi(x-U_0 t))$, $U_0=1$, $t=0.25$, CFL$=0.4$, periodic). WENO5 recovers high-order accuracy away from discontinuities; FO remains first-order.}
\label{tab:mms-smooth}
\begin{tabular}{@{}rrrrrrr@{}}\toprule
$N$ & $L_1$ FO & $p$ & $L_1$ WENO3 & $p$ & $L_1$ WENO5 & $p$ \\
\midrule
40 & 1.478e-02 & --- & 2.104e-03 & --- & 1.718e-05 & --- \\
80 & 7.617e-03 & 0.96 & 4.296e-04 & 2.29 & 1.198e-06 & 3.84 \\
160 & 3.867e-03 & 0.98 & 4.180e-05 & 3.36 & 5.613e-08 & 4.42 \\
320 & 1.948e-03 & 0.99 & 2.424e-06 & 4.11 & 3.992e-09 & 3.81 \\
640 & 9.780e-04 & 0.99 & 1.316e-07 & 4.20 & 5.035e-10 & 2.99 \\
\bottomrule\end{tabular}\end{table}

\subsection{Discontinuous self-convergence}
\label{sec:ooa-shock}

\begin{figure}[htbp]
\centering
\includegraphics[width=0.95\textwidth]{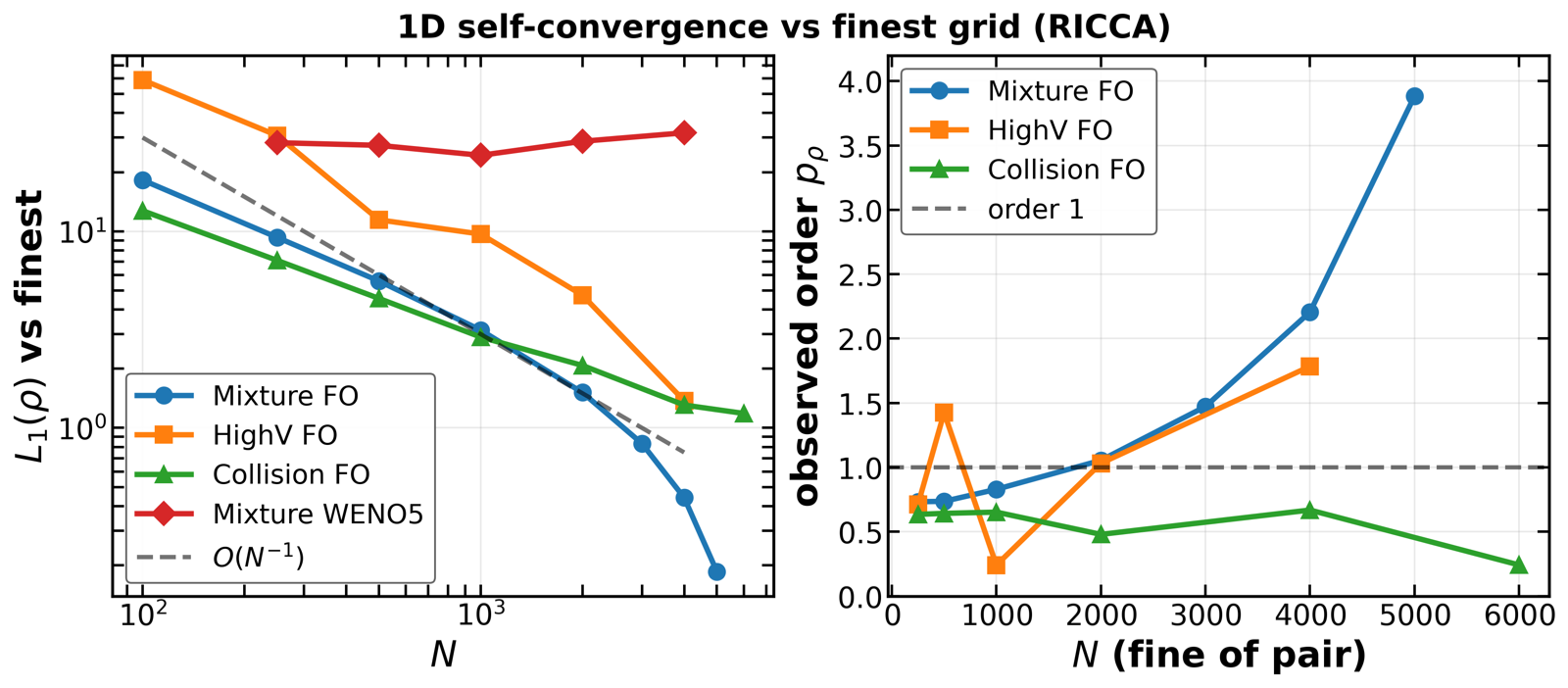}
\caption{1D self-convergence of RICCA: $L_1(\rho)$ versus $N$ (left) and
successive observed order (right). The dashed line marks $O(N^{-1})$.}
\label{fig:ooa}
\end{figure}

Because the gas--magma Riemann problems contain shocks and contacts, a global
smooth-solution Taylor argument does not apply.
We therefore measure \emph{self-convergence}: for each grid $N$, the discrete
$L_1$ difference in mixture density (and pressure) is evaluated against a fine
reference ($N=6000$ for the expansion tube), after linear interpolation of the
reference onto the coarse mesh.
The observed order between successive grids is
\begin{equation}
p
=
\frac{\log\bigl(E(N_1)/E(N_2)\bigr)}{\log(N_2/N_1)},
\qquad
E(N)=\tfrac{1}{N}\sum_i\bigl|\rho_i^{(N)}-\rho^{\mathrm{ref}}(x_i)\bigr|.
\end{equation}
Table~\ref{tab:ooa-1d} and Fig.~\ref{fig:ooa} show that first-order RICCA
converges at a mean rate $p_\rho\approx 0.7$--$1.1$ over the mid-range of
meshes ($N\le 2000$), consistent with an $O(\Delta x)$ scheme limited by
discontinuities; local rates fluctuate when wave structures realign under
refinement.
Including the intermediate grids $N=3000$ and $N=5000$ confirms that
$L_1(\rho)$ decreases monotonically through the full suite, but successive
observed orders inflate as $N$ approaches the finite $N=6000$ reference
($p_\rho\approx 1.5$--$3.9$ for $N\ge 3000$).
That late rise is the expected finite-reference pollution of
self-convergence (errors become correlated with the approximate reference),
not a change in formal accuracy; we therefore treat the mid-range rates as
the meaningful measure of global order.
WENO5 reduces the $L_1$ error relative to FO on the same $N$ but remains
globally first-order in the presence of shocks, as expected.

% Auto-generated — do not edit by hand
\begin{table}[htbp]\centering\small
\caption{1D mixture expansion: self-convergence of RICCA FO against $N=6000$ reference.}
\label{tab:ooa-1d}
\begin{tabular}{@{}rrrrr@{}}\toprule
$N$ & $L_1(\rho)$ & $p_\rho$ & $L_1(P)$ & $p_P$ \\
\midrule
100 & 1.823e+01 & --- & 2.426e+08 & --- \\
250 & 9.311e+00 & 0.73 & 1.130e+08 & 0.83 \\
500 & 5.588e+00 & 0.74 & 6.247e+07 & 0.86 \\
1000 & 3.144e+00 & 0.83 & 3.307e+07 & 0.92 \\
2000 & 1.513e+00 & 1.06 & 1.521e+07 & 1.12 \\
3000 & 8.328e-01 & 1.47 & 8.252e+06 & 1.51 \\
4000 & 4.415e-01 & 2.21 & 4.306e+06 & 2.26 \\
5000 & 1.856e-01 & 3.88 & 1.837e+06 & 3.82 \\
\bottomrule\end{tabular}\end{table}

\begin{figure}[htbp]
\centering
\begin{subfigure}{0.48\textwidth}
\includegraphics[width=\linewidth]{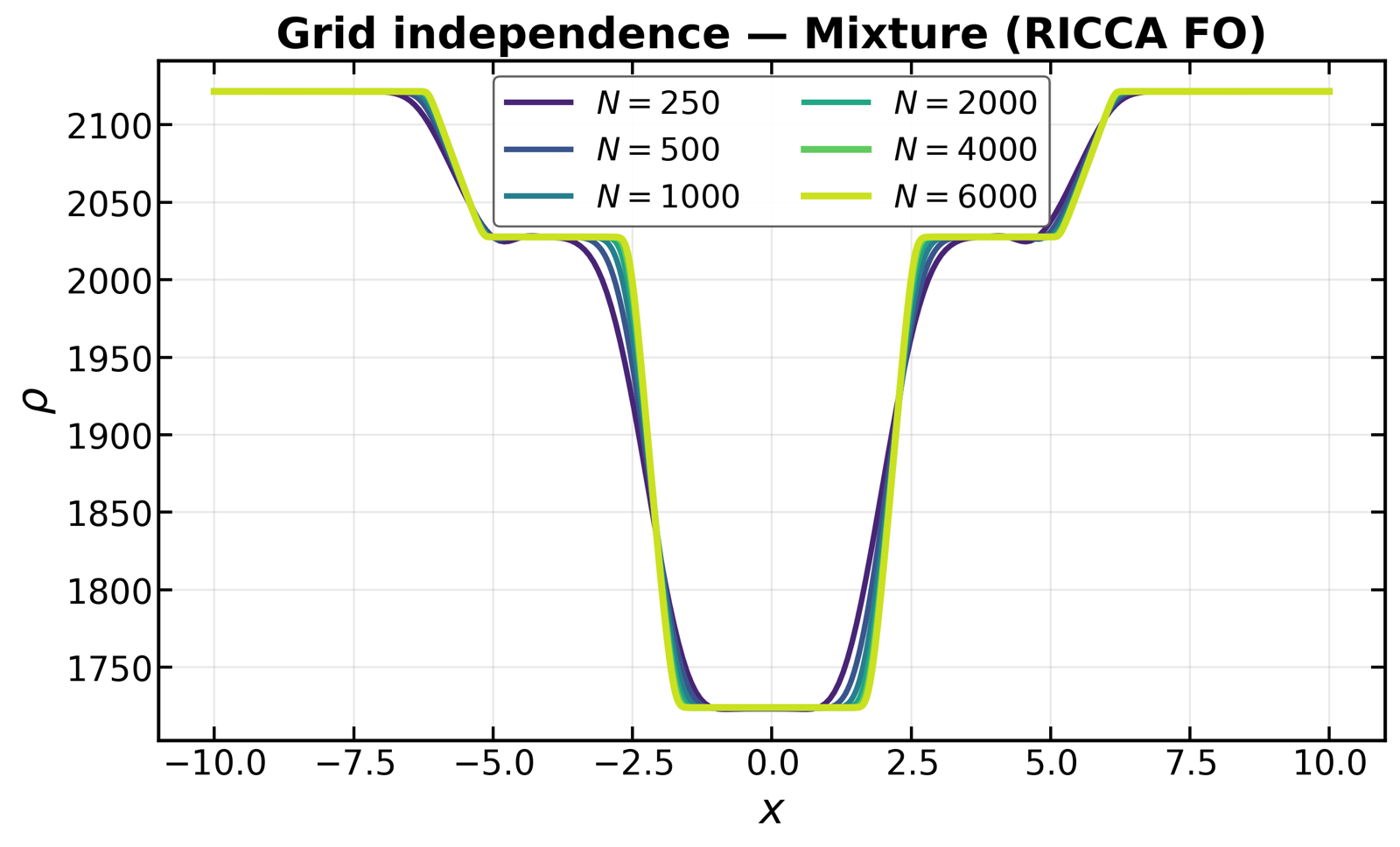}
\caption*{Expansion}
\end{subfigure}
\hfill
\begin{subfigure}{0.48\textwidth}
\includegraphics[width=\linewidth]{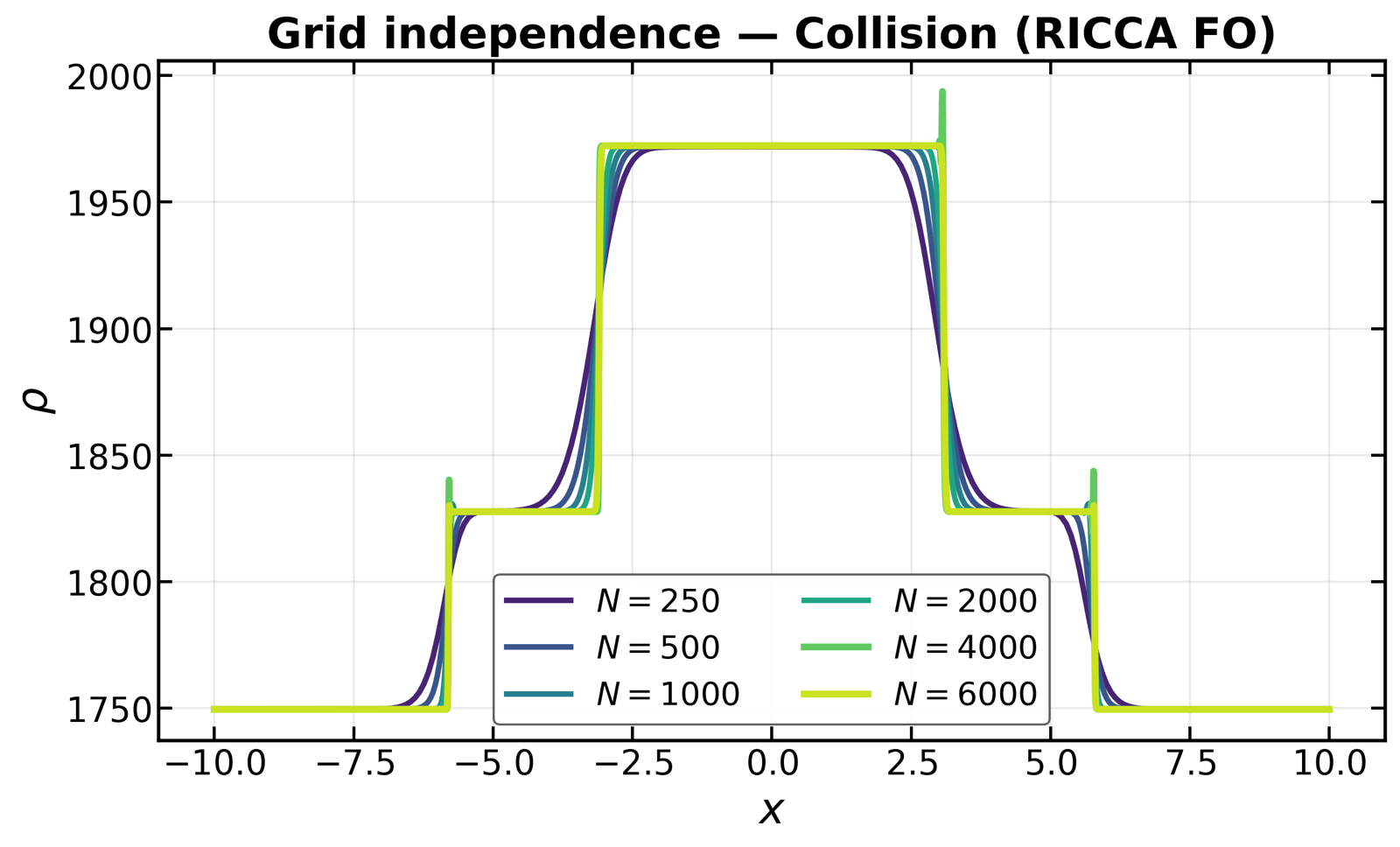}
\caption*{Collision}
\end{subfigure}
\caption{1D RICCA FO grid independence ($N=250$--$6000$).}
\label{fig:gi-1d}
\end{figure}

\begin{figure}[htbp]
\centering
\begin{subfigure}{0.48\textwidth}
\includegraphics[width=\linewidth]{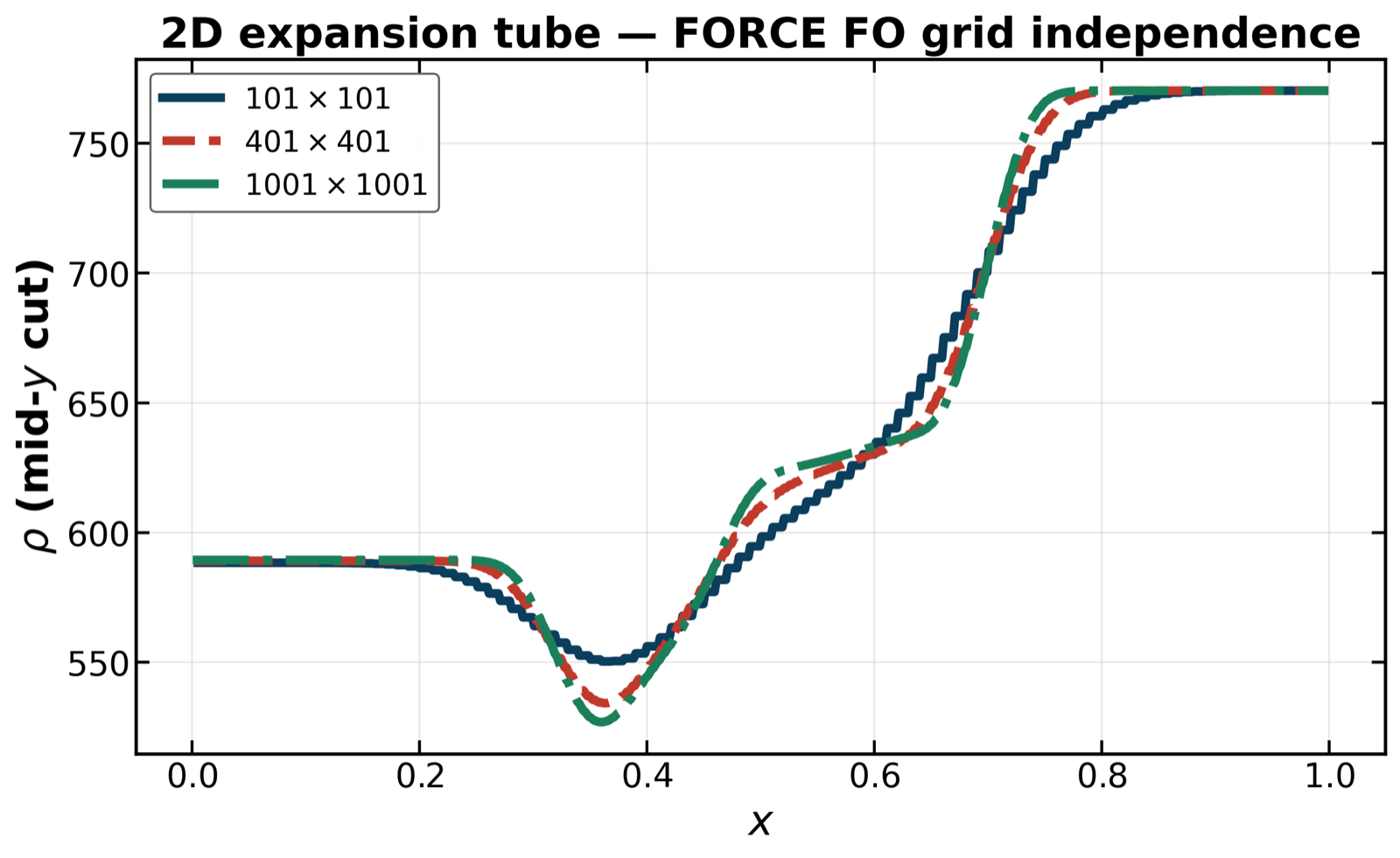}
\caption*{Expansion tube}
\end{subfigure}
\hfill
\begin{subfigure}{0.48\textwidth}
\includegraphics[width=\linewidth]{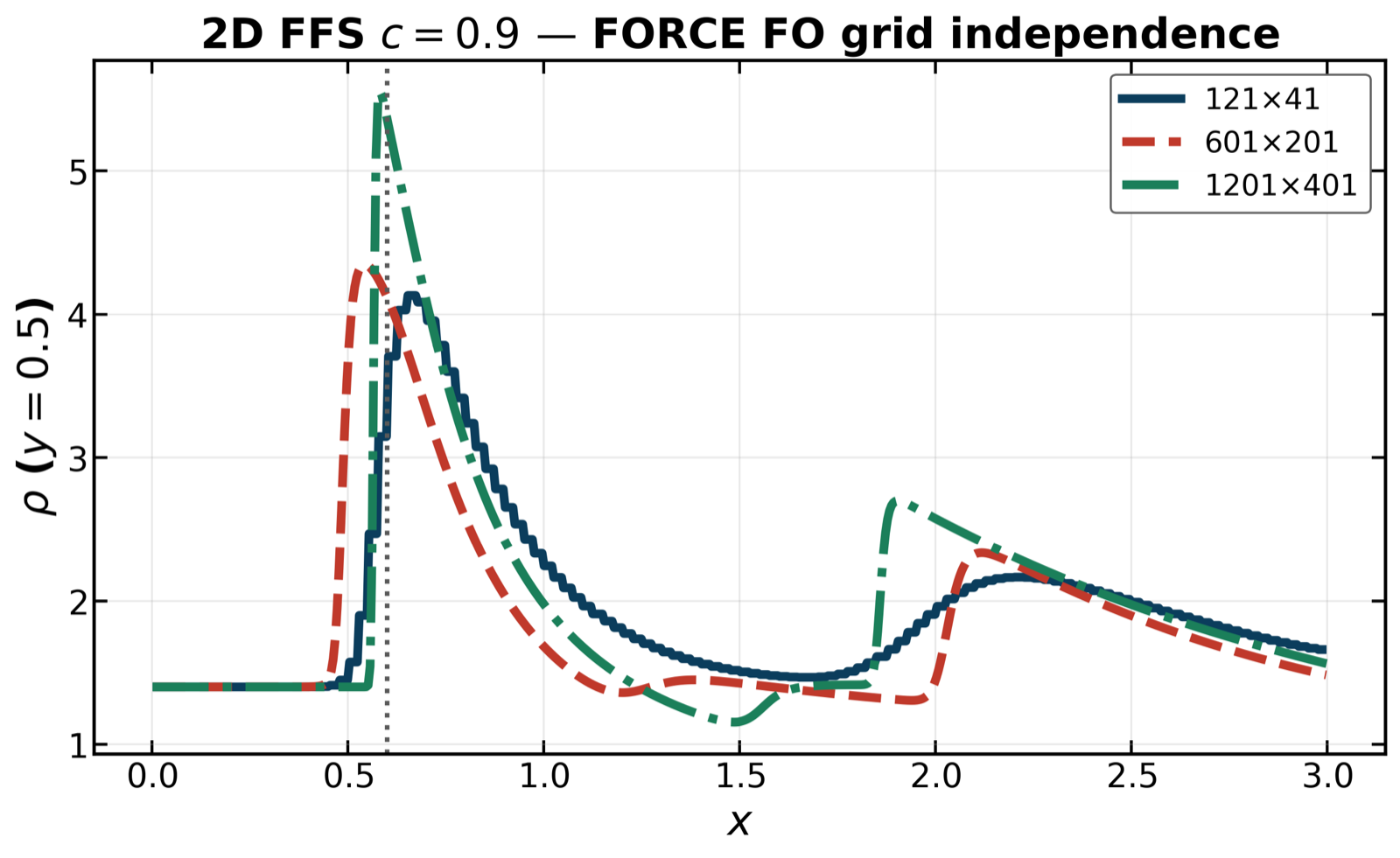}
\caption*{FFS, $c=0.9$}
\end{subfigure}
\caption{2D FORCE FO grid independence (mid-line density cuts).}
\label{fig:gi-2d}
\end{figure}

Grid independence is illustrated by overlaying density profiles on successive
meshes.
In 1D (Fig.~\ref{fig:gi-1d}) the expansion and collision solutions change
little for $N\ge 2000$--$4000$.
In 2D (Fig.~\ref{fig:gi-2d}) FORCE FO mid-line cuts for the expansion tube
($101^2$, $401^2$, $1001^2$) and the FFS at $c=0.9$
($121\times41$, $601\times201$, $1201\times401$) show progressive sharpening
of shocks/contacts with the fine-grid structure essentially established on the
paper meshes used in Section~\ref{sec:2d}.

%==============================================================================
\section{Metal GPU implementation and laptop runtimes}
\label{sec:perf}

The production solvers are accelerated with Apple's Metal API through a thin
C++ layer (\texttt{Metal\_GPU}/\texttt{MetalFlow}).
Metal is an intentional target---not an exotic niche---because Apple Silicon
laptops provide a capable GPU with unified memory, which matches the resident
2D path (device-resident $\mathbf{U}$, host BC refresh of ghosts only).
The same kernel structure ports conceptually to CUDA/HIP; we have not yet
duplicated the backend, so portability is architectural rather than
benchmarked here.
A CUDA port is the natural next step once cross-platform validation and
occupancy/bandwidth profiling are justified by community demand.
On the GPU, conserved variables and fluxes reside in device buffers; first-order
and primitive WENO3/WENO5 face reconstruction, RICCA/RICCA--RH/FORCE intercell
fluxes and
explicit updates are executed as Metal compute kernels (float32).
Boundary conditions and I/O remain on the host.
In 1D, Metal currently covers RICCA first-order; FORCE, RICCA--RH and WENO run
on the CPU (RICCA--RH is CPU-only in 1D by design).
In 2D, RICCA, RICCA--RH and FORCE (FO and WENO) share the resident Metal path.

\begin{figure}[htbp]
\centering
\includegraphics[width=0.95\textwidth]{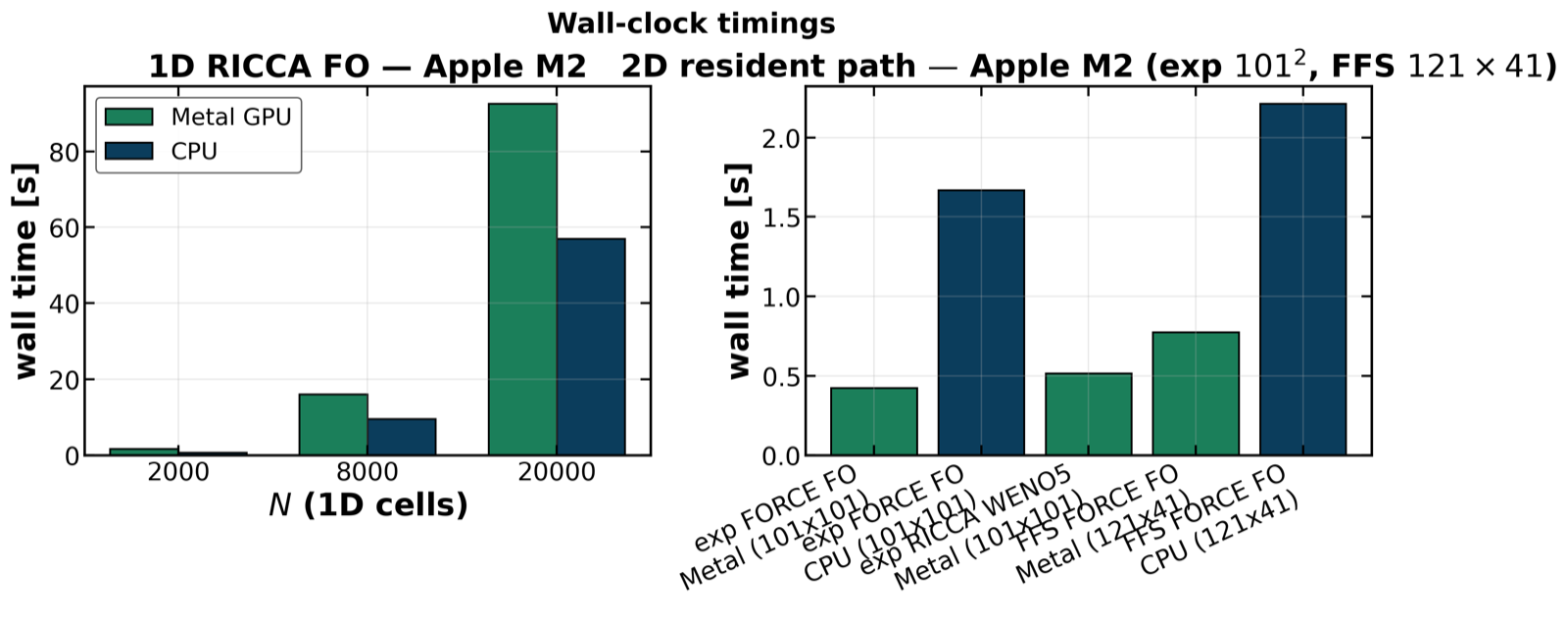}
\caption{Wall-clock times on Apple M2: 1D RICCA FO Metal vs CPU (left;
CPU faster---non-resident 1D path) and 2D resident-path timings (right;
expansion $101\times101$ and FFS $121\times41$; Metal faster).}
\label{fig:runtime}
\end{figure}

All timings in Table~\ref{tab:runtime} and Fig.~\ref{fig:runtime} were measured
as wall-clock time of the time-marching loop on a MacBook with an
\textbf{Apple M2} SoC (10-core GPU, Metal~4, 24\,GB unified memory).
Table~\ref{tab:runtime} reports paired Metal/CPU times and the speedup
$t_{\mathrm{CPU}}/t_{\mathrm{Metal}}$ (values $>1$ favour the GPU).
\paragraph{Why 1D Metal can look slower.}
The present 1D Metal FO path is \emph{not} fully resident: each step packs
host data to the device, launches kernels, synchronises, and copies the update
back for the host error/update loop.
On a short 1D residual this overhead exceeds the arithmetic gain, so the CPU
path is faster (speedup $<1$ in Table~\ref{tab:runtime}).
This is expected for the current 1D binding, not a mis-timing.
\paragraph{Where Metal wins.}
The 2D solver keeps $\mathbf{U}$ and fluxes on the device.
On the expansion tube with $101\times101$ cells, Metal FORCE FO is measured
at $3.93\times$ faster than CPU (Table~\ref{tab:runtime}); the coarse FFS
($121\times41$) FORCE FO adds a second completed baseline at $2.86\times$.
RICCA--WENO5 on Metal finishes in a comparable Metal-only wall time on the
$101\times101$ mesh, but that row still lacks a CPU counterpart.
Metal is therefore the production path for 2D; 1D Metal is kept for
algorithmic parity pending a resident 1D port.
Larger-grid ($401^2$, $801^2$) scaling and kernel occupancy / bandwidth
profiles are planned once a portable CUDA twin is available; the present
section establishes that the algorithm is GPU-amenable on commodity Apple
hardware.
For a large-scale HPC audience the Metal timings are therefore a promising
laptop-scale demonstration, not a complete cross-platform performance study.

% Auto-generated — do not edit by hand
\begin{table}[htbp]\centering\small
\caption{Wall-clock times on Apple M2 (24\,GB, Metal~4). Speedup $=t_{\mathrm{CPU}}/t_{\mathrm{Metal}}$ ($>1$ means GPU faster). 1D RICCA FO uses a non-resident Metal path (host--device traffic each step), so CPU is faster; 2D uses a resident GPU path and Metal wins.}
\label{tab:runtime}
\begin{tabular}{@{}lllrrr@{}}\toprule
Problem & Scheme & $N$ & $t_{\mathrm{Metal}}$ [s] & $t_{\mathrm{CPU}}$ [s] & Speedup \\
\midrule
1D case~1 & RICCA FO & 2000 & 1.67 & 0.66 & 0.40 \\
1D case~1 & RICCA FO & 8000 & 16.04 & 9.56 & 0.60 \\
1D case~1 & RICCA FO & 20000 & 92.52 & 56.97 & 0.62 \\
1D case~1 & RICCA WENO5 & 1000 & --- & 0.90 & --- \\
1D case~1 & RICCA WENO5 & 2000 & --- & 3.69 & --- \\
2D case~1 & FORCE FO & 101x101 & 0.42 & 1.67 & 3.93 \\
2D case~1 & RICCA WENO5 & 101x101 & 0.52 & --- & --- \\
2D case~4 & FORCE FO & 121x41 & 0.77 & 2.21 & 2.86 \\
\bottomrule\end{tabular}\end{table}

%==============================================================================
\section{Discussion: hybrid sensor design}
\label{sec:disc}
\label{sec:sensor-discussion}

The hybrid sensor is pragmatic rather than derived from characteristic fields,
wave-strength measures or entropy production.
It achieves \emph{modest} improvements in contact smoothness and
relative-velocity behaviour over the legacy pressure-only sensor on the
reported suite, and it is documented here with enough honesty that a user can
judge when retuning is warranted.

\subsection{Formulation and tuning}

Six primitive indicators are combined by a simple maximum, with different
$C$ per variable and an extra factor $1.5$ on $\phi_{u_r}$
(Section~\ref{sec:weno}).
Each indicator can detect a different discontinuity type, which is intuitive,
but the construction is empirical: the sensor itself is not derived from
characteristic wave strengths, and the $(C,w)$ weights are hand-tuned.
The blend
$\theta=1/(1+C\phi)$
is likewise hand-tuned.
Figure~\ref{fig:sensor-tuning} shows that neighbouring $C$ and $w$ values only
shift the FO/WENO transition slightly on the expansion tube, yet
transferability to other Mach numbers, extreme slip, stiff EOS, or industrial
stratified regimes is \emph{not} guaranteed.
In practice the published constants are a reproducible starting point for
Zeidan-style mixture tests; other applications should expect a short retuning
pass.

\subsection{Metrics and evaluation scope}

The primary quantitative comparison uses five sensor variants on the 1D
mixture expansion at $N\approx1000$
(Table~\ref{tab:sensor},
Figs.~\ref{fig:weno-sensor}--\ref{fig:weno-sensor-advantage}).
Roughness $\mathrm{std}(\Delta\cdot)$ and peak $|u_r-u_r^{\mathrm{FO}}|$
capture residual ringing relative to FO, but they do not fully probe
spurious interface oscillations, every non-monotone contact detail, or
long-time error accumulation.
Figure~\ref{fig:sensor-suite} extends the same metrics to HighV and
collision: combo consistently cuts density overshoot and slip roughness,
while the peak slip residual remains case-dependent.
We therefore refrain from claiming optimality beyond this suite.

\subsection{Relation to more principled designs}

Modern hybridisation is often grounded in characteristic decomposition,
TVD/TVC constraints or invariant-domain arguments, with positivity treated as
an explicit design target.
The production path now follows that spirit: WENO acts on approximate
face-normal acoustic characteristics, the combo sensor provides additional
FO blending, and a Zhang--Shu IDP limiter enforces
$(\rho,\alpha,c)\in\mathcal{A}$ at faces
(Sections~\ref{sec:weno-char}--\ref{sec:positivity}).
The characteristic transform is still approximate (mixture acoustics only;
not a full Romenski $R\lvert\Lambda\rvert R^{-1}$), and the IDP argument is
empirical rather than a proved CFL-independent update for the nonlinear
RICCA operator.
RICCA/RICCA--RH are compared to HLLC/Roe on the 1D suite
(Section~\ref{sec:1d-upwind}), but those upwind solvers themselves remain
approximate (Rusanov-stable HLLC; field-by-field Roe).
Roughness and slip-residual metrics used in Sections~\ref{sec:weno}
and~\ref{sec:1d-rh} quantify local contact quality on short-time Riemann
problems; they do not replace long-time stability studies or a full interface
oscillation analysis.
The claim supported by the data is that, on Romenski-type mixture Riemann
problems, combo+characteristic WENO with IDP is a reproducible,
GPU-friendly production baseline that is sharper than FO and competitive with
FORCE, while RICCA--RH provides a documented dual-speed alternative to the
legacy pressure sensor.

\subsection{Practical implications}

For an adopter, the hybrid is attractive because it is easy to implement and
reduces ringing without erasing WENO sharpness on the present tests.
The cost of that simplicity is the absence of a formal tuning theory and of
broad robustness proofs.
Users moving to highly stratified gas--liquid flows, dispersed clouds or
cavitating regimes should treat $(C,w)$ as free parameters, monitor the
metrics of Section~\ref{sec:weno}, and fall back toward FO (larger $C$) if
admissibility alarms appear.
A full Romenski eigenstructure and a Guermond--Popov cell-update IDP proof
remain natural refinements if stronger guarantees are required.

\subsection{RICCA--RH design implications}

The RH dual-speed experiment (Section~\ref{sec:ricca-rh}) reinforces a practical
message parallel to the hybrid-sensor discussion: on these strong
gas--magma/gas--liquid shocks, \emph{adequate} contact dissipation matters more
than a finely graded continuous switch.
Soft $\sigma\in[0,1]$ with $\kappa\ll1$ looked attractive on paper but failed
in 1D (Section~\ref{sec:1d-rh}); the production binary switch with $\kappa=1$
recovers RICCA-like profiles in 1D and 2D (Section~\ref{sec:2d-rh}) without
erasing the contact/acoustic equation split.
We therefore present RICCA--RH as a documented alternative dissipation law---not
as a claim that continuous RH weighting alone solves oscillation control on
this suite.

%==============================================================================
\section{Conclusions}
\label{sec:conc}

We have presented and evaluated a centred, Riemann-solver-free baseline for a hyperbolic two-phase mixture model with relative velocity. The baseline couples RICCA dual-speed dissipation with primitive-variable WENO3/WENO5 reconstruction and is assessed on Zeidan's 1D gas--magma suite and 2D gas--liquid benchmarks, with FORCE as a centred reference and HLLC/Roe as upwind references.

The main practical findings are:
\begin{itemize}
\item \textbf{First-order contact behaviour:} RICCA remains less dissipative than FORCE at material contacts; RICCA--RH (production binary switch) closely tracks RICCA while providing a documented alternative dissipation law.
\item \textbf{High-order regime:} with WENO5, centred flux choices (RICCA, RICCA--RH, FORCE) give near-identical smooth profiles on the reported suite, so differences are largely confined to first-order contact/roughness metrics.
\item \textbf{Robustness tooling:} the combo hybrid sensor and the operational invariant-domain (IDP) face limiter on $(\rho,\alpha,c)$ reduce ringing relative to pressure-only blending on the tested problems; a soft RH weighting with $\kappa\ll1$ is under-dissipative and can fail dramatically.
\item \textbf{GPU path:} a resident 2D Apple Metal implementation supports RICCA, RICCA--RH and FORCE and delivers up to $3.9\times$ wall-clock speedup on an Apple M2 laptop on completed CPU/Metal comparisons ($3.93\times$ on $101\times101$ FORCE FO, $2.86\times$ on coarse FFS); broader CPU baselines across grids and schemes (including RICCA--WENO5) are left to future work.
\end{itemize}

Smooth manufactured advection confirms that WENO5 recovers high-order accuracy away from discontinuities, while discontinuous self-convergence remains globally first-order, as expected for shock-capturing schemes.

\paragraph{Limitations.}
The following restrictions delimit how far the conclusions should be
extrapolated---many of them intentional for a first detailed study:
\begin{enumerate}
\item[(a)] \textbf{Scope of physics and tests.}
      Relaxation/source terms are omitted; the suite is the Zeidan Riemann /
      CFD2019 matrix rather than strongly stratified industrial regimes.
      Several 2D benchmarks use equal-EOS simplifications, and the Mach~3 FFS
      is effectively single-velocity ($c\equiv\alpha$,
      $\mathbf{u}_r=\mathbf{0}$), so two-velocity validation rests mainly on the
      expansion and collision tubes.
\item[(b)] \textbf{Heuristic numerics.}
      The combo hybrid and the RICCA--RH binary switch are empirical; soft
      continuous RH weighting with $\kappa\ll1$ failed on these shocks
      (Section~\ref{sec:1d-rh}).
      Characteristic WENO uses mixture acoustics rather than a full Romenski
      eigenstructure, and the Zhang--Shu IDP limiter is operational rather than
      a proved CFL-independent invariant-domain update for nonlinear
      WENO--RICCA.
\item[(c)] \textbf{Incomplete upwind / high-resolution comparisons.}
      HLLC and Roe are included as FO/WENO upwind references
      (Sections~\ref{sec:hllc}--\ref{sec:1d-upwind}), but production HLLC uses a
      Rusanov-stable fallback on strong shocks, and Roe is field-by-field rather
      than a full $R\lvert\Lambda\rvert R^{-1}$ eigen-decomposition.
\item[(d)] \textbf{Accuracy on shocks.}
      Global accuracy remains first-order in the presence of discontinuities;
      we do not claim higher formal order via specialised reconstruction or AMR.
\item[(e)] \textbf{GPU performance.}
      Profiling is limited to Metal wall-clock timings on a single Apple
      platform, without CUDA/HIP cross-checks or large-grid occupancy /
      bandwidth studies.
\end{enumerate}
Within these stated limits, the presented RICCA--WENO/IDP pathway (with RICCA--RH as a companion option) provides a reproducible, eigenstructure-light centred reference that is competitive with FORCE on the Zeidan suite and amenable to GPU-resident implementation.

Future work includes a full Roe eigenstructure
($R\lvert\Lambda\rvert R^{-1}$) and a stricter three-wave HLLC without Rusanov
fallback, a Guermond--Popov cell-update IDP proof for WENO--RICCA, systematic
sensor and $\kappa$/RH threshold calibration across Mach/slip/EOS regimes, an
FFS variant with nontrivial inlet slip, relaxation/source terms, larger-grid
Metal/CUDA scaling with occupancy metrics, and tighter long-time analysis of
relative-velocity oscillations across contacts.

%==============================================================================
\section*{Acknowledgements}

The authors thank colleagues and collaborators who supported the development
of the 1D/2D mixture solvers used for the computations in this paper.

%==============================================================================
\bibliographystyle{elsarticle-num}
\bibliography{references}

\end{document}